\pdfoutput=1
\documentclass[twocolumn]{jpsj3}
\usepackage{bm}
\usepackage{tabularx}
\usepackage{ascmac}
\usepackage{amsmath}
\usepackage{wrapfig}
\usepackage{graphicx}
\usepackage{float}
\usepackage[FIGBOTCAP]{subfigure}
\usepackage{color}

\usepackage{widetext}

\bmdefine{\bdi}{i}
\bmdefine{\bdj}{j}
\bmdefine{\bdx}{x}
\bmdefine{\bdy}{y}
\bmdefine{\bdr}{r}
\bmdefine{\bdR}{R}
\bmdefine{\bdS}{S}
\bmdefine{\bdL}{L}
\bmdefine{\bds}{s}
\bmdefine{\bdl}{l}
\bmdefine{\bdJ}{J}
\bmdefine{\bdA}{A}
\bmdefine{\bdE}{E}
\bmdefine{\bdD}{D}
\bmdefine{\bdQ}{Q}
\bmdefine{\bdq}{q}
\bmdefine{\bdk}{k}
\bmdefine{\bdzero}{0}
\bmdefine{\bdv}{v}
\bmdefine{\bde}{e}
\bmdefine{\bdb}{b}
\bmdefine{\bddelta}{\delta}

\title{Onsager Reciprocal Relations
  for Charge and Spin Transport
  in Periodically Driven Systems}

\author{Naoya Arakawa$^{1}$\thanks{E-mail address: arakawa@phys.chuo-u.ac.jp},
Kenji Yonemitsu$^{1,2}$}
\inst{$^{1}$The Institute of Science and Engineering,
  Chuo University, Bunkyo, Tokyo, 112-8551, Japan\\
  $^{2}$Department of Physics,
  Chuo University, Bunkyo, Tokyo 112-8551, Japan}

\abst{
  A time-periodic driving field can be used to  
  generate and control transport phenomena.
  Any transport coefficients in the linear-response regime
  are restricted by the Onsager reciprocal relations, 
  but these relations in periodically driven systems have been poorly understood.
  In particular,
  the Onsager reciprocal relation in spin transport of periodically driven systems
  is lacking 
  despite the fact that
  it guarantees the detection of a spin current via the inverse spin Hall effect.
  Here we establish the Onsager reciprocal relations
  for charge and spin transport in periodically driven systems.
  We consider
  the time-averaged charge and spin off-diagonal dc conductivities
  $\sigma_{yx}^{\textrm{C}}$ and $\sigma_{yx}^{\textrm{S}}$,
  which are the transport coefficients for the charge and spin currents, respectively,
  perpendicular to the probe electric field 
  in the nonequilibrium steady state with the pump field of light. 
  First, we argue the Onsager reciprocal relations
  for these conductivities
  with the pump field of circularly, linearly, or bicircularly polarized light.
  We show that
  $\sigma_{yx}^{\textrm{C}}$ and $\sigma_{yx}^{\textrm{S}}$ satisfy 
  the Onsager reciprocal relations in all the cases considered,
  but their main terms depend on the polarization of light. 
  In the case with circularly or linearly polarized light,
  $\sigma_{yx}^{\textrm{C}}$ is restricted to the antisymmetric or symmetric part, respectively,
  whereas
  $\sigma_{yx}^{\textrm{S}}$ is restricted to the antisymmetric part.
  Meanwhile,
  in the case with bicircularly polarized light,
  $\sigma_{yx}^{\textrm{C}}$ and $\sigma_{yx}^{\textrm{S}}$
  are not restricted 
  to either the antisymmetric or symmetric parts generally.
  Then, we numerically study the Onsager reciprocal relations
  for $\sigma_{yx}^{\textrm{C}}$ and $\sigma_{yx}^{\textrm{S}}$
  in periodically driven Sr$_{2}$RuO$_{4}$ 
  using the Floquet linear-response theory.
  Our numerical calculations validate our general arguments.
  Therefore,
  the spin current generated in periodically driven systems
  is detectable by the inverse spin Hall effect.  
  Our numerical calculations also show that 
  $\sigma_{yx}^{\textrm{C}}$ cannot necessarily be regarded as the anomalous Hall conductivity
  even with broken time-reversal symmetry,
  whereas 
  $\sigma_{yx}^{\textrm{S}}$ can be regarded as the spin Hall conductivity
  in all the cases considered.
  Our results suggest that
  it is highly required to check whether or not
  the charge and spin off-diagonal conductivities
  are dominated by the antisymmetric parts in discussing
  the anomalous Hall and spin Hall effects, respectively. 
  This study will become a cornerstone
  of theoretical and experimental studies of 
  transport phenomena in periodically driven systems. 
}

\begin{document}
\maketitle

\section{Introduction}

Periodically driven systems
have opened a new way for light-induced and light-controlled transport phenomena. 
In general, systems are periodically driven by 
a time-periodic field such as the pump field of light.
Because of the time periodicity,
the periodically driven systems are described
by the Floquet Hamiltonian~\cite{Floquet1,Floquet2}.
Since this Hamiltonian can be tuned by varying
the parameters of the pump field,
various properties of the systems can be engineered
without changing the materials. 
This is called Floquet
engineering~\cite{Floquet-review1,Floquet-review2,Floquet-review3,NA-Floquet-Mott1}. 
For example,
the pump field of circularly polarized light (CPL) can be used to induce 
the anomalous Hall effect (AHE)~\cite{Oka-PRB,Mikami,Light-AHE-exp1,Light-AHE-exp2,NA-FloquetSHE},
in which  
the charge current perpendicular to the probe field
is generated~\cite{AHE-Hall,Karplus-Luttinger,AHE-review}.
Furthermore,
the charge current generated in this AHE
can be changed in magnitude and direction
by varying
the amplitude and helicity of CPL~\cite{Oka-PRB,Mikami,Light-AHE-exp2,NA-FloquetSHE}.
Such optical control is also possible for
the spin Hall effect (SHE)~\cite{NA-FloquetSHE,NA-TMD},
in which 
the spin current, the flow of the spin angular momentum,
is generated with the probe field perpendicular to it~\cite{SHE-JETP,SHE-Hirsch,SHE-review}.

Despite various studies of transport phenomena in periodically driven systems,
the Onsager reciprocal relations in these systems
have been poorly understood.
In general,
the Onsager reciprocal relations connect two transport coefficients~\cite{Onsager1,Onsager2,Kubo}.
Therefore, these relations provide
general constraints on the symmetry of transport coefficients.
Such constraints are useful for theoretical and experimental studies of
transport phenomena. 
Moreover,
the Onsager reciprocal relations are extremely important for spin transport.
In general,
it is much harder to detect the spin current than the charge current.
Because of this,
the spin current is usually detected indirectly in experiments.
In fact,
the observation of the inverse SHE can be regarded
as the existence of the SHE~\cite{InvSHE1,InvSHE2}
because
their transport coefficients are connected by 
the Onsager reciprocal relation;
in the inverse SHE,
the charge current perpendicular to the spin current is generated~\cite{InvSHE1,InvSHE2}. 
Although there are some studies of the Onsager reciprocal relations
in periodically driven systems~\cite{Onsager-PD1,Onsager-PD2,Onsager-PD3,NA-Onsager},
that relation for spin transport has been unexplored yet.

\begin{figure}
  \begin{center}
    \includegraphics[width=86mm]{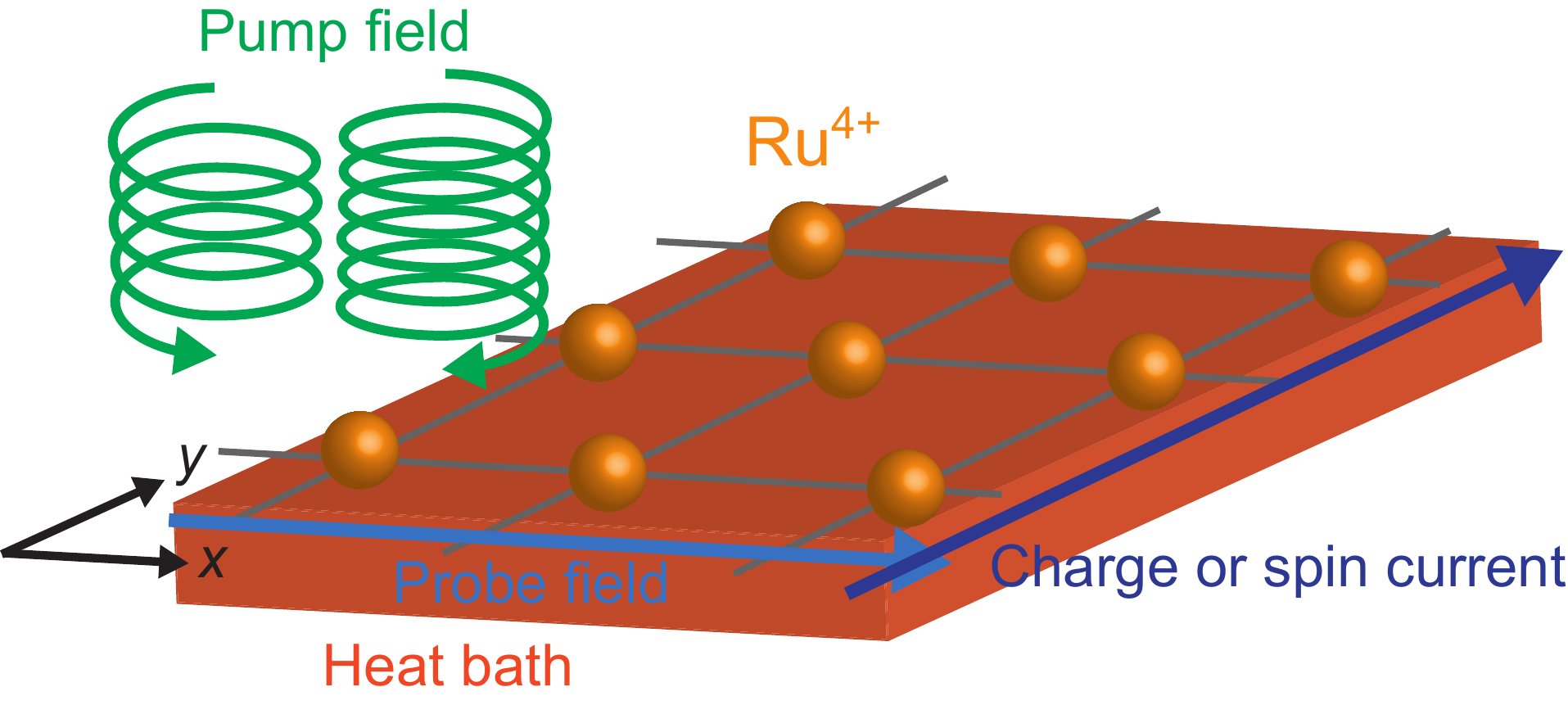}
    \end{center}
  \caption{\label{fig1}
    (Color online) The setup for the pump-probe measurements of
    $\sigma_{yx}^{\textrm{C}}$ and $\sigma_{yx}^{\textrm{S}}$
    in our periodically driven systems.
    $\sigma_{yx}^{\textrm{C}}$ and $\sigma_{yx}^{\textrm{S}}$
    are the transport coefficients for describing 
    the charge and spin currents, respectively, along the $y$ axis,
    which are perpendicular to the probe field applied along the $x$ axis
    in the nonequilibrium steady state with the pump field.
    The system is periodically driven Sr$_{2}$RuO$_{4}$
    and is coupled to the heat bath. 
    In Sr$_{2}$RuO$_{4}$, Ru$^{4+}$ ions form the square lattice.
    This panel shows the case with the pump field of BCPL;
    the cases with those of CPL and LPL are also considered.
  }
\end{figure}

\begin{figure}
  \begin{center}
    \includegraphics[width=86mm]{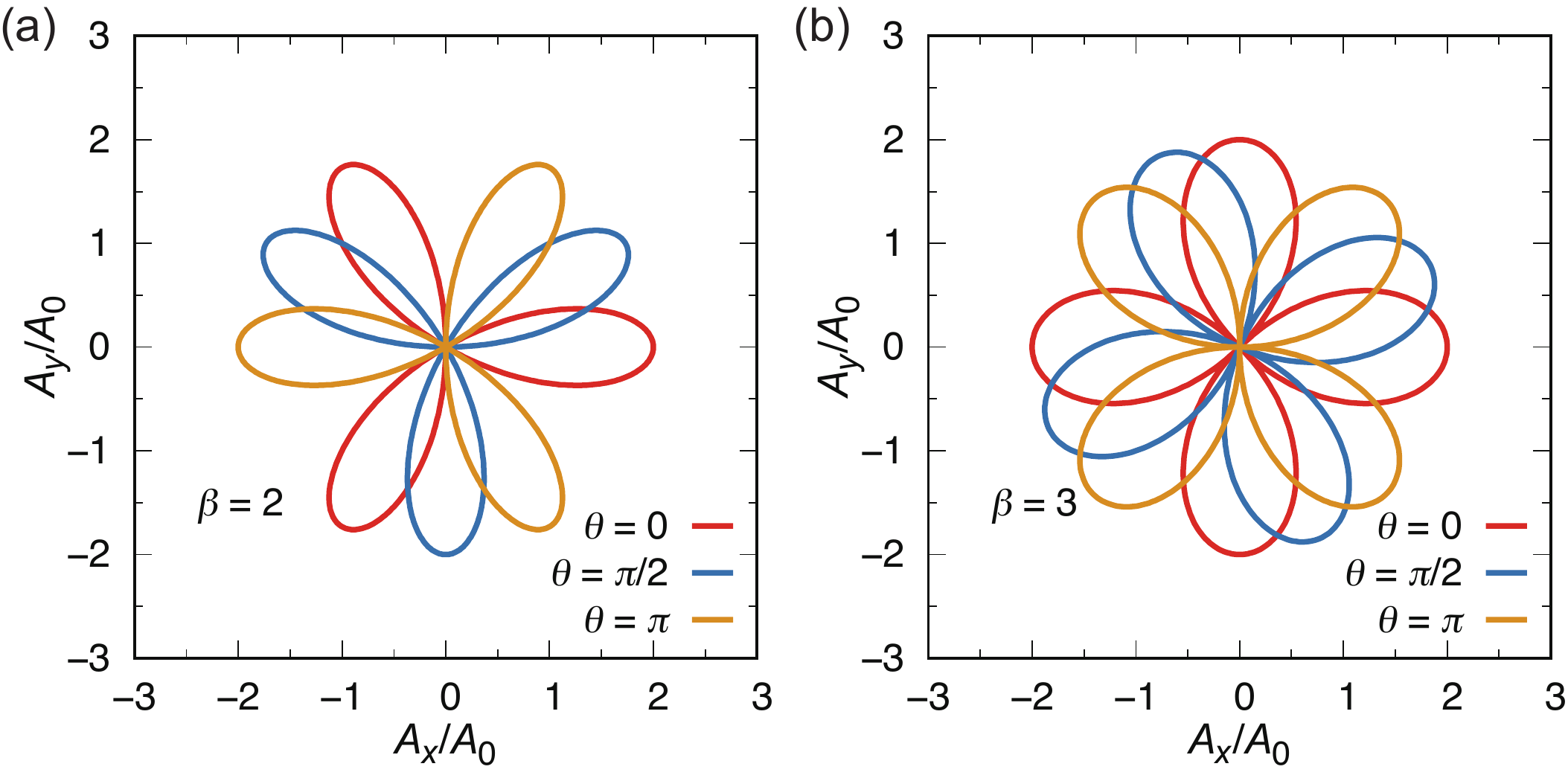}
    \end{center}
  \caption{\label{fig2}
    (Color online) The trajectories of the pump field of BCPL
    at (a) $\beta=2$ and (b) $\beta=3$
    and $\theta=0$, $\frac{\pi}{2}$, or $\pi$.
    The three or four loops with the same color are obtained per period
    of the pump field 
    at $\beta=2$ or $3$, respectively.
  }
\end{figure}

In this paper, 
we theoretically study the Onsager reciprocal relations
for charge and spin transport in periodically driven systems.
The main results are summarized in Table \ref{table1}.
We consider 
the time-averaged charge and spin off-diagonal dc conductivities
$\sigma_{yx}^{\textrm{C}}$ and $\sigma_{yx}^{\textrm{S}}$ to describe 
the charge and spin currents, respectively,
perpendicular to the probe electric field in the linear-response regime 
for the nonequilibrium steady state of an electron system
driven by the pump field of light.
Figure \ref{fig1} shows the set-up for their measurements.
We begin with general arguments about the Onsager reciprocal relations
for $\sigma_{yx}^{\textrm{C}}$ and $\sigma_{yx}^{\textrm{S}}$ 
in the systems driven by CPL, linearly polarized light (LPL),
or bicircularly polarized light (BCPL).
Here 
BCPL consists of a linear combination of
the left- and right-handed CPL with different frequencies $\Omega$ and $\beta\Omega$
and a relative phase difference $\theta$~\cite{BCPL-NatPhoto,BCPL-Oka,NA-BCPL}
(see Fig. \ref{fig2}).
We show that
$\sigma_{yx}^{\textrm{C}}$ and $\sigma_{yx}^{\textrm{S}}$ 
satisfy the Onsager reciprocal relations in all the cases considered,
although their main terms depend on the polarization of light. 
In the case with CPL or LPL, 
the main term of $\sigma_{yx}^{\textrm{C}}$ is
given by the antisymmetric or symmetric part, respectively,
whereas
that of $\sigma_{yx}^{\textrm{S}}$ is given by the antisymmetric part.
This suggests that
the antisymmetric part of $\sigma_{yx}^{\textrm{S}}$ can be finite even with time-reversal symmetry,
whereas that of $\sigma_{yx}^{\textrm{C}}$ is finite only without it. 
This difference arises from the
difference between the time-reversal symmetries of the charge and spin currents.
Meanwhile,
in the case with BCPL,
the Onsager reciprocal relations do not restrict
$\sigma_{yx}^{\textrm{C}}$ and $\sigma_{yx}^{\textrm{S}}$
to either the antisymmetric or symmetric parts generally.
This unusual property is due to
the lack of a simple relation between the pump field of BCPL
and its time-reversal counterpart.
Then,
we numerically test 
the Onsager reciprocal relations for 
$\sigma_{yx}^{\textrm{C}}$ and $\sigma_{yx}^{\textrm{S}}$
by applying the Floquet linear-response
theory~\cite{NA-FloquetSHE}
to a model of Sr$_{2}$RuO$_{4}$ driven by CPL, LPL, or BCPL
with weak coupling to a heat bath.
We demonstrate the validity of our arguments. 
Therefore,
our results indicate that, even for periodically driven systems,
the spin current can be detected by the inverse SHE.
This is useful to develop and observe many spintronics phenomena
in periodically driven systems.  
Then, our numerical calculations with BCPL 
show that
the main term of $\sigma_{yx}^{\textrm{S}}$ is given by the antisymmetric part.
Combining these results with the results with CPL or LPL,
we conclude that 
$\sigma_{yx}^{\textrm{S}}$ can be regarded as the spin Hall conductivity
in all the cases considered.
Our numerical calculations with BCPL also show that
the main term
of $\sigma_{yx}^{\textrm{C}}$ depends on the magnitude, $\beta$, and $\theta$ of
the pump field.
More precisely,
$\sigma_{yx}^{\textrm{C}}$ with BCPL for weak magnitude is dominated by the antisymmetric part,
whereas that for moderately strong magnitude is
almost vanishing in the cases of
$(\beta,\theta)=(2,0)$, $(2,\frac{\pi}{2})$, $(2,\pi)$, $(3,0)$, and $(3,\pi)$
or dominated by the symmetric part in the cases of
$(\beta,\theta)=(2,\frac{\pi}{4})$, $(2,\frac{3\pi}{4})$,
$(3,\frac{\pi}{4})$, $(3,\frac{\pi}{2})$, and $(3,\frac{3\pi}{4})$.
Therefore,
$\sigma_{yx}^{\textrm{C}}$ cannot necessarily be
regarded as the anomalous Hall conductivity
even with broken time-reversal symmetry.
Our results suggest that
it is necessary to check the main terms of $\sigma_{yx}^{\textrm{C}}$ and $\sigma_{yx}^{\textrm{S}}$
in discussing the AHE and SHE, respectively. 

The remainder of this paper is organized as follows.
In Sect. 2,
we argue the Onsager reciprocal relations
in nondriven and periodically driven systems.
After reviewing these relations in nondriven systems,
we derive the Onsager reciprocal relations for
$\sigma_{yx}^{\textrm{C}}$ and $\sigma_{yx}^{\textrm{S}}$
in the electron systems driven by CPL, LPL, or BCPL.
In Sect. 3,
we introduce the model of periodically driven Sr$_{2}$RuO$_{4}$.
In this model,
we consider the heat bath
as well as the system of Sr$_{2}$RuO$_{4}$ driven by the pump field of light
and suppose that
a nonequilibrium steady state is realized
due to the damping induced by the coupling to the heat bath.
The reason why 
we choose Sr$_{2}$RuO$_{4}$ is twofold:
its realistic model possesses the finite spin off-diagonal dc conductivity,
which is more difficult to be realized
than the finite charge off-diagonal dc conductivity;
and 
its electronic structure is so simple that
the Onsager reciprocal relations for these conductivities
can be numerically studied even in the case with BCPL. 
In Sect. 4,
we formulate $\sigma_{yx}^{\textrm{C}}$ and $\sigma_{yx}^{\textrm{S}}$
as well as their counterparts appearing in the Onsager reciprocal relations
using the Floquet linear-response theory.
We also comment on the applicability of this theory.
In Sect. 5,
we show the numerical results of the Onsager reciprocal relations
for $\sigma_{yx}^{\textrm{C}}$ and $\sigma_{yx}^{\textrm{S}}$
in Sr$_{2}$RuO$_{4}$ driven by CPL, LPL, or BCPL at $\beta=2$ and $3$.
In Sect. 6,
we discuss the origin of the characteristic $\theta$ dependences
of $\sigma_{yx}^{\textrm{C}}$ and $\sigma_{yx}^{\textrm{S}}$
in the cases with BCPL,
compare our results with other relevant studies,
and remark on the experimental realization of our results.
In Sec. VII,
we make some concluding remarks on implications and outlooks.

\begin{table}
  \caption{\label{table1}
    The Onsager reciprocal relations for
    $\sigma_{yx}^{\textrm{C}}$ and $\sigma_{yx}^{\textrm{S}}$
    in the nonequilibrium steady states of systems driven by CPL, LPL, and BCPL.
    $\sigma_{yx}^{\textrm{C}}$ and $\sigma_{yx}^{\textrm{S}}$ are the transport coefficients
    for the charge and spin currents, respectively, along the $y$ axis
    perpendicular to the probe electric field applied along the $x$ axis
    with the pump field $\mathbf{A}_{\textrm{pump}}(t)$;
    $\sigma_{xy}^{\textrm{C}}$ and $\tilde{\sigma}_{xy}^{\textrm{S}}$
    are those for the charge currents along the $x$ axis perpendicular
    to the probe electric and spin fields, respectively, along the $y$ axis
    with $\mathbf{A}_{\textrm{pump}}(t)$;
    and $\bar{\sigma}_{xy}^{\textrm{C}}$ and $\bar{\tilde{\sigma}}_{xy}^{\textrm{S}}$
    are those for the charge currents along the $x$ axis perpendicular
    to the probe electric and spin fields, respectively, along the $y$ axis
    with $\mathbf{A}_{\textrm{pump}}(-t)$.
    In the case with CPL or LPL,
    $\sigma_{yx}^{\textrm{C}}$ is restricted to
    the antisymmetric part $(\sigma_{yx}^{\textrm{C}}-\sigma_{xy}^{\textrm{C}})/2$
    (corresponding to the anomalous Hall conductivity)
    or the symmetric part $(\sigma_{yx}^{\textrm{C}}+\sigma_{xy}^{\textrm{C}})/2$, respectively,
    whereas $\sigma_{yx}^{\textrm{S}}$ is restricted to
    the antisymmetric part $(\sigma_{yx}^{\textrm{S}}-\tilde{\sigma}_{xy}^{\textrm{S}})/2$
    (corresponding to the spin Hall conductivity).
    (For the reason
    why we have defined the antisymmetric part of $\sigma_{yx}^{\textrm{S}}$ in this way,
    read the last paragraph of Sect. 2.2.1.)
    Meanwhile, in the case with BCPL,
    $\sigma_{yx}^{\textrm{C}}$ and $\sigma_{yx}^{\textrm{S}}$ are restricted
    to neither the antisymmetric nor symmetric parts generally. 
    \vspace{8pt}
  }
  \begin{center}
    \begin{tabular}{cccc}
      \hline
      & CPL & LPL & BCPL \\[0pt] \hline\\[-10pt]    
      $\sigma_{yx}^{\textrm{C}}$
      & $\sigma_{yx}^{\textrm{C}}=-\sigma_{xy}^{\textrm{C}}$
      & $\sigma_{yx}^{\textrm{C}}=\sigma_{xy}^{\textrm{C}}$
      & $\sigma_{yx}^{\textrm{C}}=\bar{\sigma}_{xy}^{\textrm{C}}$\\[4pt]
      $\sigma_{yx}^{\textrm{S}}$
      & $\sigma_{yx}^{\textrm{S}}=-\tilde{\sigma}_{xy}^{\textrm{S}}$
      & $\sigma_{yx}^{\textrm{S}}=-\tilde{\sigma}_{xy}^{\textrm{S}}$
      & $\sigma_{yx}^{\textrm{S}}=-\bar{\tilde{\sigma}}_{xy}^{\textrm{S}}$\\[3pt]
      \hline
    \end{tabular}
  \end{center}
\end{table}

\section{Onsager reciprocal relations}

We begin with general argument about the Onsager reciprocal relations
in nondriven or periodically driven systems.
We argue these relations for
the charge and spin off-diagonal dc conductivities in Sects. 2.1 and 2.2,
respectively. 
The results for periodically driven systems are summarized in Table \ref{table1}.
Although the Onsager reciprocal relations in nondriven systems may be well known,
  we will review them below
  to help the readers understand our general arguments in periodically driven systems.
  Since the essential symmetry in discussing the Onsager reciprocal relations
  is the symmetry against a time-reversal operation~\cite{Onsager1,Onsager2,Kubo},
  we will not specify whether a certain vector is a vector or pseudo-vector.

\begin{figure}
  \begin{center}
    \includegraphics[width=86mm]{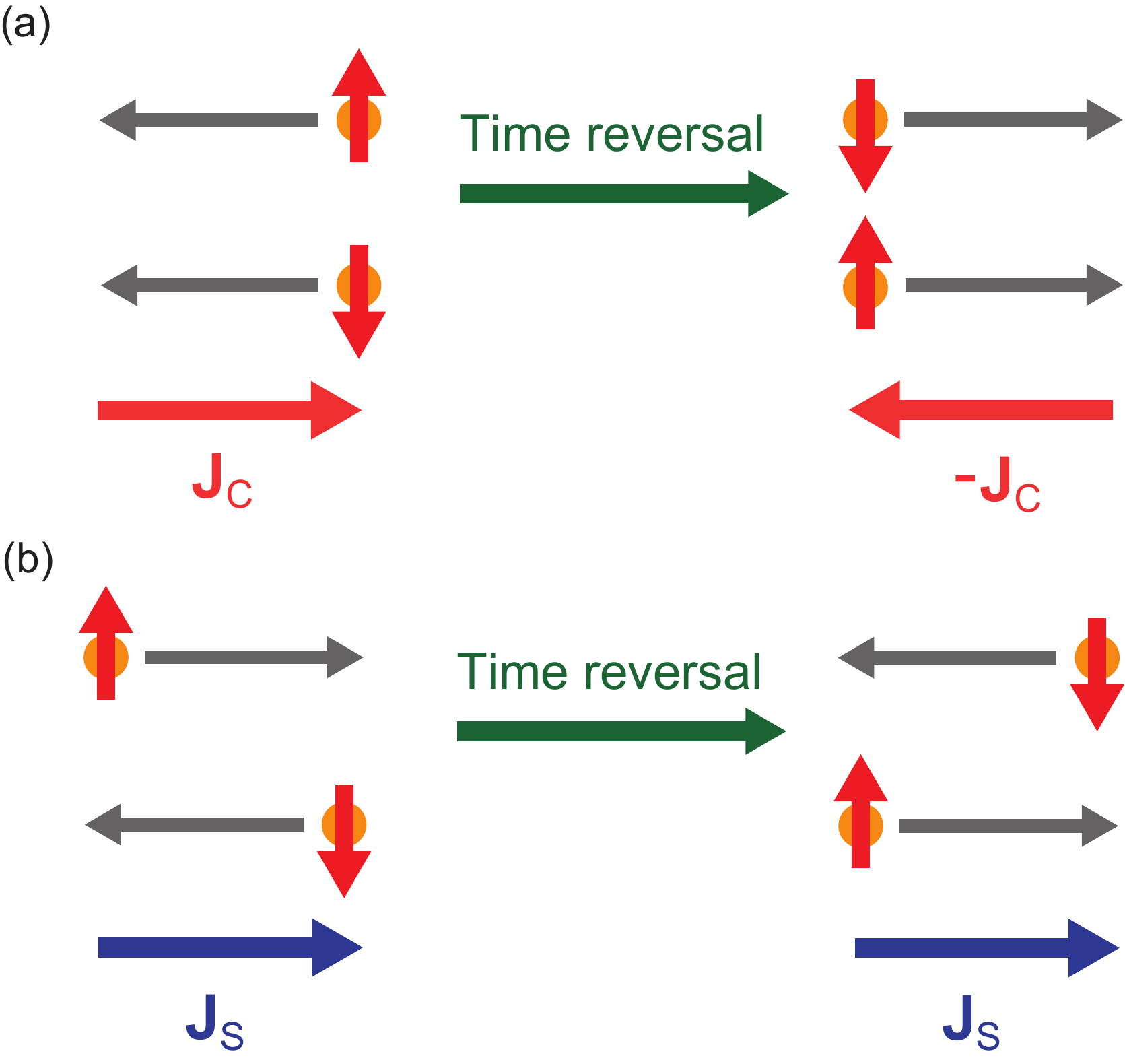}
    \end{center}
  \caption{\label{fig3}
    (Color online) The changes in (a) the charge current $\mathbf{J}_{\textrm{C}}$
    and (b) the spin current $\mathbf{J}_{\textrm{S}}$
    under the time-reversal operation.
    The up or down arrows represent the spin-up or spin-down electrons, respectively. 
    Here $\mathbf{J}_{\textrm{C}}$ and $\mathbf{J}_{\textrm{S}}$ are defined as 
    $\mathbf{J}_{\textrm{C}}=(-e)(\mathbf{J}_{\uparrow}+\mathbf{J}_{\downarrow})$
    and $\mathbf{J}_{\textrm{S}}=(\hbar/2)(\mathbf{J}_{\uparrow}-\mathbf{J}_{\downarrow})$,
    where $\mathbf{J}_{\uparrow}$ and $\mathbf{J}_{\downarrow}$
    are the currents of the spin-up and spin-down electrons, respectively.
    Under the time-reversal operation,
    $\mathbf{J}_{\uparrow}\rightarrow -\mathbf{J}_{\downarrow}$ and
    $\mathbf{J}_{\downarrow}\rightarrow -\mathbf{J}_{\uparrow}$,
    resulting in $\mathbf{J}_{\textrm{C}}\rightarrow -\mathbf{J}_{\textrm{C}}$
    and $\mathbf{J}_{\textrm{S}}\rightarrow \mathbf{J}_{\textrm{S}}$.
  }
\end{figure}

As we will show below,
the charge and spin off-diagonal dc conductivities
satisfy the Onsager reciprocal relations in all the cases considered,
although there are some essential differences
due to the different time-reversal symmetries of the charge and spin currents
(Fig. \ref{fig3}). 
The following results are valid in general
as long as
the probe field can be described in the linear-response theory.

\begin{figure*}[h]
  \begin{center}
    \includegraphics[width=160mm]{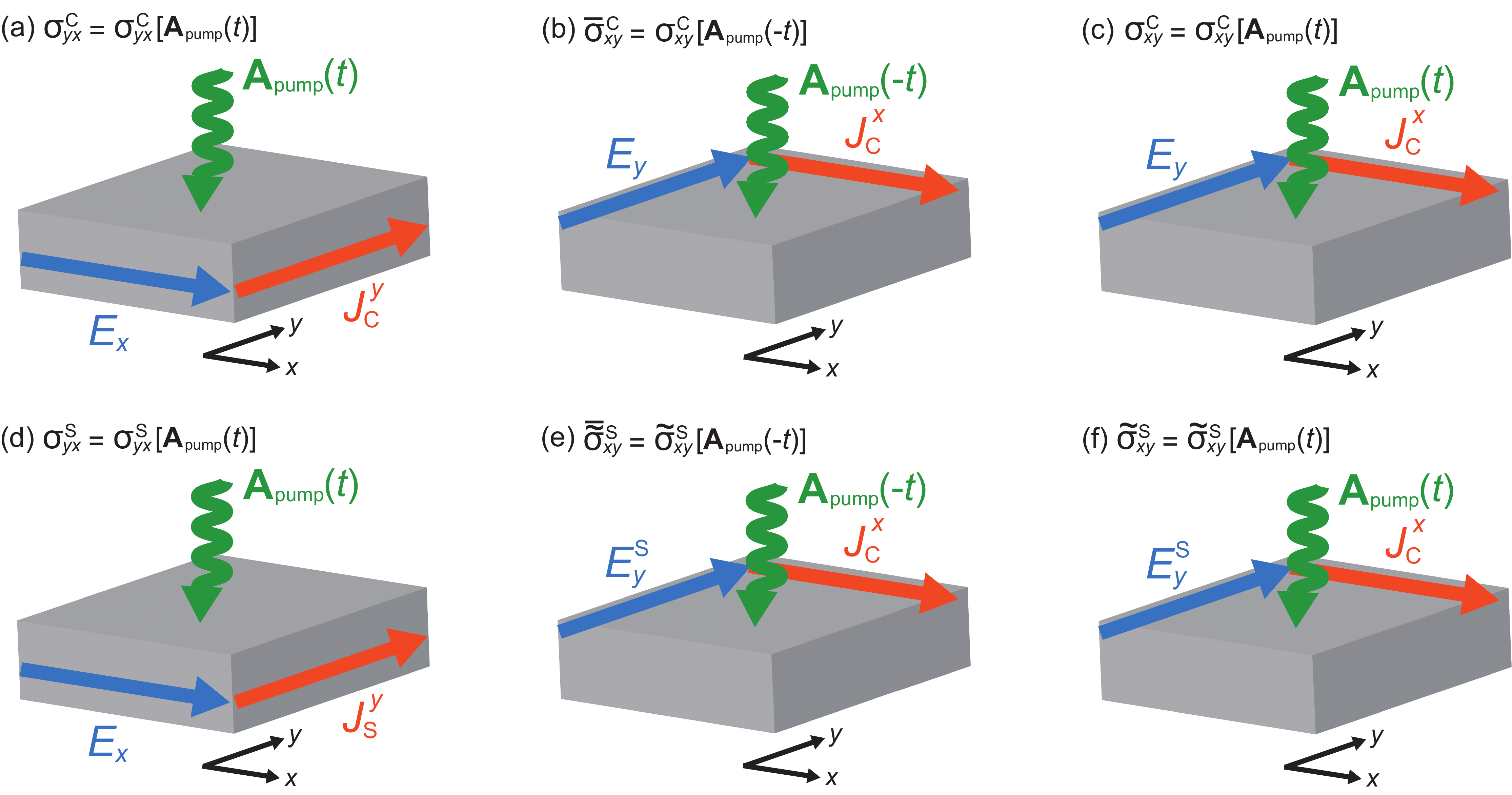}
    \end{center}
  \caption{\label{fig4}
    (Color online) The transport phenomena described by
    (a) $\sigma_{yx}^{\textrm{C}}=\sigma_{yx}^{\textrm{C}}[\mathbf{A}_{\textrm{pump}}(t)]$,
    (b) $\bar{\sigma}_{xy}^{\textrm{C}}=\sigma_{xy}^{\textrm{C}}[\mathbf{A}_{\textrm{pump}}(-t)]$,
    (c) $\sigma_{xy}^{\textrm{C}}=\sigma_{xy}^{\textrm{C}}[\mathbf{A}_{\textrm{pump}}(t)]$,
    (d) $\sigma_{yx}^{\textrm{S}}=\sigma_{yx}^{\textrm{S}}[\mathbf{A}_{\textrm{pump}}(t)]$,
    (e) $\bar{\tilde{\sigma}}_{xy}^{\textrm{S}}
    =\tilde{\sigma}_{xy}^{\textrm{S}}[\mathbf{A}_{\textrm{pump}}(-t)]$,
    and (f) $\tilde{\sigma}_{xy}^{\textrm{S}}=\tilde{\sigma}_{xy}^{\textrm{S}}[\mathbf{A}_{\textrm{pump}}(t)]$.
    The blue arrows represent the probe electric fields $E_{x}$ and $E_{y}$
    and the probe spin field $E_{y}^{\textrm{S}}$. 
    Here $E_{y}^{\textrm{S}}$ is given by, for example, 
    the gradient of either the Zeeman field or
    a spin-dependent chemical potential;
    it can induce a spin current
    as the probe electric field can induce a charge current.
    The orange arrows represent
    the charge currents $J_{\textrm{C}}^{y}$ and $J_{\textrm{C}}^{x}$
    and the spin current $J_{\textrm{S}}^{y}$.
    The green arrows represent the pump fields $\mathbf{A}_{\textrm{pump}}(t)$
    and $\mathbf{A}_{\textrm{pump}}(-t)$.
    The $x$ and $y$ axes are also drawn in these panels.
  }
\end{figure*}

\subsection{Charge off-diagonal dc conductivity}

We argue the Onsager reciprocal relations 
for the charge off-diagonal dc conductivity
in nondriven or periodically driven systems.
In the following arguments,
we suppose that 
this conductivity is given by the correlation function
between the charge current operators.
This is valid as long as the probe field is of linear response.

\subsubsection{Nondriven systems}

First, we consider the nondriven case with a magnetic field
$\mathbf{H}$.
In general,
the Onsager reciprocal relation connects 
the transport coefficient for a certain transport phenomenon
with another for the transport phenomenon
obtained by applying the time-reversal operation to
the original phenomenon~\cite{Onsager1,Onsager2,Kubo}.
Therefore, 
the Onsager reciprocal relation for the charge off-diagonal dc conductivity with $\mathbf{H}$
is given by~\cite{Onsager1,Onsager2,Kubo}
\begin{align}
  %No. XXXXIII 9/22 No. 01
  \sigma_{yx}^{\textrm{C}}(\mathbf{H})
  =(-1)^{2}\sigma_{xy}^{\textrm{C}}(-\mathbf{H}),\label{eq:Onsager-H_start}
\end{align}
where
$\sigma_{yx}^{\textrm{C}}(\mathbf{H})$
is the charge off-diagonal dc conductivity for
the charge current $J_{\textrm{C}}^{y}$ generated perpendicular
to the probe electric field $E_{x}$
in the presence of $\mathbf{H}$
[i.e., $J_{\textrm{C}}^{y}=\sigma_{yx}^{\textrm{C}}(\mathbf{H})E_{x}$],
and $\sigma_{xy}^{\textrm{C}}(-\mathbf{H})$
is that for the charge current $J_{\textrm{C}}^{x}$ generated perpendicular
to the probe electric field $E_{y}$
in the presence of $-\mathbf{H}$
[i.e., $J_{\textrm{C}}^{x}=\sigma_{xy}^{\textrm{C}}(-\mathbf{H})E_{y}$]. 
The factor $(-1)^{2}$ in Eq. (\ref{eq:Onsager-H_start})
arises from
the sign changes in the charge current operators appearing
in the charge off-diagonal dc conductivity
under the time-reversal operation [Fig. \ref{fig3}(a)].
Then,
we suppose that
the charge off-diagonal dc conductivity satisfies   
$\sigma_{xy}^{\textrm{C}}(-\mathbf{H})=-\sigma_{xy}^{\textrm{C}}(\mathbf{H})$,
which usually holds in the case with $\mathbf{H}=(0\ 0\ H)^{T}$.
By combining this relation with Eq. (\ref{eq:Onsager-H_start}),
the Onsager reciprocal relation can be reduced to
\begin{align}
  %No. XXXXIII 9/22 No. 01
  \sigma_{yx}^{\textrm{C}}(\mathbf{H})
  =-\sigma_{xy}^{\textrm{C}}(\mathbf{H}).\label{eq:Onsager-H_last}
\end{align}
This is often called the Onsager reciprocal relation, but
it has been derived from a combination of
the Onsager reciprocal relation Eq. (\ref{eq:Onsager-H_start}) and
the additional symmetric property
[i.e., $\sigma_{xy}^{\textrm{C}}(-\mathbf{H})=-\sigma_{xy}^{\textrm{C}}(\mathbf{H})$].
This point is important to discuss
whether a certain transport coefficient satisfies the Onsager reciprocal relation.
If there is no such additional symmetric property,
whether
the Onsager reciprocal relation holds 
should be discussed using the equation such as Eq. (\ref{eq:Onsager-H_start});
this is true in the case with BCPL, as we will show in Sect. 2.1.2.

Next, we consider the nondriven case with magnetization $\mathbf{M}$.
Since $\mathbf{M}$ breaks time-reversal symmetry (as $\mathbf{H}$ does), 
the similar argument is applicable to this case;
as a result, we have
\begin{align}
  %No. XXXXIII 9/22 No. 01
  \sigma_{yx}^{\textrm{C}}(\mathbf{M})
  =(-1)^{2}\sigma_{xy}^{\textrm{C}}(-\mathbf{M})
  =-\sigma_{xy}^{\textrm{C}}(\mathbf{M}).\label{eq:Onsager-M}
\end{align}
In driving this equation,
we have supposed that       
$\sigma_{xy}^{\textrm{C}}(-\mathbf{M})=-\sigma_{xy}^{\textrm{C}}(\mathbf{M})$ is satisfied.

In both cases,
the charge off-diagonal dc conductivity has only the antisymmetric part
$(\sigma_{yx}^{\textrm{C}}-\sigma_{xy}^{\textrm{C}})/2$
and thus can be regarded as the Hall conductivity,
which is described by the antisymmetric part.
This antisymmetric part
can be finite
only with broken time-reversal symmetry~\cite{Onsager1}.

\subsubsection{Periodically driven systems}

We now argue the Onsager reciprocal relations
in periodically driven systems in two dimensions. 
To do this,
we consider the nonequilibrium steady state
under the pump field $\mathbf{A}_{\textrm{pump}}(t)$ with a time period $T_{\textrm{p}}$
[i.e., $\mathbf{A}_{\textrm{pump}}(t+T_{\textrm{p}})=\mathbf{A}_{\textrm{pump}}(t)$]
and discuss the symmetric properties of
the time-averaged charge off-diagonal dc conductivity 
[for its definition, see Eq. (\ref{eq:t-av-sig}) for Q$=$C with Eq. (\ref{eq:t-av-sig-w})].
The following arguments can be extended
to any other periodically driven systems.

First, we discuss the Onsager reciprocal relation
with the pump field of left-handed CPL, $\mathbf{A}_{\textrm{LCPL}}(t)$,
where
\begin{align}
  %No. XXXXIII 9/22 No. 02
  \mathbf{A}_{\textrm{LCPL}}(t)=(A_{0}\cos\Omega t\ A_{0}\sin\Omega t)^{T},\label{eq:A_LCPL}
\end{align}
and $\Omega=2\pi/T_{\textrm{p}}$ is the light frequency.
In this case,
the Onsager reciprocal relation is given by 
\begin{align}
  %No. XXXXIII 9/22 No. 02
  \sigma_{yx}^{\textrm{C}}[\mathbf{A}_{\textrm{LCPL}}(t)]
  =(-1)^{2}\sigma_{xy}^{\textrm{C}}[\mathbf{A}_{\textrm{LCPL}}(-t)],\label{eq:Onsager_sigC-CPL_start}
\end{align}
where $\sigma_{\nu\eta}^{\textrm{C}}[\mathbf{A}_{\textrm{pump}}(t)]$
is the time-averaged charge off-diagonal dc conductivity
for the charge current $J_{\textrm{C}}^{\nu}$ generated perpendicular to
the probe field $E_{\eta}$ with $\mathbf{A}_{\textrm{pump}}(t)$ [Figs. \ref{fig4}(a){--}\ref{fig4}(c)].
Since $\mathbf{A}_{\textrm{LCPL}}(-t)$ is equal to the pump field of right-handed CPL,
\begin{align}
  %No. XXXXIII 9/22 No. 02
  \mathbf{A}_{\textrm{RCPL}}(t)=(A_{0}\cos\Omega t\ -A_{0}\sin\Omega t)^{T},
\end{align} 
Eq. (\ref{eq:Onsager_sigC-CPL_start}) is rewritten as 
\begin{align}
  %No. XXXXIII 9/22 No. 02
  \sigma_{yx}^{\textrm{C}}[\mathbf{A}_{\textrm{LCPL}}(t)]
  =\sigma_{xy}^{\textrm{C}}[\mathbf{A}_{\textrm{RCPL}}(t)].\label{eq:Onsager_sigC-CPL_next}
\end{align}
Furthermore,
we suppose that
the time-averaged charge off-diagonal dc conductivity changes its sign
by switching the helicity of light~\cite{NA-FloquetSHE,NA-Onsager}, 
\begin{align}
  %No. XXXXIII 9/22 No. 02
  \sigma_{xy}^{\textrm{C}}[\mathbf{A}_{\textrm{RCPL}}(t)]
  =-\sigma_{xy}^{\textrm{C}}[\mathbf{A}_{\textrm{LCPL}}(t)].\label{eq:sigC-helicity}
\end{align}
This property can be understood in terms of
the symmetry of the charge current under the time-reversal operation,
which switches the helicity of light~\cite{NA-FloquetSHE}.
Combining Eqs. (\ref{eq:Onsager_sigC-CPL_next}) and (\ref{eq:sigC-helicity}),
we can reduce the Onsager reciprocal relation in this case to 
\begin{align}
  %No. XXXXIII 9/22 No. 02
  \sigma_{yx}^{\textrm{C}}[\mathbf{A}_{\textrm{LCPL}}(t)]
  =-\sigma_{xy}^{\textrm{C}}[\mathbf{A}_{\textrm{LCPL}}(t)].\label{eq:Onsager_sigC-CPL}
\end{align}
Therefore,
the time-averaged charge off-diagonal dc conductivity with the pump field of CPL
is restricted to be antisymmetric.
This has been numerically confirmed in graphene driven by CPL~\cite{NA-Onsager}.

Meanwhile, 
in systems driven by LPL,
the time-averaged charge off-diagonal dc conductivity is restricted to be symmetric,
\begin{align}
  \sigma_{yx}^{\textrm{C}}[\mathbf{A}_{\textrm{LPL}}(t)]
  =\sigma_{xy}^{\textrm{C}}[\mathbf{A}_{\textrm{LPL}}(t)],\label{eq:Onsager_sigC-LPL}
\end{align}
where the pump field of LPL, $\mathbf{A}_{\textrm{LPL}}(t)$, is given by
\begin{align}
  \mathbf{A}_{\textrm{LPL}}(t)=(A_{0}\cos\Omega t\ A_{0}\cos\Omega t)^{T}.\label{eq:A_LPL}
\end{align}
Equation (\ref{eq:Onsager_sigC-LPL})
has been also numerically confirmed in graphene driven by LPL~\cite{NA-Onsager}.

The difference
between Eqs. (\ref{eq:Onsager_sigC-CPL}) and (\ref{eq:Onsager_sigC-LPL})
is due to the difference in time-reversal symmetry~\cite{NA-Onsager}.
Note that
CPL can break time-reversal symmetry~\cite{Light-TrevBreaking},
whereas LPL does not.
We will check Eqs. (\ref{eq:Onsager_sigC-CPL}) and (\ref{eq:Onsager_sigC-LPL})
for periodically driven Sr$_{2}$RuO$_{4}$ in Sect. 5.2.1.

The situation becomes different
in the presence of the pump field of BCPL~\cite{BCPL-NatPhoto,BCPL-Oka,NA-BCPL},
$\mathbf{A}_{\textrm{BCPL}}(t)=(A_{x}(t)\ A_{y}(t))^{T}$,
where 
\begin{align}
  %No. XXXXII 9/12 No. 01
  A_{x}(t)&=A_{0}[\cos\Omega t+\cos(\beta\Omega t-\theta)],\label{eq:Ax-BCPL}\\
  A_{y}(t)&=A_{0}[\sin\Omega t-\sin(\beta\Omega t-\theta)].\label{eq:Ay-BCPL}
\end{align}
Figures \ref{fig2}(a) and \ref{fig2}(b)
show the trajectories of $\mathbf{A}_{\textrm{BCPL}}(t)$ per period
at $\beta=2$ and $3$, respectively. 
In this case,
the Onsager reciprocal relation is written as 
\begin{align}
  %No. XXXXII 9/12 No. 01
  \sigma_{yx}^{\textrm{C}}[\mathbf{A}_{\textrm{BCPL}}(t)]
  =(-1)^{2}\sigma_{xy}^{\textrm{C}}[\mathbf{A}_{\textrm{BCPL}}(-t)],\label{eq:Onsager_sigC-BCPL}
\end{align}
where $\mathbf{A}_{\textrm{BCPL}}(-t)=(A_{x}(-t)\ A_{y}(-t))^{T}$
is given by
\begin{align}
  %No. XXXXII 9/12 No. 01
  A_{x}(-t)&=A_{0}[\cos\Omega t+\cos(\beta\Omega t+\theta)],\label{eq:Ax-BCPL_-t}\\
  A_{y}(-t)&=-A_{0}[\sin\Omega t-\sin(\beta\Omega t+\theta)].\label{eq:Ay-BCPL_-t}
\end{align}
Since
$\mathbf{A}_{\textrm{BCPL}}(-t)$ and $\mathbf{A}_{\textrm{BCPL}}(t)$
are not connected by a simple relation
such as $\mathbf{A}_{\textrm{LCPL}}(-t)=\mathbf{A}_{\textrm{RCPL}}(t)$,
we cannot rewrite Eq. (\ref{eq:Onsager_sigC-BCPL}) anymore in general. 
Therefore, 
in periodically driven systems with BCPL,
the time-averaged charge off-diagonal dc conductivity
is restricted to neither the antisymmetric nor symmetric part generally. 
We will validate this property numerically in Sect. 5.2.2.

\subsection{Spin off-diagonal dc conductivity}

We argue the Onsager reciprocal relations for the spin off-diagonal dc conductivity.
We suppose that 
this conductivity is given by
the correlation function between the charge current and
spin current operators.
This is valid in the linear-response regime of the probe field.
Note that this conductivity is different from
that for describing the spin current generated by the probe spin field,
which is given by the correlation function
between the spin current operators~\cite{NA-MagnonTrans1,NA-MagnonTrans2}. 
We also suppose that
the spin current remains unchanged under the time-reversal operation.
Therefore,
the following arguments are valid for any definition of the spin current 
as long as it is symmetric with respect to the time-reversal operation.
This property against the time-reversal operation may be reasonable
because the spin current is the flow of the spin angular momentum,
such as 
$\mathbf{J}_{\textrm{S}}=\frac{\hbar}{2}(\mathbf{J}_{\uparrow}-\mathbf{J}_{\downarrow})$~\cite{NA-FloquetSHE},
where $\mathbf{J}_{\sigma}$ is the current of spin-$\sigma$ electrons
[Fig. \ref{fig3}(b)].

  In this paper,
  we define the spin current as
  the flow of the $z$ component of the spin angular momentum.
  In the following arguments,
  we discuss the Onsager reciprocal relations
  about the spin off-diagonal dc conductivities
  for the $y$ component of this spin current
  perpendicular to the probe field applied along the $x$ direction.
  Since the other components of the spin current have
  the same time-reversal symmetry,
  the following arguments can be applied to the spin off-diagonal dc conductivities
  for the other components.
  Furthermore,
  they can be extended to the spin off-diagonal dc conductivities
  for another spin current,
  the flow of the $x$ or $y$ component of the spin angular momentum.
  Namely,
  the following arguments are sufficient to clarify the Onsager reciprocal relations
  about the spin off-diagonal dc conductivities.

\subsubsection{Nondriven systems}

In the nondriven case with the magnetic field $\mathbf{H}$,
the spin off-diagonal dc conductivity satisfies the Onsager reciprocal relation, 
\begin{align}
  %No. XXXXIII 9/22 No. 01
  \sigma_{yx}^{\textrm{S}}(\mathbf{H})
  =(-1)\tilde{\sigma}_{xy}^{\textrm{S}}(-\mathbf{H}),\label{eq:Onsager_sigS-H}
\end{align}
where $\sigma_{yx}^{\textrm{S}}(\mathbf{H})$ is
the spin off-diagonal dc conductivity for 
the spin current $J_{\textrm{S}}^{y}$ generated
perpendicular to the probe electric field $E_{x}$
in the presence of $\mathbf{H}$
[i.e., $J_{\textrm{S}}^{y}=\sigma_{yx}^{\textrm{S}}(\mathbf{H})E_{x}$],
and $\tilde{\sigma}_{xy}^{\textrm{S}}(-\mathbf{H})$
is another for the charge current $J_{\textrm{C}}^{x}$ generated
perpendicular to the probe spin field $E_{y}^{\textrm{S}}$
(e.g., the gradient of the Zeeman field or
of a spin-dependent chemical potential~\cite{CorrectOnsager-JS1})
in the presence of $-\mathbf{H}$
[i.e., $J_{\textrm{C}}^{x}=\tilde{\sigma}_{xy}^{\textrm{S}}(-\mathbf{H})E^{\textrm{S}}_{y}$].
The factor $(-1)$ in Eq. (\ref{eq:Onsager_sigS-H}) arises from the properties that
the spin off-diagonal dc conductivity is given by the correlation function
between the charge current and spin current operators
and that
only the charge current operator changes its sign
under the time-reversal operation (Fig. \ref{fig3}).
This minus sign has been missing
in some studies~\cite{IncorrectOnsager-JS1,IncorrectOnsager-JS2}.
Therefore,
the violation of the Onsager reciprocal relation claimed in
Ref. \cite{IncorrectOnsager-JS2} may be due to
the incorrect treatment about the difference between time-reversal symmetries 
of the charge and spin currents.
Then,
Eq. (\ref{eq:Onsager_sigS-H}) shows that
the counterpart connected by the Onsager reciprocal relation
for $\sigma_{yx}^{\textrm{S}}(\mathbf{H})$
is not $\sigma_{xy}^{\textrm{S}}(-\mathbf{H})$,
but $\tilde{\sigma}_{xy}^{\textrm{S}}(-\mathbf{H})$,
where $\sigma_{xy}^{\textrm{S}}(-\mathbf{H})$ is
the spin off-diagonal dc conductivity for the spin current $J_{\textrm{S}}^{x}$ generated
perpendicular to the probe electric field $E_{y}$
in the presence of $-\mathbf{H}$
[i.e., $J_{\textrm{S}}^{x}=\sigma_{xy}^{\textrm{S}}(-\mathbf{H})E_{y}$].
This is because
the transport coefficient for a certain transport phenomenon
is connected by the Onsager reciprocal relation
with that for the transport phenomenon obtained by
applying the time-reversal operation to
the original phenomenon~\cite{Onsager1,Onsager2,Kubo},
as explained above. 

The similar Onsager reciprocal relation holds
in the nondriven case with the magnetization $\mathbf{M}$:
\begin{align}
  %No. XXXXIII 9/22 No. 01
  \sigma_{yx}^{\textrm{S}}(\mathbf{M})
  =(-1)\tilde{\sigma}_{xy}^{\textrm{S}}(-\mathbf{M}).\label{eq:Onsager_sigS-M}
\end{align}

These results are consistent with
some previous studies~\cite{CorrectOnsager-JS1,CorrectOnsager-JS2},
although the origin of the minus sign appearing in the Onsager reciprocal relation
is different.

Because of Eq. (\ref{eq:Onsager_sigS-H}) or (\ref{eq:Onsager_sigS-M}),
the spin off-diagonal dc conductivity has the antisymmetric part
$(\sigma_{yx}^{\textrm{S}}-\tilde{\sigma}_{xy}^{\textrm{S}})/2$
even with no magnetic field 
and thus can be regarded as the spin Hall conductivity.
This is the most important difference between the spin and charge off-diagonal dc conductivities:
the former possesses the antisymmetric part even with time-reversal symmetry,
whereas the latter is restricted to the symmetric part with this symmetry.
We have defined the antisymmetric part of $\sigma_{yx}^{\textrm{S}}$
as not $(\sigma_{yx}^{\textrm{S}}-\sigma_{xy}^{\textrm{S}})/2$,
but $(\sigma_{yx}^{\textrm{S}}-\tilde{\sigma}_{xy}^{\textrm{S}})/2$
because the latter corresponds to
the spin Hall conductivity for describing
the spin current along the $y$ axis with
the probe electric field applied along the $x$ axis. 
This can be understood if we recall its expression
in nondriven systems with no dissipation; in these systems, 
the spin Hall conductivity is expressed in terms of the spin Berry curvature
the numerator of which is proportional to 
%No. XXXXV 2/29 No. 04
$\textrm{Im}[\langle \mathbf{k}\alpha|J_{\textrm{C}}^{x}|\mathbf{k}\beta\rangle
  \langle \mathbf{k}\beta|J_{\textrm{S}}^{y}|\mathbf{k}\alpha\rangle]$,
where
$\langle \mathbf{k}\alpha|J_{\textrm{C}}^{x}|\mathbf{k}\beta\rangle$
and $\langle \mathbf{k}\beta|J_{\textrm{S}}^{y}|\mathbf{k}\alpha\rangle$
are the expectation values of the charge and spin current operators, respectively, 
over the states at momentum $\mathbf{k}$
for band indices $\alpha$ and $\beta(\neq \alpha)$.
Note that
in the linear-response theory,
$\sigma_{yx}^{\textrm{S}}$ and $\tilde{\sigma}_{xy}^{\textrm{S}}$
are given by the correlation functions between
$J_{\textrm{S}}^{y}$ and $J_{\textrm{C}}^{x}$,
whereas $\sigma_{xy}^{\textrm{S}}$ is given by that between
$J_{\textrm{S}}^{x}$ and $J_{\textrm{C}}^{y}$.
Therefore,
$(\sigma_{yx}^{\textrm{S}}-\tilde{\sigma}_{xy}^{\textrm{S}})/2$
describes the SHE in the case of the spin current generated along the $y$ axis.
If the difference between the time-reversal symmetries of the charge and spin currents
were neglected,
the spin off-diagonal dc conductivity would have only the symmetric part.
The above arguments show that
a naive analogy with the Onsager reciprocal relation
for the charge transport is incorrect
to discuss the Onsager reciprocal relation for the spin transport. 

\subsubsection{Periodically driven systems}

We turn to the cases of the periodically driven systems in two dimensions.
In the following,
we discuss the symmetric properties of
the time-averaged spin off-diagonal dc conductivity 
in the nonequilibrium steady state under $\mathbf{A}_{\textrm{pump}}(t)$.
[For its definition,
see Eq. (\ref{eq:t-av-sig}) for Q$=$S with Eq. (\ref{eq:t-av-sig-w}).]
The arguments for any other periodically driven systems
can be made in a similar way.

In the case with the pump field of left-handed CPL,
the Onsager reciprocal relation for the time-averaged spin off-diagonal dc conductivity
is given by
\begin{align}
  %No. XXXXIII 9/22 No. 02
  \sigma_{yx}^{\textrm{S}}[\mathbf{A}_{\textrm{LCPL}}(t)]
  =(-1)\tilde{\sigma}_{xy}^{\textrm{S}}[\mathbf{A}_{\textrm{LCPL}}(-t)],\label{eq:Onsager_sigS-CPL_start}
\end{align}
where $\sigma_{yx}^{\textrm{S}}[\mathbf{A}_{\textrm{LPCL}}(t)]$
is the time-averaged spin off-diagonal dc conductivity
for the spin current $J_{\textrm{S}}^{y}$ generated perpendicular to
the probe electric field $E_{x}$ with $\mathbf{A}_{\textrm{LCPL}}(t)$ [Fig. \ref{fig4}(d)],
and 
$\tilde{\sigma}_{xy}^{\textrm{S}}[\mathbf{A}_{\textrm{LCPL}}(-t)]$
is another for the charge current $J_{\textrm{C}}^{x}$ generated
perpendicular to the probe spin field $E_{y}^{\textrm{S}}$
with $\mathbf{A}_{\textrm{LCPL}}(-t)$ [Fig. \ref{fig4}(e)]. 
Since $\mathbf{A}_{\textrm{LCPL}}(-t)=\mathbf{A}_{\textrm{RCPL}}(t)$,
we have
\begin{align}
  %No. XXXXIII 9/22 No. 02
  \sigma_{yx}^{\textrm{S}}[\mathbf{A}_{\textrm{LCPL}}(t)]
  =-\tilde{\sigma}_{xy}^{\textrm{S}}[\mathbf{A}_{\textrm{RCPL}}(t)].\label{eq:Onsager_sigS-CPL_next}
\end{align}
Similarly,
we have
\begin{align}
  \sigma_{yx}^{\textrm{S}}[\mathbf{A}_{\textrm{RCPL}}(t)]
  =-\tilde{\sigma}_{xy}^{\textrm{S}}[\mathbf{A}_{\textrm{LCPL}}(t)].\label{eq:Onsager_sigS-RCPL_next}
\end{align}
In addition,
the time-averaged spin off-diagonal dc conductivity is independent of
the helicity of light~\cite{NA-FloquetSHE}, 
\begin{align}
  %No. XXXXIII 9/22 No. 02
  \sigma_{yx}^{\textrm{S}}[\mathbf{A}_{\textrm{LCPL}}(t)]
  =\sigma_{yx}^{\textrm{S}}[\mathbf{A}_{\textrm{RCPL}}(t)].\label{eq:sigS-helicity}
\end{align}
This equation can be understood in terms of
the symmetry of the spin current under the time-reversal operation~\cite{NA-FloquetSHE}.
Combining Eq. (\ref{eq:sigS-helicity}) with Eqs. (\ref{eq:Onsager_sigS-CPL_next}) and
(\ref{eq:Onsager_sigS-RCPL_next}),
we obtain the Onsager reciprocal relation in this case, 
\begin{align}
  %No. XXXXIII 9/22 No. 02
  \sigma_{yx}^{\textrm{S}}[\mathbf{A}_{\textrm{LCPL}}(t)]
  =-\tilde{\sigma}_{xy}^{\textrm{S}}[\mathbf{A}_{\textrm{LCPL}}(t)].\label{eq:Onsager_sigS-CPL}
\end{align}
For $\tilde{\sigma}_{xy}^{\textrm{S}}[\mathbf{A}_{\textrm{LCPL}}(t)]$,
see Fig. \ref{fig4}(f).

The similar Onsager reciprocal relation is satisfied
with the pump field of LPL: 
\begin{align}
  %No. XXXXIII 9/22 No. 02
  \sigma_{yx}^{\textrm{S}}[\mathbf{A}_{\textrm{LPL}}(t)]
  =-\tilde{\sigma}_{xy}^{\textrm{S}}[\mathbf{A}_{\textrm{LPL}}(t)].\label{eq:Onsager_sigS-LPL}
\end{align}

Therefore,
the time-averaged spin off-diagonal dc conductivity is restricted to be antisymmetric 
in the periodically driven system with the pump field of CPL or LPL.
We will check this property numerically in Sect. 5.1.1.

In the case with the pump field of BCPL,
the Onsager reciprocal relation for the time-averaged spin off-diagonal dc conductivity
reads
\begin{align}
  %No. XXXXII 9/12
  \sigma_{yx}^{\textrm{S}}[\mathbf{A}_{\textrm{BCPL}}(t)]
  =-\tilde{\sigma}_{xy}^{\textrm{S}}[\mathbf{A}_{\textrm{BCPL}}(-t)].\label{eq:Onsager_sigS-BCPL}
\end{align}
As well as Eq. (\ref{eq:Onsager_sigC-BCPL}),
this equation cannot be rewritten anymore in general.
However, as we will show in Sect. 5.1.2,
the antisymmetric part dominates 
the time-averaged spin off-diagonal dc conductivity in
the nonequilibrium steady states with BCPL.
This is different from the result of the time-averaged charge off-diagonal dc conductivity
with BCPL (see Sect. 5.2.2).

\section{Model}

\begin{figure*}[h]
  \begin{center}
    \includegraphics[width=140mm]{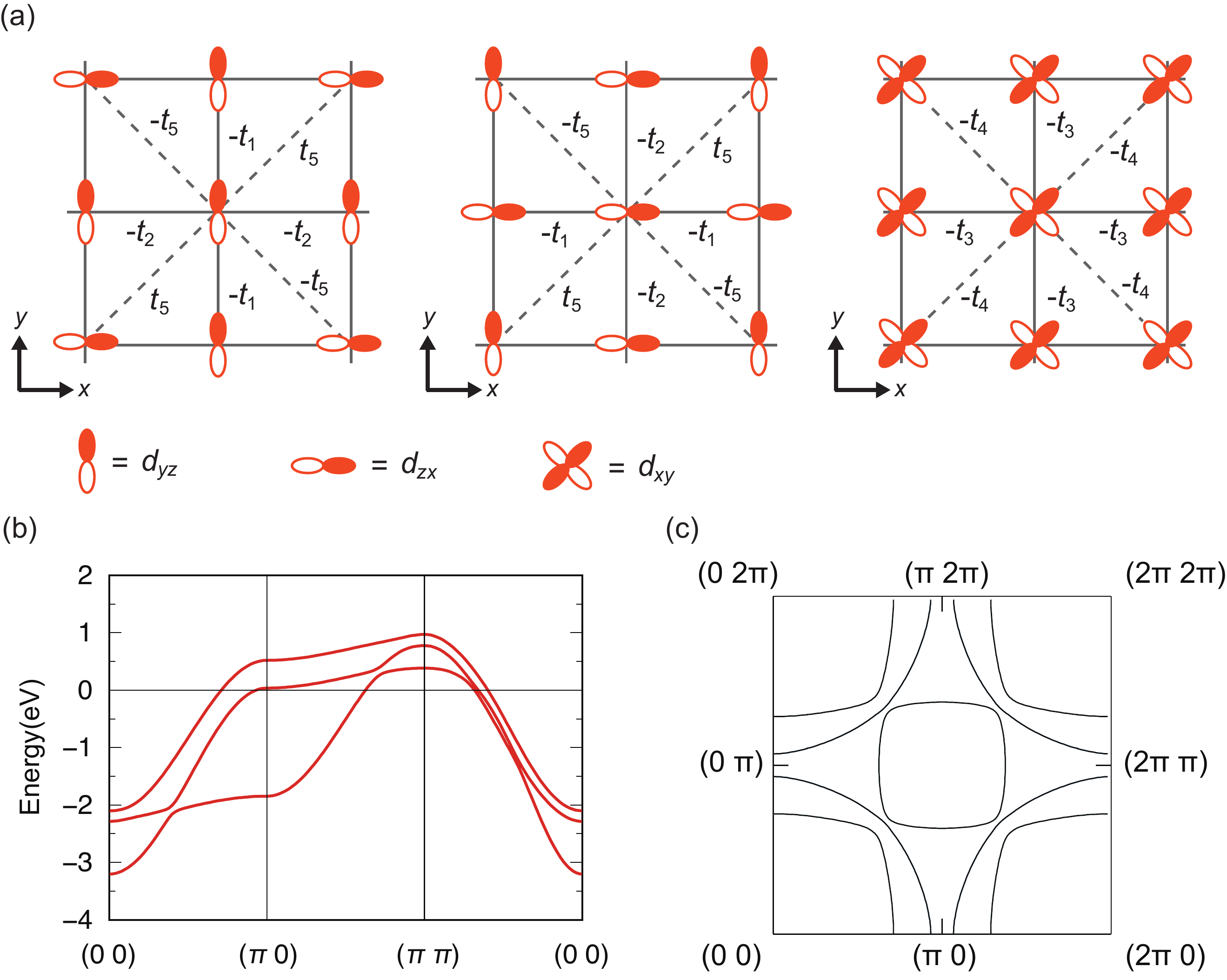}
    \end{center}
  \caption{\label{fig5}
    (Color online) (a) The nearest neighbor and next nearest neighbor hopping integrals
    for the $t_{2g}$-orbital electrons of Ru ions on the square lattice.
    The $d_{yz}$, $d_{zx}$, and $d_{xy}$ correspond to
    the $d_{yz}$, $d_{zx}$, and $d_{xy}$ orbitals, respectively. 
    (b) The band structure and (c) Fermi surface of our model
    for nondriven Sr$_{2}$RuO$_{4}$ at $T_{\textrm{b}}=0.02$ eV
    for a $N_{x}\times N_{y}$ mesh with $N_{x}=N_{y}=100$.
    In panel (b),
    the energies are measured from the chemical potential $\mu$. 
  }
\end{figure*}

To show the validity of our general arguments made in Sect. 2,
we consider a concrete model 
and analyze its time-averaged spin and charge off-diagonal dc conductivities.
Sections 3, 4, and 5 are devoted to this analysis. 

In this section,
we introduce the concrete model for periodically driven electron systems.
As the concrete model,
we consider a periodically driven multiorbital metal 
with weak coupling to a heat bath. 
Our model Hamiltonian consists of three parts~\cite{NA-FloquetSHE}:
\begin{align}
  H(t)=H_{\textrm{s}}(t)+H_{\textrm{b}}+H_{\textrm{sb}},\label{eq:Htot}
\end{align}
where $H_{\textrm{s}}(t)$ is the system Hamiltonian, 
$H_{\textrm{b}}$ is the bath Hamiltonian,
and 
$H_{\textrm{sb}}$ is the system-bath coupling Hamiltonian.
As $H_{\textrm{s}}(t)$,
we have considered the periodically driven Sr$_{2}$RuO$_{4}$
because it is suitable for achieving the finite time-averaged
spin and charge off-diagonal dc conductivities~\cite{NA-FloquetSHE}. 
In addition to $H_{\textrm{s}}(t)$,
we have considered $H_{\textrm{b}}$ and $H_{\textrm{sb}}$
because we suppose that
our periodically driven system can reach a nonequilibrium steady state
due to the coupling with the heat bath~\cite{Tsuji,Mikami,NA-FloquetSHE}.
If such a relaxation mechanism is absent,
a periodically driven system reaches an infinite-temperature state
due to the heating of the pump field~\cite{Heating1,Heating2}.
Hereafter,
we use the unit $\hbar=k_{\textrm{B}}=c=a_{\textrm{NN}}=1$,
where $a_{\textrm{NN}}$ is the distance between nearest-neighbor sites
on a square lattice. 

\subsection{System Hamiltonian $H_{\textrm{s}}(t)$}

$H_{\textrm{s}}(t)$ consists of
the hopping integrals between $t_{2g}$-orbital electrons
in the presence of a light field $\mathbf{A}(t)$, 
the chemical potential $\mu$,
and the atomic SOC (i.e., the $LS$ coupling):
\begin{align}
  %No. XXXVII 12/23 No. 01-02; No. XXXVI 12/23 No. 01-02 (SOC)
  H_{\textrm{s}}(t)
  &=\sum_{i,j}\sum_{a,b=d_{yz},d_{zx},d_{xy}}\sum_{\sigma=\uparrow,\downarrow}
  [t_{ij}^{ab}(t)-\mu\delta_{i,j}\delta_{a,b}]c_{ia\sigma}^{\dagger}c_{jb\sigma}\notag\\
  &+\sum_{i}\sum_{a,b=d_{yz},d_{zx},d_{xy}}\sum_{\sigma,\sigma^{\prime}=\uparrow,\downarrow}
  \xi_{ab}^{\sigma\sigma^{\prime}}c_{ia\sigma}^{\dagger}c_{ib\sigma^{\prime}},\label{eq:Hs-start}
\end{align}
where
$c_{ia\sigma}^{\dagger}$ and $c_{ia\sigma}$
are the creation and annihilation operators, respectively,
of an electron for orbital $a$ and spin $\sigma$ at site $i$, 
$t_{ij}^{ab}(t)$'s are the hopping integrals
with the Peierls phase factor due to $\mathbf{A}(t)$,
\begin{align}
  %No. XXXVII 12/23 No. 01-02
  t_{ij}^{ab}(t)=t_{ij}^{ab}e^{-ie(\mathbf{R}_{i}-\mathbf{R}_{j})\cdot\mathbf{A}(t)},\label{eq:hop-Peierls}
\end{align}
$t_{ij}^{ab}$'s are the hopping integrals in the nondriven system, 
and $\xi_{ab}^{\sigma\sigma^{\prime}}=(\xi_{ba}^{\sigma^{\prime}\sigma})^{\ast}$
is the coupling constant of the SOC for $t_{2g}$-orbital electrons,
the finite components of which are given by
\begin{align}
  %No. XXXVI 12/23 No. 01-02
  &\xi_{d_{yz} d_{zx}}^{\uparrow\uparrow}=\xi_{d_{zx} d_{xy}}^{\uparrow\downarrow}
  =-\xi_{d_{xy} d_{zx}}^{\uparrow\downarrow}=-\xi_{d_{yz} d_{zx}}^{\downarrow\downarrow}
  =i\xi/2,\\
  &\xi_{d_{xy}d_{yz}}^{\uparrow\downarrow}=-\xi_{d_{yz} d_{xy}}^{\uparrow\downarrow}=\xi/2.
\end{align}
Using the Fourier transformation of the operators,
we have
\begin{align}
  %No. XXXVII 12/23 No. 01-02
  H_{\textrm{s}}(t)
  &=\sum_{\mathbf{k}}\sum_{a,b=d_{yz},d_{zx},d_{xy}}\sum_{\sigma,\sigma^{\prime}=\uparrow,\downarrow}
  \bar{\epsilon}_{ab}^{\sigma\sigma^{\prime}}(\mathbf{k},t)
  c_{\mathbf{k} a\sigma}^{\dagger}c_{\mathbf{k} b\sigma^{\prime}},\label{eq:Hs}
\end{align}
where
\begin{align}
  %No. XXXVII 12/23 No. 01-02
  \bar{\epsilon}_{ab}^{\sigma\sigma^{\prime}}(\mathbf{k},t)
  =[\epsilon_{ab}(\mathbf{k},t)-\mu\delta_{a,b}]\delta_{\sigma,\sigma^{\prime}}
  +\xi_{ab}^{\sigma\sigma^{\prime}},\label{eq:ebar}
\end{align}
and 
\begin{align}
  \epsilon_{ab}(\mathbf{k},t)
  &=\sum_{j}t_{ij}^{ab}e^{-i[\mathbf{k}+e\mathbf{A}(t)]\cdot(\mathbf{R}_{i}-\mathbf{R}_{j})}.
  \label{eq:e_ab-kt}
\end{align}

We choose 
the parameters of $H_{\textrm{s}}(t)$ to reproduce
the electronic structure of Sr$_{2}$RuO$_{4}$
near the Fermi level.
The hopping integrals $t_{ij}^{ab}$'s in Eq. (\ref{eq:hop-Peierls})
are parametrized by $t_{1}$, $t_{2}$, $t_{3}$, $t_{4}$,
and $t_{5}$~\cite{NA-FloquetSHE,NA-FLEX,NA-Ru-model};
the first three ones are nearest-neighbor terms,
and the others are next-nearest-neighbor ones [Fig. \ref{fig5}(a)].
We choose their values as follows:
$(t_{1},t_{2},t_{3},t_{4},t_{5})=(0.675,0.09,0.45,0.18,0.03)$ (eV)~\cite{NA-Ru-model}.
We also set $\xi=0.17$ eV~\cite{SOC-Oguchi}.
Then, we determine $\mu$ from the condition that the electron number per site is four;
the value of $\mu$ is fixed at that determined for $\mathbf{A}(t)=\mathbf{0}$
in the analyses shown in Sect. 5.
Figures \ref{fig5}(b) and \ref{fig5}(c) show the band structure and Fermi surface
of our model for Sr$_{2}$RuO$_{4}$ with $\mathbf{A}(t)=\mathbf{0}$.
The obtained Fermi surface is consistent with
that observed experimentally in nondriven Sr$_{2}$RuO$_{4}$~\cite{Sr2RuO4-ARPES}.
Note that Sr$_{2}$RuO$_{4}$ is a material with the simple electronic structure
in which the spin off-diagonal dc conductivity is finite~\cite{Kon-SHE-Ru,Mizo}.

\subsection{Bath Hamiltonian $H_{\textrm{b}}$}

$H_{\textrm{b}}$ is the Hamiltonian for
the B\"{u}ttiker-type heat bath~\cite{HeatBath1,HeatBath2,Tsuji,Mikami,NA-FloquetSHE}:
\begin{align}
  %No. XXXVII 1/11 No. 01-02
  &H_{\textrm{b}}
  =\sum_{i}\sum_{p}
  (\epsilon_{p}-\mu_{\textrm{b}})b_{ip}^{\dagger}b_{ip},\label{eq:Hb}
\end{align}
where
$b_{ip}^{\dagger}$ and $b_{ip}$ are the creation and annihilation operators,
respectively, of a bath's fermion for mode $p$ at site $i$, 
$\epsilon_{p}$ is the energy of a bath's fermion,
and $\mu_{\textrm{b}}$ is its chemical potential.
$\mu_{\textrm{b}}$ is determined in order that there is no current between
the system and bath.
This heat bath is supposed to be in equilibrium at temperature $T_{\textrm{b}}$. 

\subsection{System-bath coupling Hamiltonian $H_{\textrm{sb}}$}

$H_{\textrm{sb}}$ describes the coupling
between the system and bath~\cite{Tsuji,Mikami,NA-FloquetSHE}:
\begin{align}
  %No. XXXVII 1/11 No. 01-02
  &H_{\textrm{sb}}
  =\sum_{i}\sum_{p}
  \sum_{a=d_{yz},d_{zx},d_{xy}}\sum_{\sigma=\uparrow,\downarrow}
  V_{pa\sigma}(c_{ia\sigma}^{\dagger}b_{ip}+b_{ip}^{\dagger}c_{ia\sigma}),\label{eq:Hsb}
\end{align}
where $V_{pa\sigma}$ is the coupling constant.
In this study,
we treat the effects of $H_{\textrm{sb}}$ as second-order perturbation.
As a result,
its main effect can be reduced to
the damping~\cite{Tsuji,Mikami,NA-FloquetSHE}.
Because of this damping,
a nonequilibrium steady state could be
achieved~\cite{Tsuji,Mikami,NA-FloquetSHE}.

\section{Floquet linear-response theory}

In this section,
we formulate the time-averaged spin and charge off-diagonal dc conductivities
in the nonequilibrium steady states of our periodically driven systems
by using the Floquet linear-response
theory~\cite{Tsuji,Mikami,NA-FloquetSHE,NA-Onsager,NA-BCPL,Eckstein}.
In this theory,
we set $\mathbf{A}(t)=\mathbf{A}_{\textrm{pump}}(t)+\mathbf{A}_{\textrm{prob}}(t)$
and treat 
the effects of $\mathbf{A}_{\textrm{pump}}(t)$
in the Floquet theory~\cite{Floquet1,Floquet2}
and those of $\mathbf{A}_{\textrm{prob}}(t)$
in the linear-response theory~\cite{Kubo}.
Therefore,
this is a theory for the pump-probe measurements~\cite{Opt-review}
in which
the spin or charge current is generated
perpendicular to the probe field with the pump field (Fig. \ref{fig1}).
In addition,
this theory is a natural extension of the Kubo formula~\cite{Kubo} for nondriven systems
to the periodically driven systems.

\subsection{Time-averaged spin and charge off-diagonal dc conductivities $\sigma_{yx}^{\textrm{S}}$ and $\sigma_{yx}^{\textrm{C}}$}

Since the effects of $\mathbf{A}_{\textrm{prob}}(t)$ are
treated in the linear-response theory~\cite{Kubo},
the spin and charge off-diagonal dc conductivities as functions of time,
$\sigma_{yx}^{\textrm{S}}(t,t^{\prime})$ and $\sigma_{yx}^{\textrm{C}}(t,t^{\prime})$,
are given by~\cite{NA-FloquetSHE}
\begin{align}
  %No. XXXV 10/28 No. 01-02
  \sigma_{yx}^{\textrm{S}}(t,t^{\prime})
  &=\frac{1}{i\omega}
  \frac{\delta \langle j_{\textrm{S}}^{y}(t)\rangle}{\delta A_{\textrm{prob}}^{x}(t^{\prime})},
  \label{eq:SHC-start}\\
  \sigma_{yx}^{\textrm{C}}(t,t^{\prime})
  &=\frac{1}{i\omega}
  \frac{\delta \langle j_{\textrm{C}}^{y}(t)\rangle}{\delta A_{\textrm{prob}}^{x}(t^{\prime})},
  \label{eq:AHC-start}
\end{align}
where
$\langle j_{\textrm{S}}^{y}(t)\rangle$
and $\langle j_{\textrm{C}}^{y}(t)\rangle$
are the expectation values of the operators of
the spin current and charge current densities, respectively.
These expectation values should be taken over
the states with both the pump and probe fields. 
Then,
the operators of the spin and charge currents are given by~\cite{NA-FloquetSHE}
\begin{align}
  %No. XXXV 10/29 No. 01-02
  J_{\textrm{S}}^{y}(t)
  &=\sum_{\mathbf{k}}\sum_{a,b}\sum_{\sigma}v_{ab\sigma}^{(\textrm{S})y}(\mathbf{k},t)
  c_{\mathbf{k} a\sigma}^{\dagger}(t)c_{\mathbf{k} b\sigma}(t),\label{eq:JS}\\
  J_{\textrm{C}}^{y}(t)
  &=\sum_{\mathbf{k}}\sum_{a,b}\sum_{\sigma}v_{ab\sigma}^{(\textrm{C})y}(\mathbf{k},t)
  c_{\mathbf{k} a\sigma}^{\dagger}(t)c_{\mathbf{k} b\sigma}(t),\label{eq:JC}
\end{align}
where
\begin{align}
  v_{ab\sigma}^{(\textrm{S})\nu}(\mathbf{k},t)
  &=\frac{1}{2}\textrm{sgn}(\sigma)
  \frac{\partial \epsilon_{ab}(\mathbf{k},t)}{\partial k_{\nu}},\label{eq:v^S}\\
  v_{ab\sigma}^{(\textrm{C})\nu}(\mathbf{k},t)
  &=(-e)\frac{\partial \epsilon_{ab}(\mathbf{k},t)}{\partial k_{\nu}},\label{eq:v^C}
\end{align}
and
$\textrm{sgn}(\sigma)=1$ or $-1$ for $\sigma=\uparrow$ or $\downarrow$,
respectively.
Note that Eq. (\ref{eq:JS}) and (\ref{eq:JC})
can be derived by using
the continuity equations~\cite{NA-FloquetSHE}.
We also note that $J_{\textrm{S}}^{y}(t)=Nj_{\textrm{S}}^{y}(t)$
and $J_{\textrm{C}}^{y}(t)=Nj_{\textrm{C}}^{y}(t)$ in our model
due to $V=Na_{\textrm{NN}}^{2}=N$,
where $V$ is the volume of the system. 

$\sigma_{yx}^{\textrm{S}}(t,t^{\prime})$ and $\sigma_{yx}^{\textrm{C}}(t,t^{\prime})$
can be expressed in terms of nonequilibrium Green's functions~\cite{Eckstein,NA-FloquetSHE}.
Substituting Eqs. (\ref{eq:JS}) and (\ref{eq:JC})
into Eqs. (\ref{eq:SHC-start}) and (\ref{eq:AHC-start}), respectively, 
we have
\begin{align}
  %No. XXXIV 11/1 No. 02
  \sigma_{yx}^{\textrm{S}}(t,t^{\prime})
  &=\sigma_{yx}^{\textrm{S}(1)}(t,t^{\prime})+\sigma_{yx}^{\textrm{S}(2)}(t,t^{\prime}),\label{eq:sigStt'}\\
  \sigma_{yx}^{\textrm{C}}(t,t^{\prime})
  &=\sigma_{yx}^{\textrm{C}(1)}(t,t^{\prime})+\sigma_{yx}^{\textrm{C}(2)}(t,t^{\prime}),\label{eq:sigCtt'}
\end{align}
where $\sigma_{yx}^{\textrm{Q}(1)}(t,t^{\prime})$
and $\sigma_{yx}^{\textrm{Q}(2)}(t,t^{\prime})$ (Q$=$S or C) are given by
\begin{align}
  %No. XXXIV 11/1 No. 02
  &\sigma_{yx}^{\textrm{Q}(1)}(t,t^{\prime})
  =\frac{-1}{\omega N}\sum_{\mathbf{k}}\sum_{a,b}\sum_{\sigma}
  \frac{\delta v_{ab\sigma}^{(\textrm{Q})y}(\mathbf{k},t)}{\delta A_{\textrm{prob}}^{x}(t^{\prime})}
  G_{b\sigma a\sigma}^{<}(\mathbf{k};t,t),\label{eq:sigDM-start}\\
  &\sigma_{yx}^{\textrm{Q}(2)}(t,t^{\prime})
  =\frac{-1}{\omega N}\sum_{\mathbf{k}}\sum_{a,b}\sum_{\sigma}
  v_{ab\sigma}^{(\textrm{Q})y}(\mathbf{k},t)
  \frac{\delta G_{b\sigma a\sigma}^{<}(\mathbf{k};t,t)}{\delta A_{\textrm{prob}}^{x}(t^{\prime})},
  \label{eq:sigPM-start}
\end{align}
and $G_{b\sigma^{\prime} a\sigma}^{<}(\mathbf{k};t,t^{\prime})$ is
the lesser Green's function~\cite{Mahan,Keldysh,Kadanoff-Baym,NA-FloquetSHE},
\begin{align}
  %No. XXXIV 11/1 No. 01-02
  G_{b\sigma^{\prime} a\sigma}^{<}(\mathbf{k};t,t^{\prime})
  =i\langle c_{\mathbf{k} a\sigma}^{\dagger}(t^{\prime})c_{\mathbf{k} b\sigma^{\prime}}(t)\rangle.
\end{align}
We emphasize that
the group velocities and Green's function appearing in the right-hand sides of
Eqs. (\ref{eq:sigDM-start}) and (\ref{eq:sigPM-start})
are those with only the pump field (i.e., no probe field).
This is an advantage of the linear-response theory~\cite{Kubo}.
In periodically driven cases,
these quantities and the transport coefficients are of nonequilibrium.
Meanwhile, 
in nondriven cases,
these quantities are of equilibrium,
although the transport coefficients are of nonequilibrium;
this is sometimes misunderstood as if the transport coefficients were also
of equilibrium.
To rewrite Eq. (\ref{eq:sigPM-start}),
we use the equation, 
\begin{align}
  %No. XXXV 11/4 No. 02
  \frac{\delta G_{b\sigma a\sigma}^{<}(\mathbf{k};t,t)}{\delta A_{\textrm{prob}}^{x}(t^{\prime})}
  &=-\sum_{c,d}\sum_{\sigma^{\prime}}v_{cd\sigma^{\prime}}^{(\textrm{C})x}(\mathbf{k},t^{\prime})\notag\\
  &\times
  \Bigl[
    G_{b\sigma c\sigma^{\prime}}^{\textrm{R}}(\mathbf{k};t,t^{\prime})
    G_{d\sigma^{\prime}a\sigma}^{<}(\mathbf{k};t^{\prime},t)\notag\\
    &\ \ \ +G_{b\sigma c\sigma^{\prime}}^{<}(\mathbf{k};t,t^{\prime})
    G_{d\sigma^{\prime}a\sigma}^{\textrm{A}}(\mathbf{k};t^{\prime},t)
    \Bigr],\label{eq:rewrite}
\end{align}
where $G_{a\sigma b\sigma^{\prime}}^{\textrm{R}}(\mathbf{k};t,t^{\prime})$
and $G_{a\sigma b\sigma^{\prime}}^{\textrm{A}}(\mathbf{k};t,t^{\prime})$
are the retarded and advanced Green's
functions~\cite{Mahan,Keldysh,Kadanoff-Baym,NA-FloquetSHE}, 
\begin{align}
  %No. XXXV 11/2 No. 01
  G_{a\sigma b\sigma^{\prime}}^{\textrm{R}}(\mathbf{k};t,t^{\prime})
  &=-i\theta(t-t^{\prime})
  \langle \{c_{\mathbf{k} a\sigma}(t),c_{\mathbf{k} b\sigma^{\prime}}^{\dagger}(t^{\prime})\}\rangle,\\
  G_{a\sigma b\sigma^{\prime}}^{\textrm{A}}(\mathbf{k};t,t^{\prime})
  &=i\theta(t^{\prime}-t)
  \langle \{c_{\mathbf{k} a\sigma}(t),c_{\mathbf{k} b\sigma^{\prime}}^{\dagger}(t^{\prime})\}\rangle.
\end{align}
Note that Eq. (\ref{eq:rewrite}) is obtained
by using the Dyson equation of the Green's functions
and the Langreth rule~\cite{Noneq-G-text,Mikami,Langreth}.
Substituting Eq. (\ref{eq:rewrite}) into Eq. (\ref{eq:sigPM-start}),
we get
\begin{align}
  %No. XXXV 11/4 No. 02
  \sigma_{yx}^{\textrm{Q}(2)}(t,t^{\prime})
  &=\frac{1}{\omega N}\sum_{\mathbf{k}}\sum_{a,b,c,d}\sum_{\sigma,\sigma^{\prime}}
  v_{ab\sigma}^{(\textrm{Q})y}(\mathbf{k},t)v_{cd\sigma^{\prime}}^{(\textrm{C})x}(\mathbf{k},t^{\prime})\notag\\
  &\times 
  \Bigl[G_{b\sigma c\sigma^{\prime}}^{\textrm{R}}(\mathbf{k};t,t^{\prime})
    G_{d\sigma^{\prime} a\sigma}^{<}(\mathbf{k};t^{\prime},t)\notag\\
  &\ \ \ +G_{b\sigma c\sigma^{\prime}}^{<}(\mathbf{k};t,t^{\prime})
    G_{d\sigma^{\prime} a\sigma}^{\textrm{A}}(\mathbf{k};t^{\prime},t)
    \Bigr].\label{eq:sigPM-rewrite}
\end{align}

The spin transport and charge transport
in a nonequilibrium steady state of our periodically driven system
can be described by
the time-averaged spin and charge off-diagonal dc conductivities
$\sigma_{yx}^{\textrm{S}}$ and $\sigma_{yx}^{\textrm{C}}$, respectively
[see Figs. \ref{fig4}(d) and \ref{fig4}(a)].  
$\sigma_{yx}^{\textrm{Q}}$ (Q$=$S or C) is defined as
\begin{align}
  %No. XXXIV 11/8 No. 04
  \sigma_{yx}^{\textrm{Q}}
  &=\lim_{\omega\rightarrow 0}\sigma_{yx}^{\textrm{Q}}(\omega),\label{eq:t-av-sig}
\end{align}
where 
\begin{align}
  %No. XXXIV 11/8 No. 04
  &\sigma_{yx}^{\textrm{Q}}(\omega)
  =\textrm{Re}\int_{0}^{T_{\textrm{p}}}\frac{dt_{\textrm{av}}}{T_{\textrm{p}}}
  \int_{-\infty}^{\infty}dt_{\textrm{rel}}e^{i\omega t_{\textrm{rel}}}
  \sigma_{yx}^{\textrm{Q}}(t,t^{\prime}),\label{eq:t-av-sig-w}
\end{align}
$t_{\textrm{rel}}=t-t^{\prime}$, and $t_{\textrm{av}}=(t+t^{\prime})/2$.
Note that
  the dc limit $\omega\rightarrow 0$ can be appropriately taken
  in the same way as for nondriven systems
  because the probe frequency $\omega$ is coupled to
  $t_{\textrm{rel}}$ in the Fourier transformation,
  whereas the time average is taken with respect to $t_{\textrm{av}}$.
In Eq. (\ref{eq:t-av-sig-w}),
  we have considered only the real part because
  we focus on 
  the time-averaged dc conductivities in this paper.
To calculate $\sigma_{yx}^{\textrm{Q}}(\omega)$,
we use the Floquet representation of the Green's functions~\cite{Tsuji-PRB,NA-FloquetSHE}.
In general, the Green's functions in a periodically driven system depend
not only on $t_{\textrm{rel}}$, but also on $t_{\textrm{av}}$.
Because of this property,
we should perform two transformations to convert the Green's functions
into the frequency functions:
\begin{align}
  %No. XXXIV 11/5 No. 02
  G^{r}_{a\sigma b\sigma^{\prime};n}(\mathbf{k};\omega)
  &=\int_{-\infty}^{\infty}dt_{\textrm{rel}}e^{i\omega t_{\textrm{rel}}}
  \int_{0}^{T_{\textrm{p}}}\frac{dt_{\textrm{av}}}{T_{\textrm{p}}}e^{in\Omega t_{\textrm{av}}}\notag\\
  &\times 
  G^{r}_{a\sigma b\sigma^{\prime}}
  (\mathbf{k};t_{\textrm{av}}+\frac{t_{\textrm{rel}}}{2},t_{\textrm{av}}-\frac{t_{\textrm{rel}}}{2}),
\end{align}
where 
$G^{r}_{a\sigma b\sigma^{\prime};n}(\mathbf{k};\omega)$'s ($r=$ R, A, $<$)
are defined in the range of $-\infty < \omega <\infty$. 
Since our Hamiltonian has the discrete time translation symmetry,
we can restrict the range of frequency
to $-\frac{\Omega}{2}\leq \omega <\frac{\Omega}{2}$.
This property can be taken into account by performing another transformation:
\begin{align}
  %No. XXXIV 11/5 No. 02
  [G^{r}_{a\sigma b\sigma^{\prime}}(\mathbf{k},\omega)]_{mn}
  =G^{r}_{a\sigma b\sigma^{\prime};m-n}(\mathbf{k};\omega+\frac{m+n}{2}\Omega),
\end{align}
where $[G^{r}_{a\sigma b\sigma^{\prime}}(\mathbf{k},\omega)]_{mn}$'s
($-\frac{\Omega}{2}\leq \omega <\frac{\Omega}{2}$)
are the Green's functions in the Floquet representation~\cite{Tsuji-PRB,NA-FloquetSHE}.
By combining Eqs. (\ref{eq:t-av-sig-w}), (\ref{eq:sigStt'}){--}(\ref{eq:sigDM-start}),
and (\ref{eq:sigPM-rewrite}) and using the Floquet representation of the Green's functions,
we obtain
\begin{align}
  %No. XXXVI 11/30 No. 03-04
  &\sigma_{yx}^{\textrm{Q}}(\omega)
  =\frac{1}{N}\sum_{\mathbf{k}}\sum_{a,b,c,d}\sum_{\sigma,\sigma^{\prime}}
  \int_{-\Omega/2}^{\Omega/2}\frac{d\omega^{\prime}}{2\pi}\notag\\
  &\times 
  \{
  \textrm{tr}[v_{ab\sigma}^{(\textrm{Q})y}(\mathbf{k})
    \frac{G_{b\sigma c\sigma^{\prime}}^{\textrm{R}}(\mathbf{k},\omega^{\prime}+\omega)
      -G_{b\sigma c\sigma^{\prime}}^{\textrm{R}}(\mathbf{k},\omega^{\prime}-\omega)}{2\omega}\notag\\
  &\ \ \ \ \ \ \times v_{cd\sigma^{\prime}}^{(\textrm{C})x}(\mathbf{k})
    G_{d\sigma^{\prime}a\sigma}^{<}(\mathbf{k},\omega^{\prime})]\notag\\
  &-\textrm{tr}[v_{ab\sigma}^{(\textrm{Q})y}(\mathbf{k})
    G_{b\sigma c\sigma^{\prime}}^{<}(\mathbf{k},\omega^{\prime})
    v_{cd\sigma^{\prime}}^{(\textrm{C})x}(\mathbf{k})\notag\\
  &\ \ \ \ \ 
    \times \frac{G_{d\sigma^{\prime}a\sigma}^{\textrm{A}}(\mathbf{k},\omega^{\prime}+\omega)
      -G_{d\sigma^{\prime}a\sigma}^{\textrm{A}}(\mathbf{k},\omega^{\prime}-\omega)}{2\omega}]
  \},\label{eq:sigyx^Q-w}
\end{align}
where
$\textrm{tr}(ABCD)=\sum_{m,l,n,q=-\infty}^{\infty}[A]_{ml}[B]_{ln}[C]_{nq}[D]_{qm}$,
and $m$, $l$, $n$, and $q$ are indices in the Floquet representation.
For the derivation of Eq. (\ref{eq:sigyx^Q-w}), see Appendix A. 
Note that $[v_{ab\sigma}^{(\textrm{Q})\nu}(\mathbf{k})]_{mn}$ ($\nu=x,y$) is given by
\begin{align}
  %No. XXXV 11/10 No. 13-14
  [v_{ab\sigma}^{(\textrm{Q})\nu}(\mathbf{k})]_{mn}
  &=\int_{0}^{T_{\textrm{p}}}\frac{dt}{T_{\textrm{p}}}e^{i(m-n)\Omega t}v_{ab\sigma}^{(\textrm{Q})\nu}(\mathbf{k},t),
\end{align}
where $v_{ab\sigma}^{(\textrm{S})\nu}(\mathbf{k},t)$ and $v_{ab\sigma}^{(\textrm{C})\nu}(\mathbf{k},t)$
have been defined in Eqs. (\ref{eq:v^S}) and (\ref{eq:v^C}), respectively.
Combining Eq. (\ref{eq:sigyx^Q-w}) with Eq. (\ref{eq:t-av-sig}),
we get
\begin{align}
  %No. XXXVI 11/30 No. 03-04
  &\sigma_{yx}^{\textrm{S}}
  =\frac{1}{N}\sum_{\mathbf{k}}\sum_{a,b,c,d}\sum_{\sigma,\sigma^{\prime}}
  \int_{-\Omega/2}^{\Omega/2}\frac{d\omega^{\prime}}{2\pi}\notag\\
  &\times \{
  \textrm{tr}[v_{ab\sigma}^{(\textrm{S})y}(\mathbf{k})
    \frac{\partial G_{b\sigma c\sigma^{\prime}}^{\textrm{R}}(\mathbf{k},\omega^{\prime})}
         {\partial \omega^{\prime}}
    v_{cd\sigma^{\prime}}^{(\textrm{C})x}(\mathbf{k})
    G_{d\sigma^{\prime}a\sigma}^{<}(\mathbf{k},\omega^{\prime})]\notag\\
  &-\textrm{tr}[v_{ab\sigma}^{(\textrm{S})y}(\mathbf{k})
    G_{b\sigma c\sigma^{\prime}}^{<}(\mathbf{k},\omega^{\prime})
    v_{cd\sigma^{\prime}}^{(\textrm{C})x}(\mathbf{k})
    \frac{\partial G_{d\sigma^{\prime}a\sigma}^{\textrm{A}}(\mathbf{k},\omega^{\prime})}
         {\partial \omega^{\prime}}]
  \},\label{eq:sig_yx^S}
\end{align}
and
\begin{align}
  %No. XXXVI 11/30 No. 03-04
  &\sigma_{yx}^{\textrm{C}}
  =\frac{1}{N}\sum_{\mathbf{k}}\sum_{a,b,c,d}\sum_{\sigma,\sigma^{\prime}}
  \int_{-\Omega/2}^{\Omega/2}\frac{d\omega^{\prime}}{2\pi}\notag\\
  &\times \{
  \textrm{tr}[v_{ab\sigma}^{(\textrm{C})y}(\mathbf{k})
    \frac{\partial G_{b\sigma c\sigma^{\prime}}^{\textrm{R}}(\mathbf{k},\omega^{\prime})}
         {\partial \omega^{\prime}}
    v_{cd\sigma^{\prime}}^{(\textrm{C})x}(\mathbf{k})
    G_{d\sigma^{\prime}a\sigma}^{<}(\mathbf{k},\omega^{\prime})]\notag\\
  &-\textrm{tr}[v_{ab\sigma}^{(\textrm{C})y}(\mathbf{k})
    G_{b\sigma c\sigma^{\prime}}^{<}(\mathbf{k},\omega^{\prime})
    v_{cd\sigma^{\prime}}^{(\textrm{C})x}(\mathbf{k})
    \frac{\partial G_{d\sigma^{\prime}a\sigma}^{\textrm{A}}(\mathbf{k},\omega^{\prime})}
         {\partial \omega^{\prime}}]
  \}.\label{eq:sig_yx^C}
\end{align}

The Green's functions appearing
in Eqs. (\ref{eq:sig_yx^S}) and (\ref{eq:sig_yx^C}) can be determined
in the following way.
For our periodically driven system,
the Green's functions are determined from the Dyson equation
in the matrix form~\cite{Keldysh,Keldysh2,NA-FloquetSHE},
\begin{align}
  %No. XXXVII 11/18 No. 01
  G=G_{0}+G_{0}\Sigma G,\label{eq:Dyson}
\end{align}
where
$G$ is the matrix of the Green's functions with $H_{\textrm{sb}}$,
\begin{align}
  %No. XXXVII 11/18 No. 01
  G=\left(
  \begin{array}{@{\,}cc@{\,}}
    G^{\textrm{R}} & G^{\textrm{K}}\\[3pt]
    0 & G^{\textrm{A}}
  \end{array}
  \right),\label{eq:G-matrix}
\end{align}
$G_{0}$ is the matrix of the Green's functions without $H_{\textrm{sb}}$,
\begin{align}
  %No. XXXVII 11/18 No. 01
  G_{0}=\left(
  \begin{array}{@{\,}cc@{\,}}
    G^{\textrm{R}}_{0} & G^{\textrm{K}}_{0}\\[3pt]
    0 & G^{\textrm{A}}_{0}
  \end{array}
  \right),\label{eq:G0-matrix}
\end{align}
and $\Sigma$ is the matrix of
the self-energies due to the second-order perturbation
with respect to $H_{\textrm{sb}}$,
\begin{align}
  %No. XXXVII 11/18 No. 01
  \Sigma=\left(
  \begin{array}{@{\,}cc@{\,}}
    \Sigma^{\textrm{R}} & \Sigma^{\textrm{K}}\\[3pt]
    0 & \Sigma^{\textrm{A}}
  \end{array} 
  \right)\label{eq:Sigma-matrix}.
\end{align}
In Eqs. (\ref{eq:G-matrix}){--}(\ref{eq:Sigma-matrix}), 
the superscripts R, A, and K represent
the retarded, advanced, and Keldysh components, respectively.
These three components are related to the lesser component as follows:
\begin{align}
  %No. XXXVII 11/18 No. 02
  G^{<}=\frac{1}{2}(G^{\textrm{K}}-G^{\textrm{R}}+G^{\textrm{A}}).\label{eq:G^<-relation}
\end{align}
The retarded, advanced, and Keldysh components
of the self-energies can be obtained by using the second-order perturbation theory;
the results are
\begin{align}
  %No. XXXVII 1/11 No. 03-04
  [\Sigma^{\textrm{R}}_{a\sigma b\sigma^{\prime}}(\mathbf{k},\omega)]_{mn}
  &=
  -i\delta_{m,n}\delta_{a,b}\delta_{\sigma,\sigma^{\prime}}\Gamma,\label{eq:Sig^R}\\
  [\Sigma^{\textrm{A}}_{a\sigma b\sigma^{\prime}}(\mathbf{k},\omega)]_{mn}
  &=+i\delta_{m,n}\delta_{a,b}\delta_{\sigma,\sigma^{\prime}}\Gamma,\label{eq:Sig^A}\\
  [\Sigma^{\textrm{K}}_{a\sigma b\sigma^{\prime}}(\mathbf{k},\omega)]_{mn}
  &=-2i\delta_{m,n}\delta_{a,b}\delta_{\sigma,\sigma^{\prime}}\Gamma
  \tanh\frac{\omega+m\Omega}{2T_{\textrm{b}}},\label{eq:Sig^K}
\end{align}
where $\Gamma$ is the damping. 
In deriving these equations, 
we have omitted the real parts 
and replaced
%No. XXXVI 1/11 No. 03-04
$\pi\sum_{p}V_{pa\sigma}V_{pb\sigma^{\prime}}\delta(\omega+m\Omega-\epsilon_{p}+\mu_{\textrm{b}})$
by $\Gamma\delta_{a,b}\delta_{\sigma,\sigma^{\prime}}$ for simplicity~\cite{NA-FloquetSHE}.
Such simplification may be sufficient
because the main effect of $H_{\textrm{sb}}$ is to induce the damping,
which makes the system the nonequilibrium steady state~\cite{Tsuji,Mikami,NA-FloquetSHE}.
Then, Eq. (\ref{eq:Dyson}) can be rewritten as
\begin{align}
  %No. XXXVII 11/18 No. 01
  G^{-1}=G_{0}^{-1}-\Sigma,\label{eq:Dyson-rewrite}
\end{align}
where
\begin{align}
  %No. XXXVI 11/19 No. 03-04
  G^{-1}=\left(
  \begin{array}{@{\,}cc@{\,}}
    (G^{-1})^{\textrm{R}} & (G^{-1})^{\textrm{K}}\\[3pt]
    0 & (G^{-1})^{\textrm{A}}
  \end{array}
  \right),\\
  G_{0}^{-1}=\left(
  \begin{array}{@{\,}cc@{\,}}
    (G_{0}^{-1})^{\textrm{R}} & (G_{0}^{-1})^{\textrm{K}}\\[3pt]
    0 & (G_{0}^{-1})^{\textrm{A}}
  \end{array}
  \right).
\end{align}
For our model, 
the retarded, advanced, and Keldysh components of the matrix $G^{-1}$ are given by
\begin{align}
  %No. XXXVII 1/6 No. 01
  [(G^{-1})^{\textrm{R}}_{a\sigma b\sigma^{\prime}}(\mathbf{k},\omega)]_{mn}
  &=(\omega+\mu+m\Omega+i\Gamma)\delta_{m,n}\delta_{a,b}\delta_{\sigma,\sigma^{\prime}}\notag\\
  &-\xi_{ab}^{\sigma\sigma^{\prime}}\delta_{m,n}
  -[\epsilon_{ab}(\mathbf{k})]_{mn}\delta_{\sigma,\sigma^{\prime}},\label{eq:G^-1R}\\
  [(G^{-1})^{\textrm{A}}_{a\sigma b\sigma^{\prime}}(\mathbf{k},\omega)]_{mn}
  &=(\omega+\mu+m\Omega-i\Gamma)\delta_{m,n}\delta_{a,b}\delta_{\sigma,\sigma^{\prime}}\notag\\
  &-\xi_{ab}^{\sigma\sigma^{\prime}}\delta_{m,n}
  -[\epsilon_{ab}(\mathbf{k})]_{mn}\delta_{\sigma,\sigma^{\prime}},\label{eq:G^-1A}\\
  %No. XXXVII 11/18 No. 02
  [(G^{-1})^{\textrm{K}}_{a\sigma b\sigma^{\prime}}(\mathbf{k},\omega)]_{mn}
  &=2i\delta_{m,n}\delta_{a,b}\delta_{\sigma,\sigma^{\prime}}\Gamma
  \tanh\frac{\omega+m\Omega}{2T_{\textrm{b}}},\label{eq:G^-1K}
\end{align}
where
\begin{align}
  %No. XXXVI 12/27 No. 02-03
  [\epsilon_{ab}(\mathbf{k})]_{mn}
  =\int_{0}^{T_{\textrm{p}}}\frac{dt}{T_{\textrm{p}}}e^{i(m-n)\Omega t}
  \epsilon_{ab}(\mathbf{k},t).\label{eq:epsilon_mn}
\end{align}
For $[\epsilon_{ab}(\mathbf{k})]_{mn}$ of Sr$_{2}$RuO$_{4}$ driven by BCPL, 
see Appendix B;
for that of Sr$_{2}$RuO$_{4}$ driven by CPL or LPL, see Ref. [\cite{NA-FloquetSHE}].
In deriving Eq. (\ref{eq:G^-1K}),
  we have chosen the Keldysh component of the matrix $G_{0}^{-1}$ to be zero
  because it contains the information about the initial condition. 
  This treatment may be valid to describe 
  a nonequilibrium steady state
  with finite damping~\cite{Oka-PRB,Mikami,Tsuji,NA-FloquetSHE}
  because
  a nonequilibrium steady state should be independent of
  a choice of the initial condition.
  Note that the retarded and advanced components of the matrix $G_{0}^{-1}$
  are obtained by replacing $\Gamma$ in Eqs. (\ref{eq:G^-1R}) and (\ref{eq:G^-1A}),
  respectively, by a positive infinitesimal.
Because of the matrix relation $G^{-1}G=1$,
the components of the inverse matrices satisfy 
\begin{align}
  %No. XXXVII 11/18 No.02
  &(G^{\textrm{R}})^{-1}=(G^{-1})^{\textrm{R}},\label{eq:G^R-inv}\\
  &(G^{\textrm{A}})^{-1}=(G^{-1})^{\textrm{A}},\label{eq:G^A-inv}\\
  &G^{\textrm{K}}=-G^{\textrm{R}}(G^{-1})^{\textrm{K}}G^{\textrm{A}}.\label{eq:G^K-inv}
\end{align}
Therefore,
we can obtain the retarded and advanced Green's functions with $H_{\textrm{sb}}$
by using Eqs. (\ref{eq:G^-1R}) and (\ref{eq:G^R-inv})
and Eqs. (\ref{eq:G^-1A}) and (\ref{eq:G^A-inv}), respectively.
We can also calculate the Keldysh Green's function with $H_{\textrm{sb}}$
by combining Eq. (\ref{eq:G^K-inv}) with Eq. (\ref{eq:G^-1K})
and the obtained retarded and advanced Green's functions.
Using Eq. (\ref{eq:G^<-relation}) and the obtained three Green's functions,
we finally obtain the lesser Green's function with $H_{\textrm{sb}}$.

\subsection{The other off-diagonal dc conductivities $\tilde{\sigma}_{xy}^{\textrm{S}}$, $\sigma_{xy}^{\textrm{C}}$, $\bar{\tilde{\sigma}}_{xy}^{\textrm{S}}$, and $\bar{\sigma}_{xy}^{\textrm{C}}$}

To discuss the Onsager reciprocal relations,
we need to consider
$\tilde{\sigma}_{xy}^{\textrm{S}}$ and $\sigma_{xy}^{\textrm{C}}$. 
Here 
$\tilde{\sigma}_{xy}^{\textrm{S}}$ or $\sigma_{xy}^{\textrm{C}}$
is the time-averaged off-diagonal dc conductivity for 
the charge current along the $x$ axis perpendicular to
the probe spin field or electric field, respectively, applied along the $y$ axis
with the pump field $\mathbf{A}_{\textrm{pump}}(t)$ [see Figs. \ref{fig4}(f) and \ref{fig4}(c)]. 
These conductivities are given by
\begin{align}
  %No. XXXVI 11/30 No. 03-04; No. XXXXII 9/12 No. 02
  &\tilde{\sigma}_{xy}^{\textrm{S}}
  =\frac{1}{N}\sum_{\mathbf{k}}\sum_{a,b,c,d}\sum_{\sigma,\sigma^{\prime}}
  \int_{-\Omega/2}^{\Omega/2}\frac{d\omega^{\prime}}{2\pi}\notag\\
  &\times \{
  \textrm{tr}[v_{ab\sigma}^{(\textrm{C})x}(\mathbf{k})
    \frac{\partial G_{b\sigma c\sigma^{\prime}}^{\textrm{R}}(\mathbf{k},\omega^{\prime})}
         {\partial \omega^{\prime}}
    v_{cd\sigma^{\prime}}^{(\textrm{S})y}(\mathbf{k})
    G_{d\sigma^{\prime}a\sigma}^{<}(\mathbf{k},\omega^{\prime})]\notag\\
  &-\textrm{tr}[v_{ab\sigma}^{(\textrm{C})x}(\mathbf{k})
    G_{b\sigma c\sigma^{\prime}}^{<}(\mathbf{k},\omega^{\prime})
    v_{cd\sigma^{\prime}}^{(\textrm{S})y}(\mathbf{k})
    \frac{\partial G_{d\sigma^{\prime}a\sigma}^{\textrm{A}}(\mathbf{k},\omega^{\prime})}
         {\partial \omega^{\prime}}]
  \},\label{eq:tilde-sig_xy^S}
\end{align}
and
\begin{align}
  %No. XXXVI 11/30 No. 03-04
  &\sigma_{xy}^{\textrm{C}}
  =\frac{1}{N}\sum_{\mathbf{k}}\sum_{a,b,c,d}\sum_{\sigma,\sigma^{\prime}}
  \int_{-\Omega/2}^{\Omega/2}\frac{d\omega^{\prime}}{2\pi}\notag\\
  &\times \{
  \textrm{tr}[v_{ab\sigma}^{(\textrm{C})x}(\mathbf{k})
    \frac{\partial G_{b\sigma c\sigma^{\prime}}^{\textrm{R}}(\mathbf{k},\omega^{\prime})}
         {\partial \omega^{\prime}}
    v_{cd\sigma^{\prime}}^{(\textrm{C})y}(\mathbf{k})
    G_{d\sigma^{\prime}a\sigma}^{<}(\mathbf{k},\omega^{\prime})]\notag\\
  &-\textrm{tr}[v_{ab\sigma}^{(\textrm{C})x}(\mathbf{k})
    G_{b\sigma c\sigma^{\prime}}^{<}(\mathbf{k},\omega^{\prime})
    v_{cd\sigma^{\prime}}^{(\textrm{C})y}(\mathbf{k})
    \frac{\partial G_{d\sigma^{\prime}a\sigma}^{\textrm{A}}(\mathbf{k},\omega^{\prime})}
         {\partial \omega^{\prime}}]
  \}.\label{eq:sig_xy^C}
\end{align}
Here Eq. (\ref{eq:tilde-sig_xy^S}) has been obtained by replacing
$v_{ab\sigma}^{(\textrm{S})y}(\mathbf{k})$'s and $v_{cd\sigma^{\prime}}^{(\textrm{C})x}(\mathbf{k})$'s
in Eq. (\ref{eq:sig_yx^S}) by
$v_{ab\sigma}^{(\textrm{C})x}(\mathbf{k})$'s and $v_{cd\sigma^{\prime}}^{(\textrm{S})y}(\mathbf{k})$'s,
respectively;
and Eq. (\ref{eq:sig_xy^C}) has been obtained by replacing
$v_{ab\sigma}^{(\textrm{C})y}(\mathbf{k})$'s and $v_{cd\sigma^{\prime}}^{(\textrm{C})x}(\mathbf{k})$'s
in Eq. (\ref{eq:sig_yx^C}) by
$v_{ab\sigma}^{(\textrm{C})x}(\mathbf{k})$'s and $v_{cd\sigma^{\prime}}^{(\textrm{C})y}(\mathbf{k})$'s,
respectively.
The quantities appearing in Eqs. (\ref{eq:tilde-sig_xy^S}) and (\ref{eq:sig_xy^C})
can be determined in the same way as
those appearing in Eqs. (\ref{eq:sig_yx^S}) and (\ref{eq:sig_yx^C}).

In addition,
we need to consider
$\bar{\tilde{\sigma}}_{xy}^{\textrm{S}}$ and $\bar{\sigma}_{xy}^{\textrm{C}}$.
Here  
$\bar{\tilde{\sigma}}_{xy}^{\textrm{S}}$ or $\bar{\sigma}_{xy}^{\textrm{C}}$
is the time-averaged off-diagonal dc conductivity for
the charge current along the $x$ axis perpendicular to
the probe spin field or electric field, respectively, applied along the $y$ axis
with the pump field $\mathbf{A}_{\textrm{pump}}(-t)$
[see Figs. \ref{fig4}(e) and \ref{fig4}(b)].
These conductivities are given by
\begin{align}
  %No. XXXVI 11/30 No. 03-04; No. XXXXII 9/12 No. 02
  &\bar{\tilde{\sigma}}_{xy}^{\textrm{S}}
  =\frac{1}{N}\sum_{\mathbf{k}}\sum_{a,b,c,d}\sum_{\sigma,\sigma^{\prime}}
  \int_{-\Omega/2}^{\Omega/2}\frac{d\omega^{\prime}}{2\pi}\notag\\
  &\times \{
  \textrm{tr}[\bar{v}_{ab\sigma}^{(\textrm{C})x}(\mathbf{k})
    \frac{\partial \bar{G}_{b\sigma c\sigma^{\prime}}^{\textrm{R}}(\mathbf{k},\omega^{\prime})}
         {\partial \omega^{\prime}}
    \bar{v}_{cd\sigma^{\prime}}^{(\textrm{S})y}(\mathbf{k})
    \bar{G}_{d\sigma^{\prime}a\sigma}^{<}(\mathbf{k},\omega^{\prime})]\notag\\
  &-\textrm{tr}[\bar{v}_{ab\sigma}^{(\textrm{C})x}(\mathbf{k})
    \bar{G}_{b\sigma c\sigma^{\prime}}^{<}(\mathbf{k},\omega^{\prime})
    \bar{v}_{cd\sigma^{\prime}}^{(\textrm{S})y}(\mathbf{k})
    \frac{\partial \bar{G}_{d\sigma^{\prime}a\sigma}^{\textrm{A}}(\mathbf{k},\omega^{\prime})}
         {\partial \omega^{\prime}}]
  \},\label{eq:bar-tilde-sig_xy^S}
\end{align}
and
\begin{align}
  %No. XXXVI 11/30 No. 03-04
  &\bar{\sigma}_{xy}^{\textrm{C}}
  =\frac{1}{N}\sum_{\mathbf{k}}\sum_{a,b,c,d}\sum_{\sigma,\sigma^{\prime}}
  \int_{-\Omega/2}^{\Omega/2}\frac{d\omega^{\prime}}{2\pi}\notag\\
  &\times \{
  \textrm{tr}[\bar{v}_{ab\sigma}^{(\textrm{C})x}(\mathbf{k})
    \frac{\partial \bar{G}_{b\sigma c\sigma^{\prime}}^{\textrm{R}}(\mathbf{k},\omega^{\prime})}
         {\partial \omega^{\prime}}
    \bar{v}_{cd\sigma^{\prime}}^{(\textrm{C})y}(\mathbf{k})
    \bar{G}_{d\sigma^{\prime}a\sigma}^{<}(\mathbf{k},\omega^{\prime})]\notag\\
  &-\textrm{tr}[\bar{v}_{ab\sigma}^{(\textrm{C})x}(\mathbf{k})
    \bar{G}_{b\sigma c\sigma^{\prime}}^{<}(\mathbf{k},\omega^{\prime})
    \bar{v}_{cd\sigma^{\prime}}^{(\textrm{C})y}(\mathbf{k})
    \frac{\partial \bar{G}_{d\sigma^{\prime}a\sigma}^{\textrm{A}}(\mathbf{k},\omega^{\prime})}
         {\partial \omega^{\prime}}]
  \}.\label{eq:bar-sig_xy^C}
\end{align}
Here $[\bar{v}_{ab\sigma}^{(\textrm{Q})\nu}(\mathbf{k})]_{mn}$ ($\nu=x,y$) is given by
\begin{align}
  %No. XXXV 11/10 No. 13-14
  [\bar{v}_{ab\sigma}^{(\textrm{Q})\nu}(\mathbf{k})]_{mn}
  &=\int_{0}^{T_{\textrm{p}}}\frac{dt}{T_{\textrm{p}}}e^{i(m-n)\Omega t}
  v_{ab\sigma}^{(\textrm{Q})\nu}(\mathbf{k},-t),
\end{align}
and $[\bar{G}^{r}_{a\sigma b\sigma^{\prime}}(\mathbf{k},\omega)]_{mn}$'s are determined
by replacing Eqs. (\ref{eq:G^-1R}){--}(\ref{eq:G^K-inv}) by
\begin{align}
  %No. XXXVII 1/6 No. 01
  [(\bar{G}^{-1})^{\textrm{R}}_{a\sigma b\sigma^{\prime}}(\mathbf{k},\omega)]_{mn}
  &=(\omega+\mu+m\Omega+i\Gamma)\delta_{m,n}\delta_{a,b}\delta_{\sigma,\sigma^{\prime}}\notag\\
  &-\xi_{ab}^{\sigma\sigma^{\prime}}\delta_{m,n}
  -[\bar{\epsilon}_{ab}(\mathbf{k})]_{mn}\delta_{\sigma,\sigma^{\prime}},\label{eq:bar-G^-1R}\\
  [(\bar{G}^{-1})^{\textrm{A}}_{a\sigma b\sigma^{\prime}}(\mathbf{k},\omega)]_{mn}
  &=(\omega+\mu+m\Omega-i\Gamma)\delta_{m,n}\delta_{a,b}\delta_{\sigma,\sigma^{\prime}}\notag\\
  &-\xi_{ab}^{\sigma\sigma^{\prime}}\delta_{m,n}
  -[\bar{\epsilon}_{ab}(\mathbf{k})]_{mn}\delta_{\sigma,\sigma^{\prime}},\label{eq:bar-G^-1A}\\
  %No. XXXVII 11/18 No. 02
  [(\bar{G}^{-1})^{\textrm{K}}_{a\sigma b\sigma^{\prime}}(\mathbf{k},\omega)]_{mn}
  &=2i\delta_{m,n}\delta_{a,b}\delta_{\sigma,\sigma^{\prime}}\Gamma
  \tanh\frac{\omega+m\Omega}{2T_{\textrm{b}}},\label{eq:bar-G^-1K}\\
  %No. XXXVI 12/27 No. 02-03
  [\bar{\epsilon}_{ab}(\mathbf{k})]_{mn}&
  =\int_{0}^{T_{\textrm{p}}}\frac{dt}{T_{\textrm{p}}}e^{i(m-n)\Omega t}
  \epsilon_{ab}(\mathbf{k},-t),\label{eq:bar-epsilon_mn}\\
  %No. XXXVII 11/18 No.02
  (\bar{G}^{\textrm{R}})^{-1}&=(\bar{G}^{-1})^{\textrm{R}},\label{eq:bar-G^R-inv}\\
  (\bar{G}^{\textrm{A}})^{-1}&=(\bar{G}^{-1})^{\textrm{A}},\label{eq:bar-G^A-inv}\\
  \bar{G}^{\textrm{K}}&=-\bar{G}^{\textrm{R}}(\bar{G}^{-1})^{\textrm{K}}\bar{G}^{\textrm{A}},
  \label{eq:bar-G^K-inv}
\end{align}
and performing the similar procedures to those used to determine 
$[G^{r}_{a\sigma b\sigma^{\prime}}(\mathbf{k},\omega)]_{mn}$'s.
Namely,
the group velocities and Green's functions
appearing in Eqs. (\ref{eq:bar-tilde-sig_xy^S}) and (\ref{eq:bar-sig_xy^C}) 
are calculated with $\mathbf{A}_{\textrm{pump}}(-t)$
in the similar way to those with $\mathbf{A}_{\textrm{pump}}(t)$.
For $[\bar{\epsilon}_{ab}(\mathbf{k})]_{mn}$ of Sr$_{2}$RuO$_{4}$ driven by BCPL, 
see Appendix C.

\subsection{General remarks about the applicability}

First, 
the Floquet linear-response theory is applicable to
the periodically driven systems under the application of the pump field.
This is because
the discrete time translational symmetry is utilized in the Floquet theory.

Then,
our theory has wider applicability than
the theories using a high-frequency expansion.
This expansion has been often used to analyze
many periodically driven systems~\cite{Floquet-review1,Highw1,Light-AHE1,Mikami,NA-BCPL}.
It may be sufficient for a periodically driven electron system 
if the light frequency is much larger than
the bandwidth of the system.
Meanwhile,
our theory does not have such a restriction 
because the Floquet theory used here is free from the constraint on the frequency.

The applicability of our theory is also wider than
that of the theories in which
the time-averaged off-diagonal dc conductivities are expressed in terms of the Berry curvature.
In the limit $\Gamma\rightarrow 0$,
these conductivities 
could be linked to the Berry curvature~\cite{Oka-PRB,Mikami}.
This is similar to the AHE and SHE in nondriven systems~\cite{Kon-AHE,Kon-SHE-Pt,NA-SCD}.
However, for finite $\Gamma$,
the anomalous Hall or spin Hall conductivity in nondriven systems contains
not only the Berry-curvature term,
but also the others,
including the so-called Fermi-surface term~\cite{Kon-AHE,Kon-SHE-Pt,NA-SCD,Streda},
which is distinct from the Berry-curvature term. 
Furthermore,
the Fermi-surface term dominates
the intrinsic AHE and SHE at finite $\Gamma$~\cite{Kon-AHE,Kon-SHE-Pt,NA-SCD,Mizo}.
The similar crossover can be realized for the Hall conductivity with the magnetic field:
the Hall conductivity in the strong-field case $\omega_{\textrm{c}}\tau \gg 1$
can be approximated by the Berry-curvature term~\cite{TKNN},
whereas that in the weak-field case $\omega_{\textrm{c}}\tau \ll 1$
is described by the Fermi-surface term~\cite{Fukuyama}.
Here $\omega_{\textrm{c}}$ represents the energy gap between Landau levels,
and $\tau$ is inversely proportional to the damping.
Namely,
the Berry-curvature term is dominant
if the band splitting which contributes to the Hall conductivity
is much larger than the broadening in the single-particle spectrum;
otherwise, the Fermi-surface term is dominant. 
Importantly,
these terms are automatically included in the conductivities
derived from the Kubo formula without any additional approximation. 
Moreover,
the limit $\Gamma\rightarrow 0$ is unrealistic in periodically driven systems
because the finite $\Gamma$ is required to realize a nonequilibrium steady state
with the heating due to the pump field.
Therefore,
we will study the time-averaged spin and charge off-diagonal dc conductivities 
without simplification using the Berry curvature.

\section{Numerical results}

In this section,
we show the time-averaged spin and charge off-diagonal dc conductivities
calculated numerically for Sr$_{2}$RuO$_{4}$ driven by CPL, LPL, or BCPL at $\beta=2$ or $3$.
In Sect. 5.1,
we focus on the time-averaged spin off-diagonal dc conductivities.
In all the cases considered,
the Onsager reciprocal relations argued in Sect. 2.2.2 are satisfied
and the main terms are given by the antisymmetric parts. 
In Sect. 5.2,
we turn to the time-averaged charge off-diagonal dc conductivities.
Although the Onsager reciprocal relations argued in Sect. 2.1.2 are satisfied,
their main terms depend on the polarization of light.
In the case with CPL or LPL,
the main term is the antisymmetric or symmetric part, respectively.
Then, 
in the cases with BCPL at
$(\beta,\theta)=(2,\frac{\pi}{4})$, $(2,\frac{3\pi}{4})$,
$(3,\frac{\pi}{4})$, $(3,\frac{\pi}{2})$, and $(3,\frac{3\pi}{4})$,
the main term is the antisymmetric part
in the range of $0\leq u \leq 0.4$,
whereas it is the symmetric part in the range of $0.5\leq u \leq 1$.
Here $u=eA_{0}$ is the dimensionless quantity. 
Meanwhile,
in the cases with BCPL at
$(\beta,\theta)=(2,0)$, $(2,\frac{\pi}{2})$, $(2,\pi)$,
$(3,0)$, and $(3,\pi)$,
$\sigma_{yx}^{\textrm{C}}$ is almost vanishing in the range of $0.5\leq u \leq 1$,
although its main term is the antisymmetric part in the range of $0\leq u \leq 0.4$.
These results are consistent with our general arguments made in Sect. 2.

We numerically evaluated Eqs. (\ref{eq:sig_yx^S}), (\ref{eq:sig_yx^C}),
and (\ref{eq:tilde-sig_xy^S}){--}(\ref{eq:bar-sig_xy^C})
using the following four procedures.
First,
we calculated the momentum summation
by dividing the first Brillouin zone into a $N_{x}\times N_{y}$ mesh.
Second,
we performed the frequency integration
using 
%No. XXXVI 12/17 No. 01-02
$\int_{-\Omega/2}^{\Omega/2}d\omega^{\prime}F(\omega^{\prime})\approx \sum_{s=0}^{W-1}\Delta\omega^{\prime} F(\omega^{\prime}_{s})$, where $\omega^{\prime}_{s}=-\Omega/2+s\Delta\omega^{\prime}$,
and $\omega^{\prime}_{W}=\Omega/2$.
Third,
we calculated the frequency derivatives of the Green's functions
by using 
$\frac{\partial F(\omega^{\prime})}{\partial \omega^{\prime}}
\approx \frac{F(\omega^{\prime}+\Delta\omega^{\prime})-F(\omega^{\prime}-\Delta\omega^{\prime})}{2\Delta\omega^{\prime}}$.
Fourth,
we replaced the summations over the Floquet indices, $\sum_{m,l,n,q=-\infty}^{\infty}$,
by $\sum_{m,l,n,q=-n_{\textrm{max}}}^{n_{\textrm{max}}}$.

In the actual calculations,
we set
$N_{x}=N_{y}=100$,
$\Delta\omega^{\prime}=0.005$ eV,
$\Gamma=0.01$ eV, $T_{\textrm{b}}=0.02$ eV,
and $\Omega=6$ eV.
This $\Omega$ is larger than the bandwidth of the nondriven system
[Fig. \ref{fig5}(b)],
which means that the light is off-resonant.
We also set $n_{\textrm{max}}=1$ except the results for $n_{\textrm{max}}=0$.
Note that the calculations with $n_{\textrm{max}}=0$ include
the light-induced corrections due to only the zeroth-order Bessel functions,
whereas the calculations with $n_{\textrm{max}}=1$ include
those due to not only the zeroth-order and first-order Bessel functions,
but also the higher-order Bessel functions (see Appendix B).
In Appendix D,
we show that 
the results obtained for $n_{\textrm{max}}=1$ and $2$
are qualitatively the same
in the case with BCPL at $\beta=2$, $\theta=\frac{\pi}{4}$, and $\Omega=6$ eV.
The similar property holds in the case with CPL~\cite{NA-FloquetSHE}.
Therefore, setting $n_{\textrm{max}}=1$ may be reasonable
in periodically driven Sr$_{2}$RuO$_{4}$ at $\Omega=6$ eV.

The reasons for presenting the numerical results are three fold.
First, the consistency between our general arguments and numerical results
supports the correctness of the results obtained in them.
Second,
the numerical results for a specific model may help to better understand 
the Onsager reciprocal relations in the periodically driven systems.
Third,
the numerical results can determine whether
the time-averaged spin and charge off-diagonal dc conductivities
are dominated by the symmetric or antisymmetric parts
even in the cases with BCPL.
In the cases with BCPL, 
the dominant terms cannot be determined from the general arguments
due to the lack of a simple relation between
the pump field and its time-reversal counterpart,
as we have explained in Sects. 2.1.2 and 2.2.2.

\begin{figure}
  \begin{center}
    \includegraphics[width=86mm]{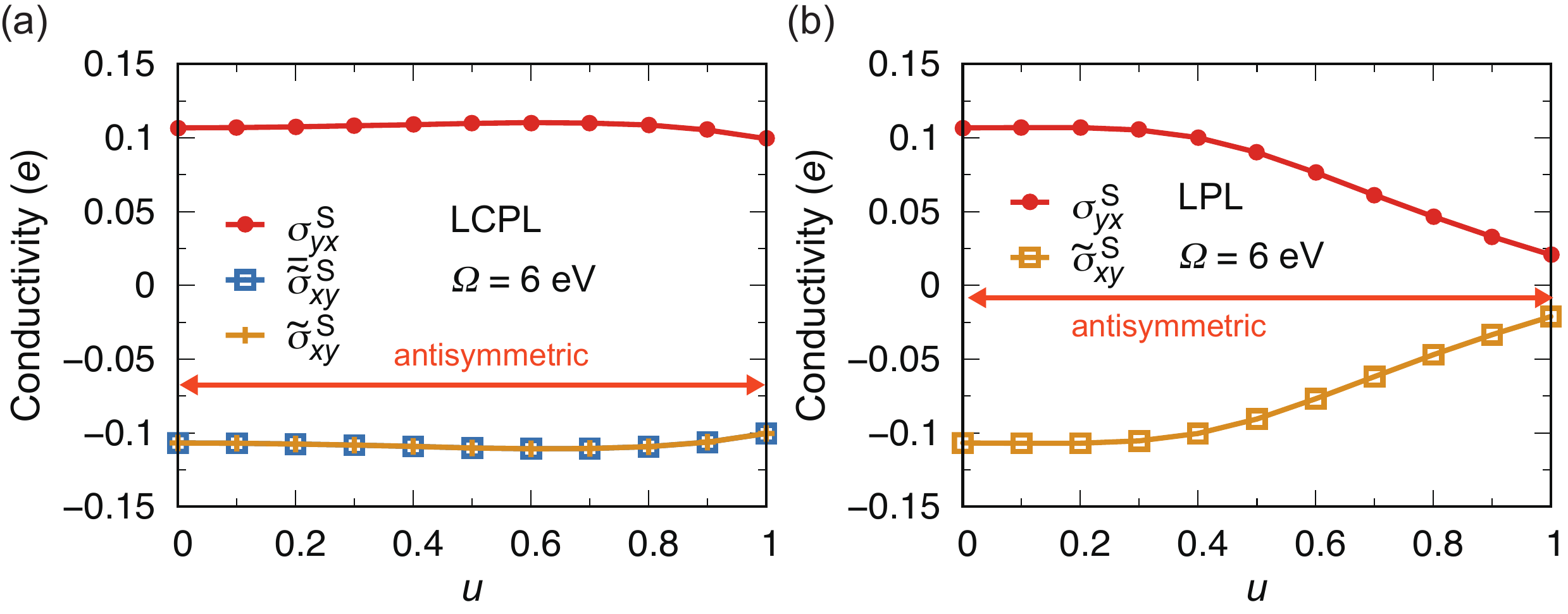}
  \end{center}
  \caption{\label{fig6}
    (Color online) The $u(=eA_{0})$ dependences of
    $\sigma_{yx}^{\textrm{S}}$, $\bar{\tilde{\sigma}}_{xy}^{\textrm{S}}$,
    and $\tilde{\sigma}_{xy}^{\textrm{S}}$
    for Sr$_{2}$RuO$_{4}$ driven by (a) LCPL
    and (b) LPL
    at $\Omega=6$ eV.
    The pump fields of LCPL and LPL have been defined
    in Eqs. (\ref{eq:A_LCPL}) and (\ref{eq:A_LPL}), respectively.
    For $\sigma_{yx}^{\textrm{S}}$, $\bar{\tilde{\sigma}}_{xy}^{\textrm{S}}$,
    and $\tilde{\sigma}_{xy}^{\textrm{S}}$,
    see Figs. \ref{fig4}(d), \ref{fig4}(e), and \ref{fig4}(f),
    respectively.
    In the case with LPL,
    $\bar{\tilde{\sigma}}_{xy}^{\textrm{S}}=\tilde{\sigma}_{xy}^{\textrm{S}}$.
  }
\end{figure}

\begin{figure*}[h]
  \begin{center}
    \includegraphics[width=140mm]{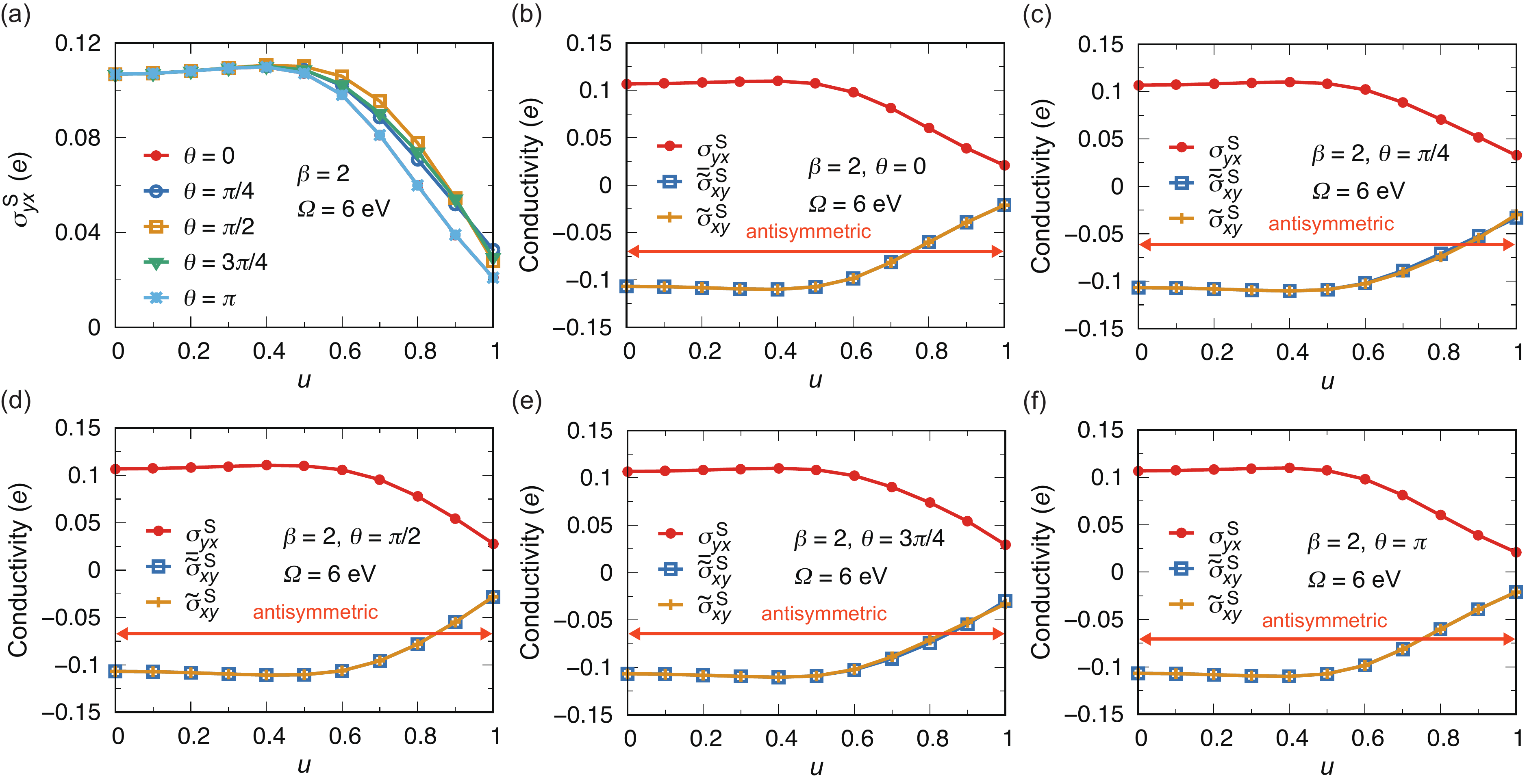}
    \end{center}
  \caption{\label{fig7}
    (Color online) (a) The $u(=eA_{0})$ dependences of $\sigma_{yx}^{\textrm{S}}$
    for Sr$_{2}$RuO$_{4}$ driven by BCPL at $\beta=2$ and $\Omega=6$ eV
    with $\theta=0$, $\frac{\pi}{4}$, $\frac{\pi}{2}$, $\frac{3\pi}{4}$, and $\pi$. 
    The pump field of BCPL has been defined in
    Eqs. (\ref{eq:Ax-BCPL}) and (\ref{eq:Ay-BCPL}).
    For $\sigma_{yx}^{\textrm{S}}$, see Fig. \ref{fig4}(d).
    The $u$ dependences of $\sigma_{yx}^{\textrm{S}}$, $\bar{\tilde{\sigma}}_{xy}^{\textrm{S}}$,
    and $\tilde{\sigma}_{xy}^{\textrm{S}}$
    for Sr$_{2}$RuO$_{4}$ driven by BCPL at $\beta=2$ and $\Omega=6$ eV 
    with (b) $\theta=0$, (c) $\theta=\frac{\pi}{4}$, (d) $\theta=\frac{\pi}{2}$,
    (e) $\theta=\frac{3\pi}{4}$, and (f) $\theta=\pi$.
    For $\bar{\tilde{\sigma}}_{xy}^{\textrm{S}}$,
    and $\tilde{\sigma}_{xy}^{\textrm{S}}$,
    see Figs. \ref{fig4}(e) and \ref{fig4}(f), respectively.
  }
\end{figure*}

\begin{figure}
  \begin{center}
    \includegraphics[width=86mm]{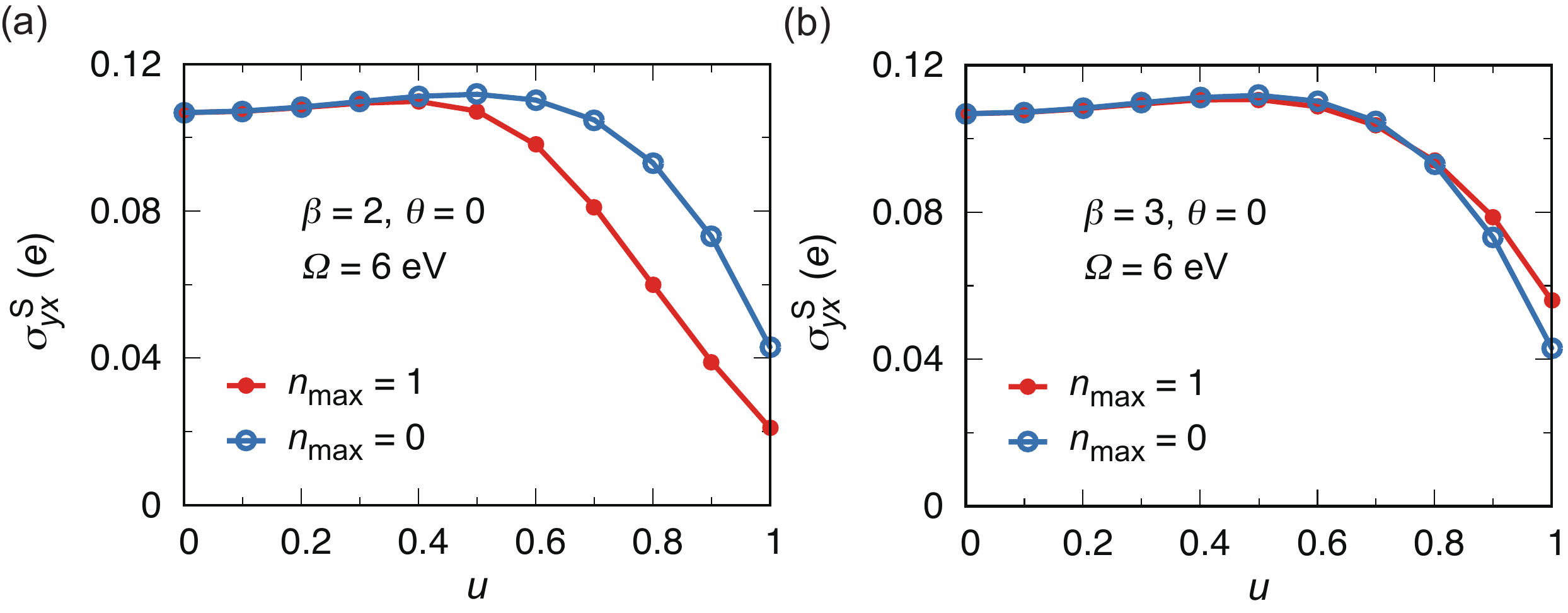}
    \end{center}
  \caption{\label{fig8}
    (Color online) The $u(=eA_{0})$ dependences of $\sigma_{yx}^{\textrm{S}}$
    for Sr$_{2}$RuO$_{4}$ driven by BCPL 
    at (a) $\beta=2$ and (b) $\beta=3$, $\theta=0$, and $\Omega=6$ eV
    with $n_{\textrm{max}}=1$ and $0$.
    Here $n_{\textrm{max}}$ is the upper limit of
    the summation over the Floquet indices, i.e., $\sum_{m,l,n,q=-n_{\textrm{max}}}^{n_{\textrm{max}}}$. 
  }
\end{figure}

\subsection{Time-averaged spin off-diagonal dc conductivities}

We compare the numerically calculated
$\sigma_{yx}^{\textrm{S}}$, $\bar{\tilde{\sigma}}_{xy}^{\textrm{S}}$,
and $\tilde{\sigma}_{xy}^{\textrm{S}}$
for Sr$_{2}$RuO$_{4}$ driven by CPL, LPL, or BCPL.
The Onsager reciprocal relations are satisfied
in all the cases considered.
Furthermore,
the main term of $\sigma_{yx}^{\textrm{S}}$
is given by the antisymmetric part.
Therefore,
$\sigma_{yx}^{\textrm{S}}$ can be regarded as the spin Hall conductivity
in all the cases considered. 

\subsubsection{Case with CPL or LPL}

Figure \ref{fig6}(a) shows
the $u$ dependences of $\sigma_{yx}^{\textrm{S}}$, $\bar{\tilde{\sigma}}_{xy}^{\textrm{S}}$,
and $\tilde{\sigma}_{xy}^{\textrm{S}}$
for Sr$_{2}$RuO$_{4}$ driven by LCPL.
These conductivities satisfy 
$\sigma_{yx}^{\textrm{S}}=-\bar{\tilde{\sigma}}_{xy}^{\textrm{S}}$
and $\sigma_{yx}^{\textrm{S}}=-\tilde{\sigma}_{xy}^{\textrm{S}}$,
which correspond to Eqs. (\ref{eq:Onsager_sigS-CPL_start})
and (\ref{eq:Onsager_sigS-CPL}), respectively. 
The former means that
$\sigma_{yx}^{\textrm{S}}$ satisfies 
the Onsager reciprocal relation, 
whereas the latter means that it is antisymmetric.
Namely, these results demonstrate the validity of Eqs. (\ref{eq:Onsager_sigS-CPL_start})
and (\ref{eq:Onsager_sigS-CPL}).

Then,
Fig. \ref{fig6}(b) shows
the $u$ dependences of $\sigma_{yx}^{\textrm{S}}$ and $\tilde{\sigma}_{xy}^{\textrm{S}}$
for Sr$_{2}$RuO$_{4}$ driven by LPL.
Note that in the case of LPL,
$\bar{\tilde{\sigma}}_{xy}^{\textrm{S}}=\tilde{\sigma}_{xy}^{\textrm{S}}$ holds.
As well as the case with CPL,
the Onsager reciprocal relation is satisfied with LPL,
and $\sigma_{yx}^{\textrm{S}}$ is given by the antisymmetric part. 
Therefore,
Eq. (\ref{eq:Onsager_sigS-LPL}) is also validated.  

\subsubsection{Cases with BCPL}

\begin{figure*}[h]
  \begin{center}
    \includegraphics[width=140mm]{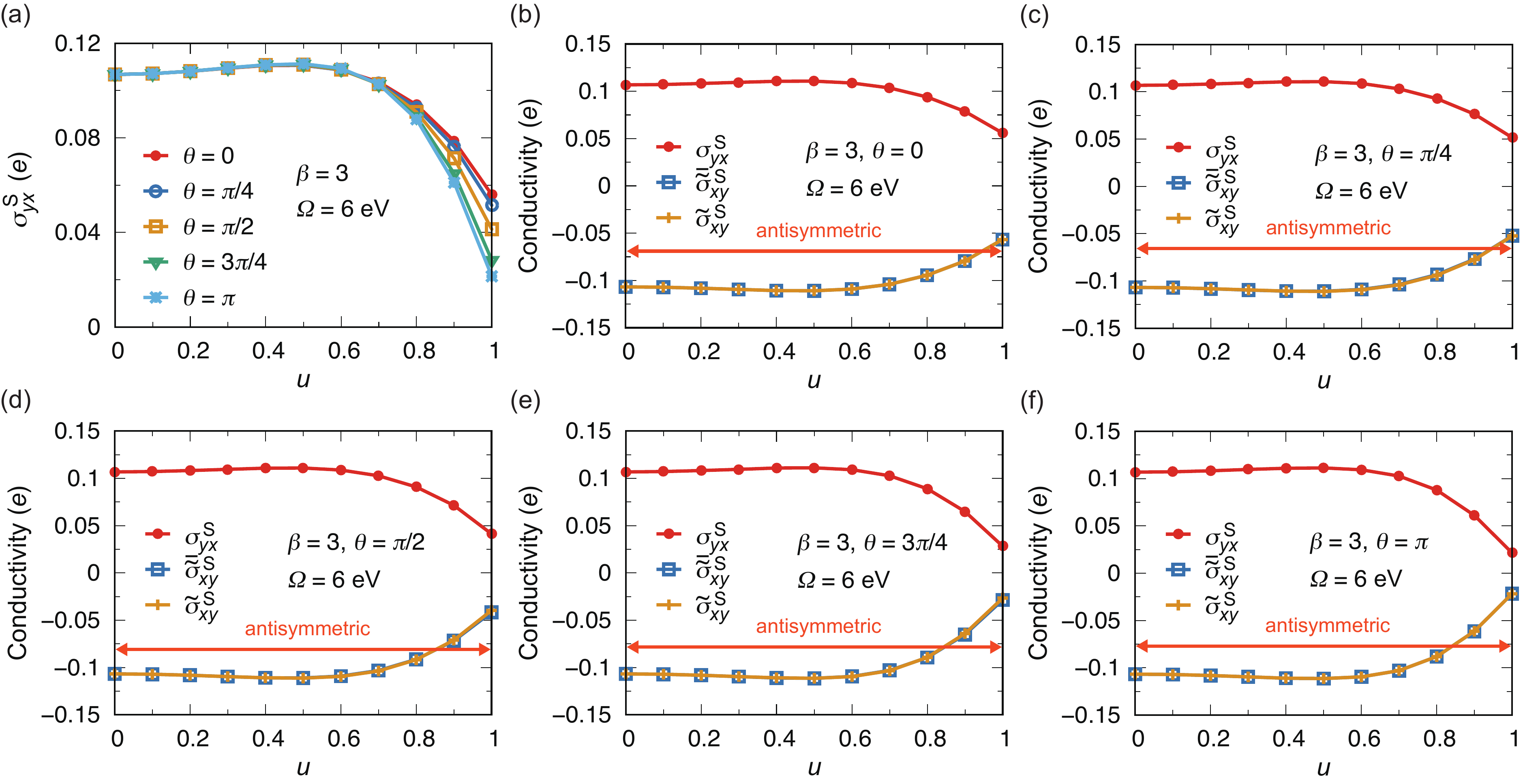}
    \end{center}
  \caption{\label{fig9}
    (Color online) (a) The $u(=eA_{0})$ dependences of $\sigma_{yx}^{\textrm{S}}$
    for Sr$_{2}$RuO$_{4}$ driven by BCPL at $\beta=3$ and $\Omega=6$ eV
    with $\theta=0$, $\frac{\pi}{4}$, $\frac{\pi}{2}$, $\frac{3\pi}{4}$, and $\pi$. 
    The pump field of BCPL has been defined in
    Eqs. (\ref{eq:Ax-BCPL}) and (\ref{eq:Ay-BCPL}).
    For $\sigma_{yx}^{\textrm{S}}$, see Fig. \ref{fig4}(d).
    The $u$ dependences of $\sigma_{yx}^{\textrm{S}}$, $\bar{\tilde{\sigma}}_{xy}^{\textrm{S}}$,
    and $\tilde{\sigma}_{xy}^{\textrm{S}}$
    for Sr$_{2}$RuO$_{4}$ driven by BCPL at $\beta=3$ and $\Omega=6$ eV
    with (b) $\theta=0$, (c) $\theta=\frac{\pi}{4}$, (d) $\theta=\frac{\pi}{2}$,
    (e) $\theta=\frac{3\pi}{4}$, and (f) $\theta=\pi$.
    For $\bar{\tilde{\sigma}}_{xy}^{\textrm{S}}$ and $\tilde{\sigma}_{xy}^{\textrm{S}}$,
    see Figs. \ref{fig4}(e) and \ref{fig4}(f), respectively.
  }
\end{figure*}

\begin{figure}
  \begin{center}
    \includegraphics[width=86mm]{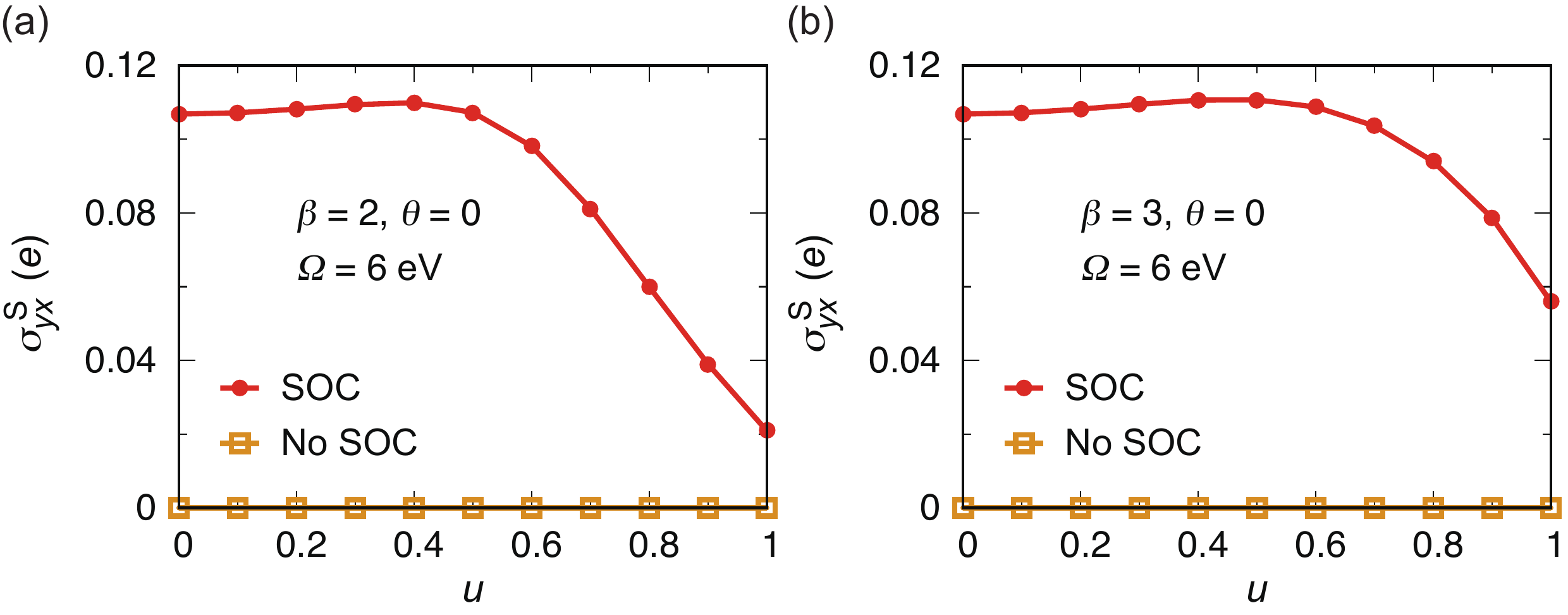}
    \end{center}
  \caption{\label{fig10}
    (Color online) The $u(=eA_{0})$ dependences of $\sigma_{yx}^{\textrm{S}}$
    for Sr$_{2}$RuO$_{4}$ driven by BCPL 
    at (a) $\beta=2$ and (b) $\beta=3$, $\theta=0$, and $\Omega=6$ eV
    with and without the SOC.
    Note that in the cases with and without the SOC,
    $\xi=0.17$ and $0$ eV, respectively.
  }
\end{figure}

\begin{figure}
  \begin{center}
    \includegraphics[width=86mm]{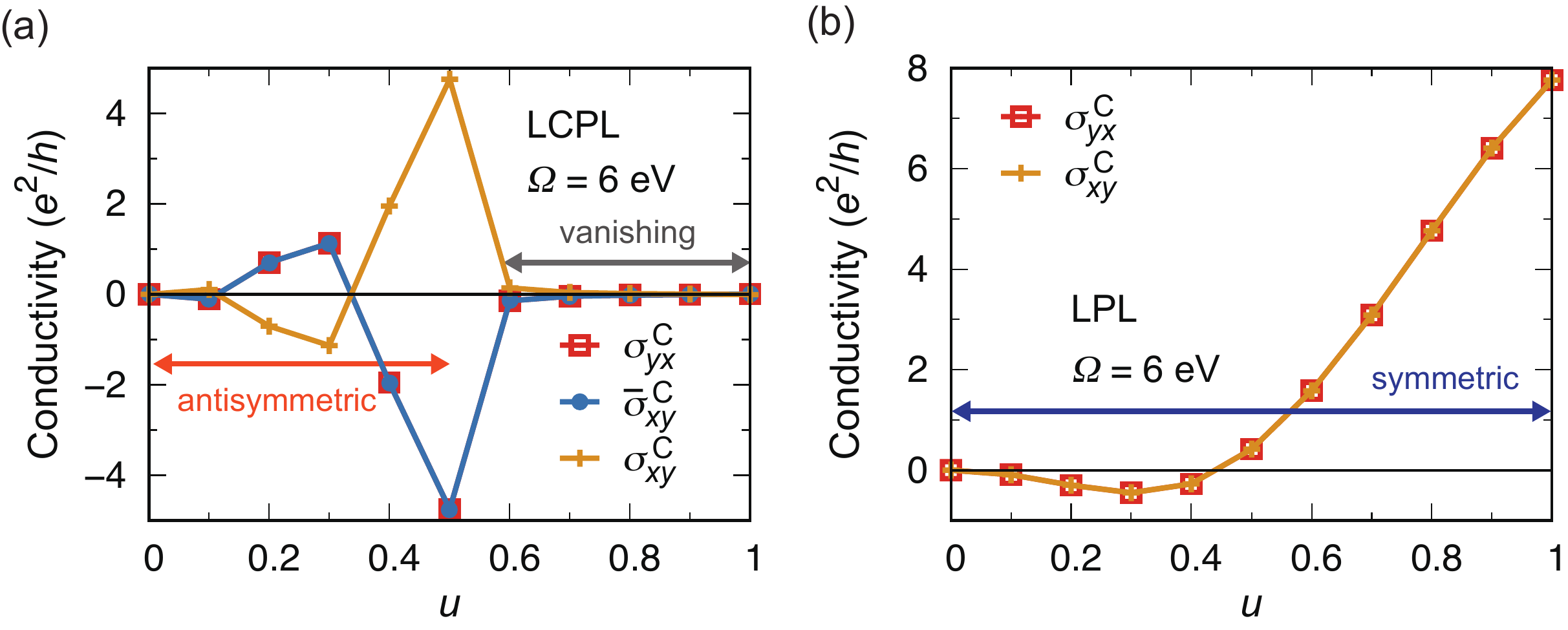}
    \end{center}
  \caption{\label{fig11}
    (Color online) The $u(=eA_{0})$ dependences of
    $\sigma_{yx}^{\textrm{C}}$, $\bar{\sigma}_{xy}^{\textrm{C}}$,
    and $\sigma_{xy}^{\textrm{C}}$
    for Sr$_{2}$RuO$_{4}$ driven by (a) LCPL and (b) LPL
    at $\Omega=6$ eV.
    The pump fields of LCPL and LPL have been defined
    in Eqs. (\ref{eq:A_LCPL}) and (\ref{eq:A_LPL}), respectively.
    For $\sigma_{yx}^{\textrm{C}}$, $\bar{\sigma}_{xy}^{\textrm{C}}$,
    and $\sigma_{xy}^{\textrm{C}}$,
    see Figs. \ref{fig4}(a), \ref{fig4}(b), and \ref{fig4}(c),
    respectively.
    In the case with LPL,
    $\bar{\sigma}_{xy}^{\textrm{C}}=\sigma_{xy}^{\textrm{C}}$.
  }
\end{figure}

\begin{figure*}[h]
  \begin{center}
    \includegraphics[width=140mm]{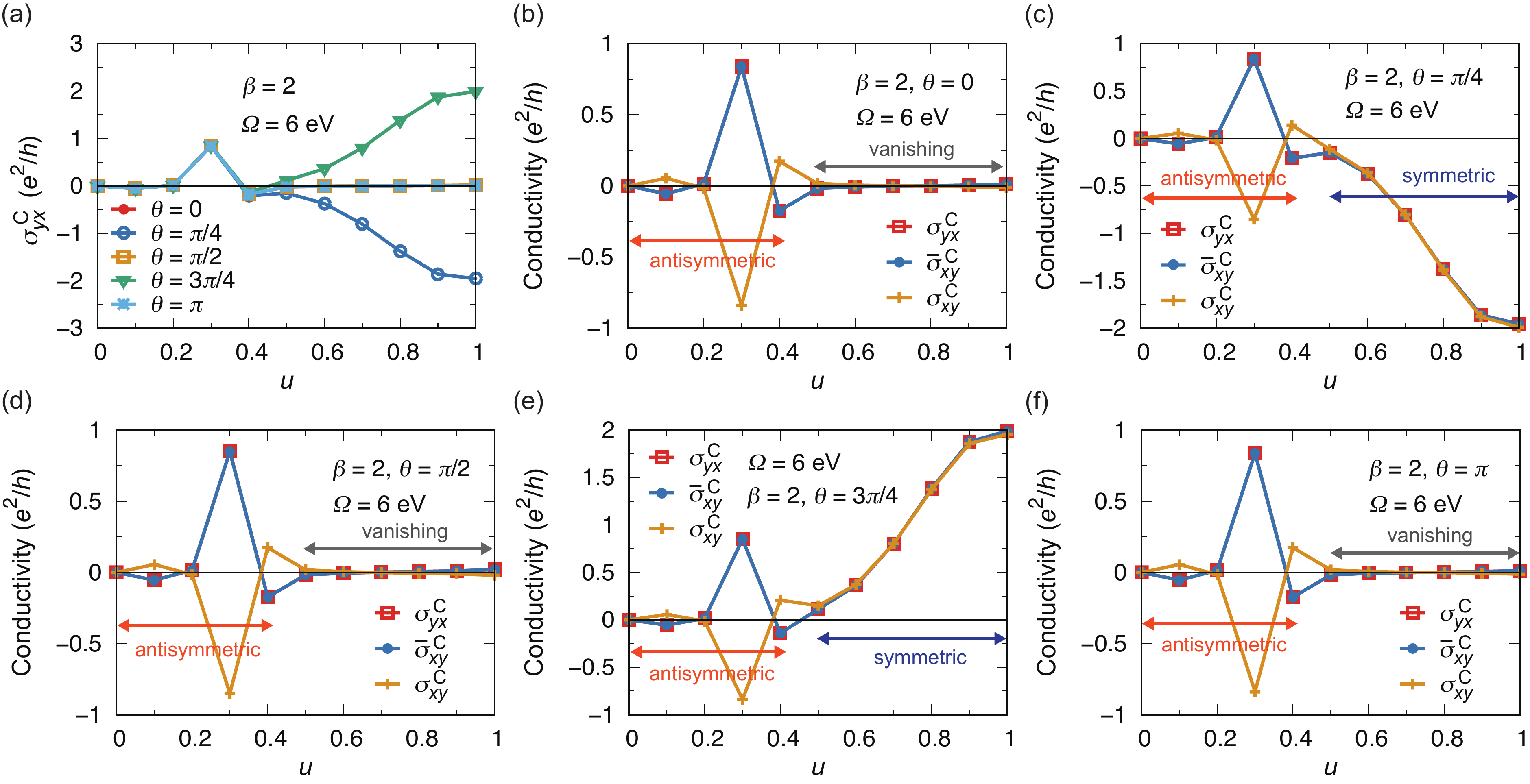}
    \end{center}
  \caption{\label{fig12}
    (Color online) (a) The $u(=eA_{0})$ dependences of $\sigma_{yx}^{\textrm{C}}$
    for Sr$_{2}$RuO$_{4}$ driven by BCPL at $\beta=2$ and $\Omega=6$ eV
    with $\theta=0$, $\frac{\pi}{4}$, $\frac{\pi}{2}$, $\frac{3\pi}{4}$, and $\pi$. 
    The pump field of BCPL has been defined in
    Eqs. (\ref{eq:Ax-BCPL}) and (\ref{eq:Ay-BCPL}).
    For $\sigma_{yx}^{\textrm{C}}$, see Fig. \ref{fig4}(a).
    The $u$ dependences of $\sigma_{yx}^{\textrm{C}}$, $\bar{\sigma}_{xy}^{\textrm{C}}$,
    and $\sigma_{xy}^{\textrm{C}}$
    for Sr$_{2}$RuO$_{4}$ driven by BCPL at $\beta=2$ and $\Omega=6$ eV
    with (b) $\theta=0$, (c) $\theta=\frac{\pi}{4}$, (d) $\theta=\frac{\pi}{2}$,
    (e) $\theta=\frac{3\pi}{4}$, and (f) $\theta=\pi$.
    For $\bar{\sigma}_{xy}^{\textrm{C}}$ and $\sigma_{xy}^{\textrm{C}}$,
    see Figs. \ref{fig4}(b) and \ref{fig4}(c), respectively.
  }
\end{figure*}

\begin{figure}
  \begin{center}
    \includegraphics[width=86mm]{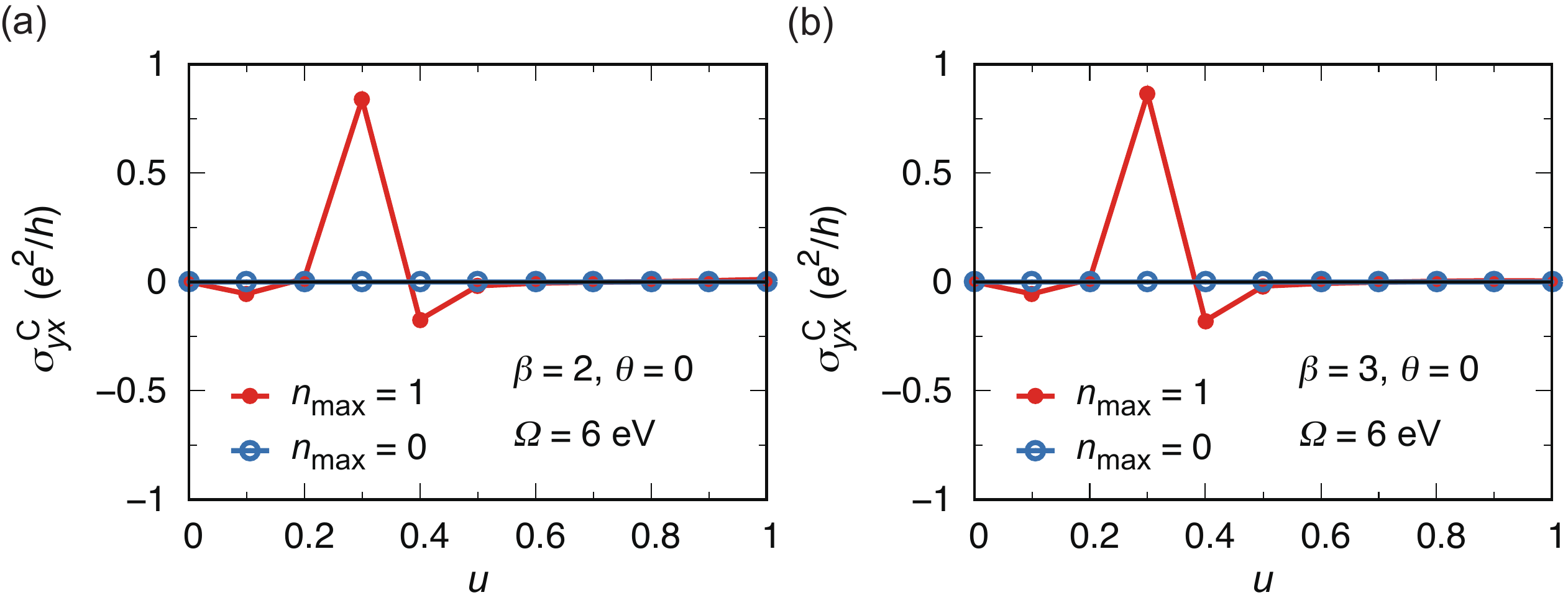}
    \end{center}
  \caption{\label{fig13}
    (Color online) The $u(=eA_{0})$ dependences of $\sigma_{yx}^{\textrm{C}}$
    for Sr$_{2}$RuO$_{4}$ driven by BCPL 
    at (a) $\beta=2$ and (b) $\beta=3$, $\theta=0$, and $\Omega=6$ eV 
    with $n_{\textrm{max}}=1$ and $0$.
    Here $n_{\textrm{max}}$ is the upper limit of
    the summation over the Floquet indices (i.e., $\sum_{m,l,n,q=-n_{\textrm{max}}}^{n_{\textrm{max}}}$). 
  }
\end{figure}

\begin{figure*}[h]
  \begin{center}
    \includegraphics[width=140mm]{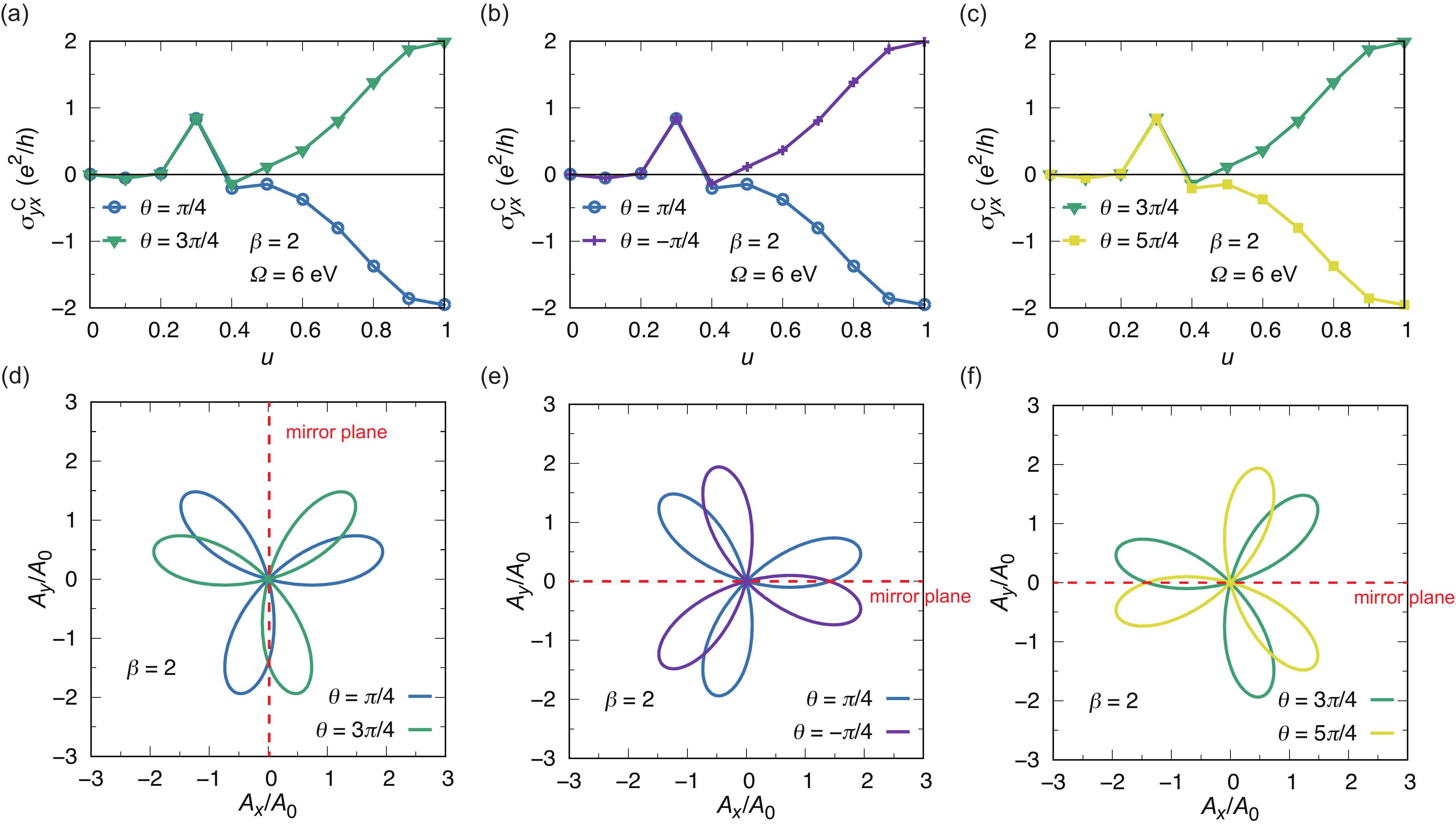}
    \end{center}
  \caption{\label{fig14}
    (Color online) The $u(=eA_{0})$ dependences of $\sigma_{yx}^{\textrm{C}}$
    for Sr$_{2}$RuO$_{4}$ driven by BCPL at $\beta=2$ and $\Omega=6$ eV
    with (a) $\theta=\frac{\pi}{4}$ and $\frac{3\pi}{4}$,
    (b) $\theta=\frac{\pi}{4}$ and $-\frac{\pi}{4}$,
    and (c) $\theta=\frac{3\pi}{4}$ and $\frac{5\pi}{4}$.
    The trajectories of the pump fields of BCPL per period $T_{\textrm{p}}$ at $\beta=2$
    with 
    (d) $\theta=\frac{\pi}{4}$ and $\frac{3\pi}{4}$,
    (e) $\theta=\frac{\pi}{4}$ and $-\frac{\pi}{4}$,
    and (f) $\theta=\frac{3\pi}{4}$ and $\frac{5\pi}{4}$.
    The two trajectories in panel (d) are connected by
    the mirror operation about the $A_{x}=0$ plane,
    whereas those in panel (e) or (f) are connected by
    the mirror operation about the $A_{y}=0$ plane. 
  }
\end{figure*}

\begin{figure*}[h]
  \begin{center}
    \includegraphics[width=140mm]{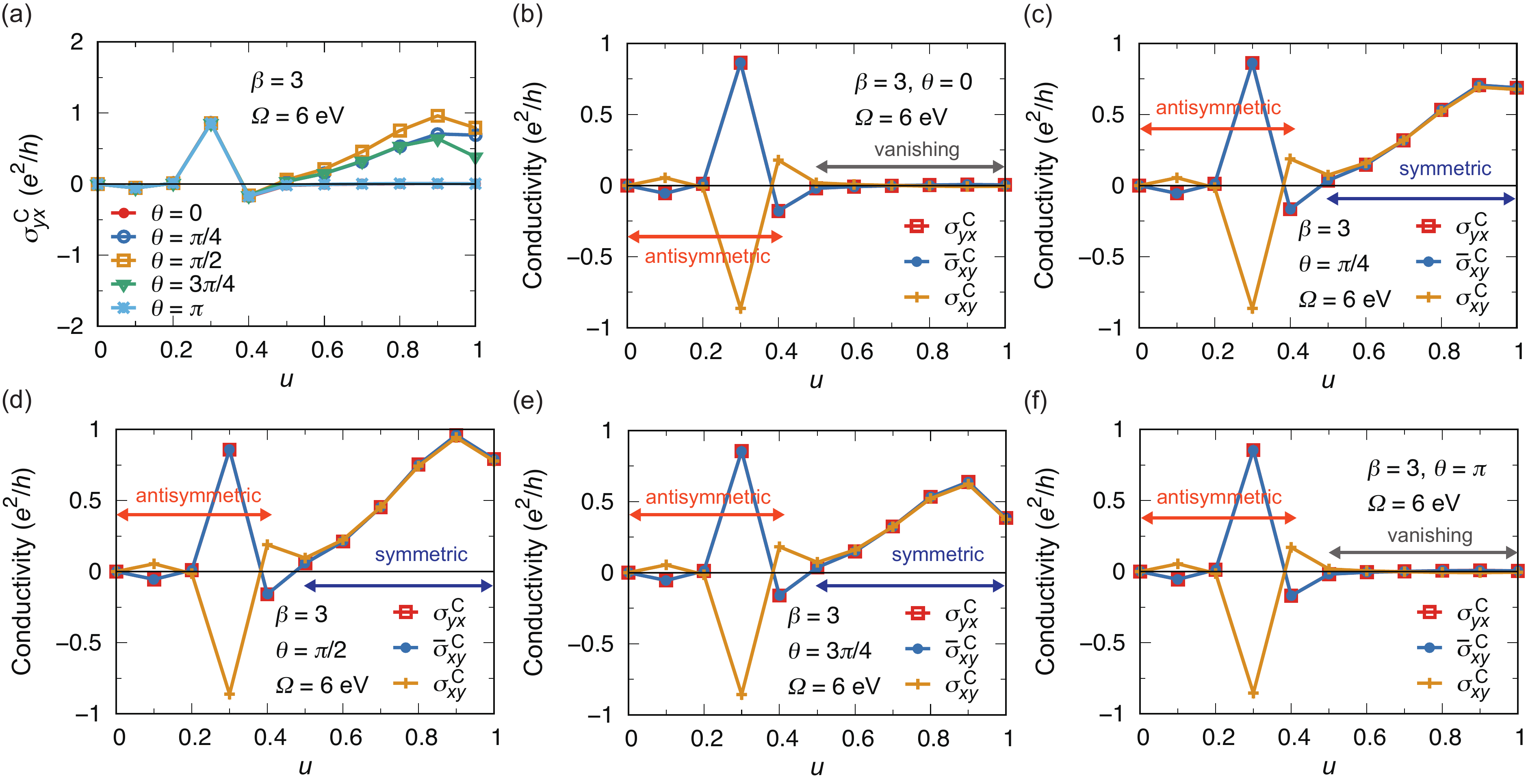}
    \end{center}
  \caption{\label{fig15}
    (Color online) (a) The $u(=eA_{0})$ dependences of $\sigma_{yx}^{\textrm{C}}$
    for Sr$_{2}$RuO$_{4}$ driven by BCPL at $\beta=3$ and $\Omega=6$ eV
    with $\theta=0$, $\frac{\pi}{4}$, $\frac{\pi}{2}$, $\frac{3\pi}{4}$, and $\pi$. 
    $A_{0}$, $\beta$, and $\theta$ of BCPL have been defined in
    Eqs. (\ref{eq:Ax-BCPL}) and (\ref{eq:Ay-BCPL}).
    For $\sigma_{yx}^{\textrm{C}}$, see Fig. \ref{fig4}(a).
    The $u$ dependences of $\sigma_{yx}^{\textrm{C}}$, $\bar{\sigma}_{xy}^{\textrm{C}}$,
    and $\sigma_{xy}^{\textrm{C}}$
    for Sr$_{2}$RuO$_{4}$ driven by BCPL at $\beta=3$ and $\Omega=6$ eV
    with (b) $\theta=0$, (c) $\theta=\frac{\pi}{4}$, (d) $\theta=\frac{\pi}{2}$,
    (e) $\theta=\frac{3\pi}{4}$, and (f) $\theta=\pi$.
    For $\bar{\sigma}_{xy}^{\textrm{C}}$ and $\sigma_{xy}^{\textrm{C}}$,
    see Figs. \ref{fig4}(b) and \ref{fig4}(c), respectively.
  }
\end{figure*}

\begin{figure*}[h]
  \begin{center}
    \includegraphics[width=140mm]{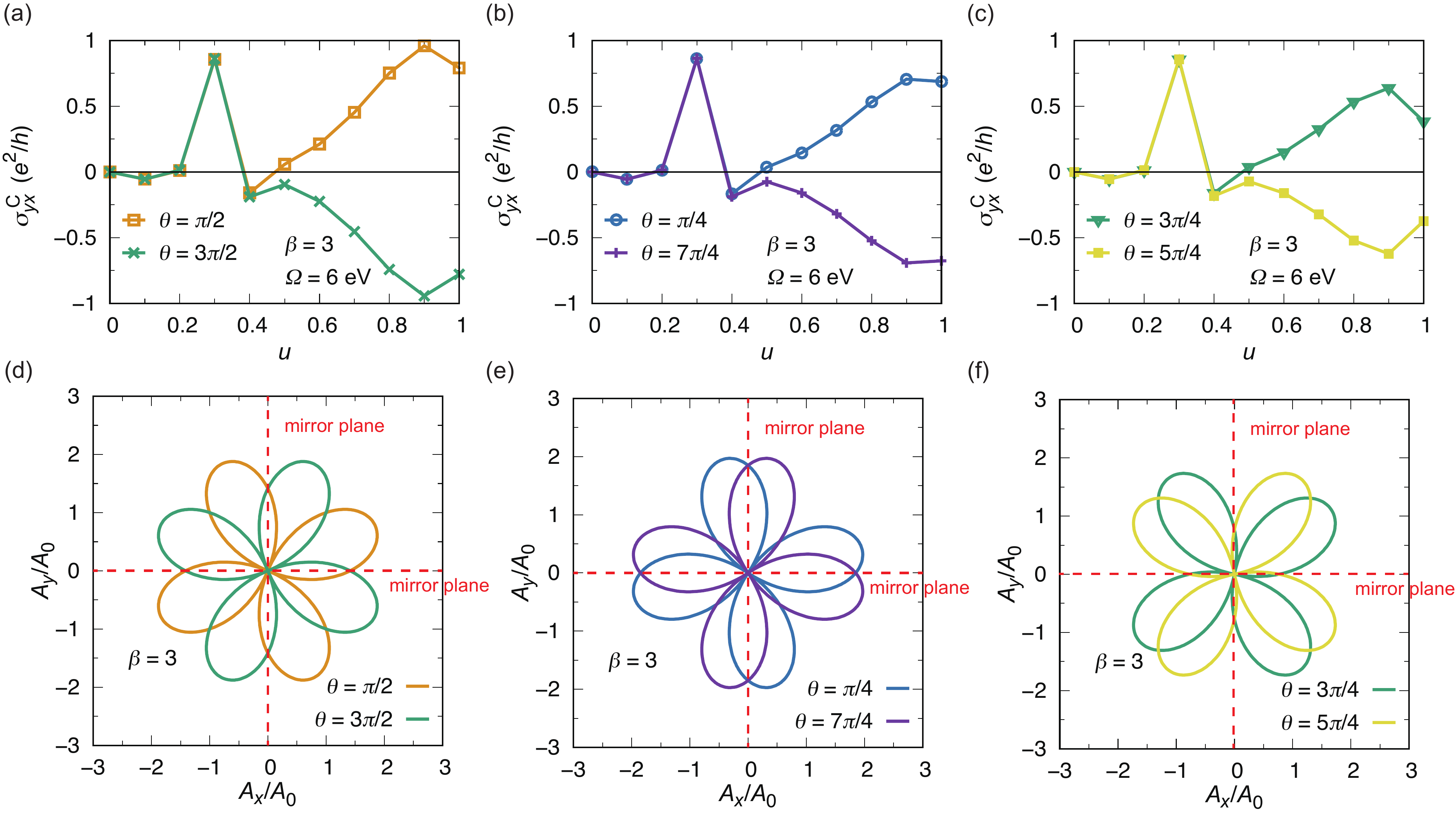}
    \end{center}
  \caption{\label{fig16}
    (Color online) The $u(=eA_{0})$ dependences of $\sigma_{yx}^{\textrm{C}}$
    for Sr$_{2}$RuO$_{4}$ driven by BCPL at $\beta=3$ and $\Omega=6$ eV
    with (a) $\theta=\frac{\pi}{2}$ and $\frac{3\pi}{2}$,
    (b) $\theta=\frac{\pi}{4}$ and $\frac{7\pi}{4}$,
    and (c) $\theta=\frac{3\pi}{4}$ and $\frac{5\pi}{4}$.
    The trajectories of the pump fields of BCPL per period $T_{\textrm{p}}$ at $\beta=3$ 
    with (d) $\theta=\frac{\pi}{2}$ and $\frac{3\pi}{2}$,
    (e) $\theta=\frac{\pi}{4}$ and $\frac{7\pi}{4}$,
    and (f) $\theta=\frac{3\pi}{4}$ and $\frac{5\pi}{4}$.
    The two trajectories in each panel are connected by
    the mirror operation about the $A_{x}=0$ or $A_{y}=0$ plane. 
  }
\end{figure*}

\begin{figure}
  \begin{center}
    \includegraphics[width=86mm]{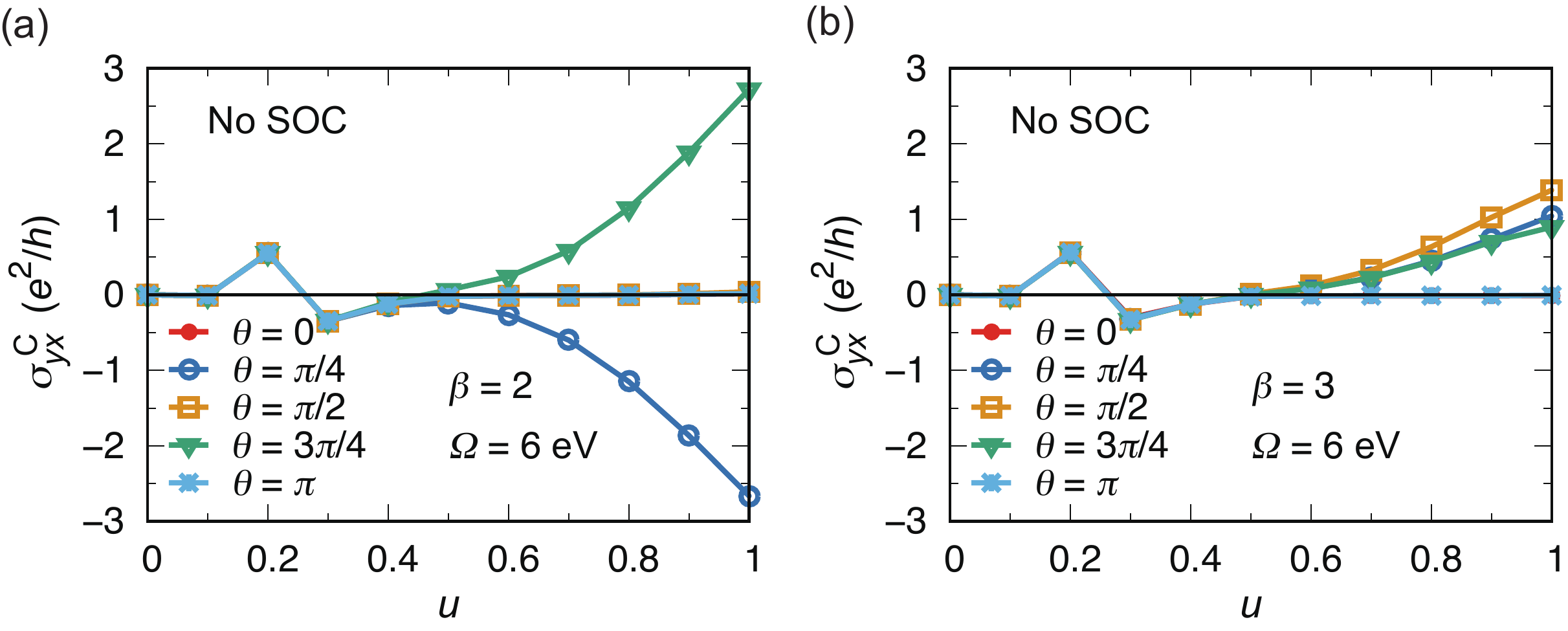}
    \end{center}
  \caption{\label{fig17}
    (Color online) The $u(=eA_{0})$ dependences of $\sigma_{yx}^{\textrm{C}}$
    for Sr$_{2}$RuO$_{4}$ driven by BCPL 
    at (a) $\beta=2$ and (b) $\beta=3$ and $\Omega=6$ eV 
    with $\theta=0$, $\frac{\pi}{4}$, $\frac{\pi}{2}$, $\frac{3\pi}{4}$, and $\pi$ 
    in the absence of SOC.
  }
\end{figure}

Figure \ref{fig7}(a) 
shows the $u$ dependences of $\sigma_{yx}^{\textrm{S}}$
for Sr$_{2}$RuO$_{4}$ driven by BCPL at $\beta=2$
and $\theta=0$, $\frac{\pi}{4}$, $\frac{\pi}{2}$, $\frac{3\pi}{4}$, and $\pi$.
$\sigma_{yx}^{\textrm{S}}$ is almost independent of $\theta$
in the range of $0\leq u \leq 0.4$,
whereas it depends on $\theta$ in the range of $0.5\leq u \leq 1$.
This $\theta$ dependence may arise from the light-induced corrections due to
the higher-order Bessel functions for moderately large $u$,
as we will discuss in Sect. 6.1.
The nearly monotonically decreasing $u$ dependences may be due to
the dynamical localization,
the reduction in the kinetic energy due to the zeroth-order Bessel function
in the Peierls phase factor.
Note that the dynamical localization can be described by the time-averaged Hamiltonian,
in which 
the zeroth-order Bessel function leads to the correction to the hopping integrals.
In fact,
the almost monotonically decreasing $u$ dependence 
can be reproduced by
the $\sigma_{yx}^{\textrm{S}}$ calculated numerically
with $n_{\textrm{max}}=0$,
in which 
only the zeroth-order Bessel function gives the light-induced corrections [Fig. \ref{fig8}(a)].

In addition,
Figs. \ref{fig7}(b){--}\ref{fig7}(f)
show the $u$ dependences of
$\sigma_{yx}^{\textrm{S}}$, $\bar{\tilde{\sigma}}_{xy}^{\textrm{S}}$,
and $\tilde{\sigma}_{xy}^{\textrm{S}}$
for Sr$_{2}$RuO$_{4}$ driven by BCPL at $\beta=2$
and $\theta=0$, $\frac{\pi}{4}$, $\frac{\pi}{2}$, $\frac{3\pi}{4}$,
and $\pi$.  
As well as the cases with CPL and LPL,
the Onsager reciprocal relation
$\sigma_{yx}^{\textrm{S}}=-\bar{\tilde{\sigma}}_{xy}^{\textrm{S}}$ holds.
Furthermore,
the main term is the antisymmetric part.
This may be surprising because
the Onsager reciprocal relation in this case
does not restrict $\sigma_{yx}^{\textrm{S}}$
to the antisymmetric part [see Eq. (\ref{eq:Onsager_sigS-BCPL})].

The similar results are obtained at $\beta=3$.
Figure \ref{fig9}(a) shows the $u$ dependences of $\sigma_{yx}^{\textrm{S}}$
in Sr$_{2}$RuO$_{4}$ driven by BCPL 
at $\beta=3$ and $\theta=0$, $\frac{\pi}{4}$, $\frac{\pi}{2}$, $\frac{3\pi}{4}$,
and $\pi$. 
$\sigma_{yx}^{\textrm{S}}$ is almost independent of $\theta$
in the range of $0\leq u\leq 0.7$,
whereas it depends on $\theta$ in the range of $0.8\leq u \leq 1$.
As we will discuss in Sect. 6.1,
we can understand this $\theta$ dependence in a similar way to that obtained at $\beta=2$.
Then,
Figs. \ref{fig9}(b){--}\ref{fig9}(f) show the relations
among $\sigma_{yx}^{\textrm{S}}$, $\bar{\tilde{\sigma}}_{xy}^{\textrm{S}}$,
and $\tilde{\sigma}_{xy}^{\textrm{S}}$ as functions of $u$
at $\beta=3$ and $\theta=0$, $\frac{\pi}{4}$, $\frac{\pi}{2}$, $\frac{3\pi}{4}$,
and $\pi$.
At these $\theta$'s, 
the Onsager reciprocal relation holds
and the antisymmetric part gives the main contribution.
Note that the almost monotonically decreasing $u$ dependence
is reproducible by the $n_{\textrm{max}}=0$ terms [Fig. \ref{fig8}(b)].

Before showing the results of the time-averaged charge off-diagonal dc conductivities, 
we comment on the role of the SOC.
Figure \ref{fig10}(a) 
compares the $u$ dependences of $\sigma_{yx}^{\textrm{S}}$
for Sr$_{2}$RuO$_{4}$ driven by BCPL at $\beta=2$ and $\theta=0$
with and without the SOC.
[Note that in our model, the SOC is the $LS$ coupling,
  as shown in Eq. (\ref{eq:Hs-start}).]
This result suggests that
the SOC is vital for achieving the finite $\sigma_{yx}^{\textrm{S}}$.
This is the same as the property for Sr$_{2}$RuO$_{4}$ driven by CPL~\cite{NA-FloquetSHE}.
In addition,
the same conclusion is obtained with BCPL at $\beta=3$ and $\theta=0$,
as shown in Fig. \ref{fig10}(b).
Therefore,
we conclude that
the SOC plays the vital role in the SHE of
the periodically driven multiorbital metals. 

\subsection{Time-averaged charge off-diagonal dc conductivities}

We now compare $\sigma_{yx}^{\textrm{C}}$, $\bar{\sigma}_{xy}^{\textrm{C}}$,
and $\sigma_{xy}^{\textrm{C}}$
for Sr$_{2}$RuO$_{4}$ driven by CPL, LPL, or BCPL.
In all the cases considered, 
the Onsager reciprocal relations are satisfied.
In addition, in the case with CPL or LPL, 
$\sigma_{yx}^{\textrm{C}}$ is given by the antisymmetric or symmetric part, respectively.
Therefore, 
$\sigma_{yx}^{\textrm{C}}$ with CPL can be regarded as the anomalous Hall conductivity. 
Then, in the cases with BCPL
at $(\beta,\theta)=(2,\frac{\pi}{4})$, $(2,\frac{3\pi}{4})$,
$(3,\frac{\pi}{4})$, $(3,\frac{\pi}{2})$, and $(3,\frac{3\pi}{4})$,
the main term is given by the antisymmetric part for small $u$ and
by the symmetric part for moderately large $u$. 
Meanwhile, 
in the cases with BCPL at
$(\beta,\theta)=(2,0)$, $(2,\frac{\pi}{2})$, $(2,\pi)$,
$(3,0)$, and $(3,\pi)$,
$\sigma_{yx}^{\textrm{C}}$ for moderately large $u$ is almost vanishing,
although its main term for small $u$ is given by the antisymmetric part. 
The unusual crossover between the antisymmetric and symmetric parts
may result from
the mixture of these parts
and the accidentally small value of the antisymmetric part
for the moderately strong magnitude of BCPL.

\subsubsection{Case with CPL or LPL}

Figure \ref{fig11}(a) shows the $u$ dependences of
$\sigma_{yx}^{\textrm{C}}$, $\bar{\sigma}_{xy}^{\textrm{C}}$,
and $\sigma_{xy}^{\textrm{C}}$
for Sr$_{2}$RuO$_{4}$ driven by LCPL at $\Omega=6$ eV.
Equations (\ref{eq:Onsager_sigC-CPL_start}) and (\ref{eq:Onsager_sigC-CPL}),
i.e., $\sigma_{yx}^{\textrm{C}}=\bar{\sigma}_{xy}^{\textrm{C}}$
and $\sigma_{yx}^{\textrm{C}}=-\sigma_{xy}^{\textrm{C}}$, hold. 
These results indicate that 
$\sigma_{yx}^{\textrm{C}}$ satisfies the Onsager reciprocal relation
and is given by the antisymmetric part.
This is consistent with the numerical result obtained
in graphene driven by CPL~\cite{NA-Onsager}.

Then,
Fig. \ref{fig11}(b) shows the $u$ dependences of
$\sigma_{yx}^{\textrm{C}}$ and $\sigma_{xy}^{\textrm{C}}$
for Sr$_{2}$RuO$_{4}$ driven by LPL at $\Omega=6$ eV.
Note that in this case $\bar{\sigma}_{xy}^{\textrm{C}}=\sigma_{xy}^{\textrm{C}}$.
In contrast to the case with CPL,
$\sigma_{yx}^{\textrm{C}}$ is given by the symmetric part,
although it satisfies the Onsager reciprocal relation
[i.e., Eq. (\ref{eq:Onsager_sigC-LPL})].
This result also agrees with that obtained
in graphene driven by LPL~\cite{NA-Onsager}.

\subsubsection{Cases with BCPL}

Figure \ref{fig12}(a) shows the $u$ dependences of $\sigma_{yx}^{\textrm{C}}$
for Sr$_{2}$RuO$_{4}$ driven by BCPL at $\beta=2$ with
$\theta=0$, $\frac{\pi}{4}$, $\frac{\pi}{2}$, $\frac{3\pi}{4}$, and $\pi$.
$\sigma_{yx}^{\textrm{C}}$ is almost $\theta$-independent in the range of $0\leq u \leq 0.4$,
whereas it is $\theta$-dependent in the range of $0.5\leq u \leq 1$.
This $\theta$ dependence can be understood
in a similar way to the origin of the $\theta$-dependent $\sigma_{yx}^{\textrm{S}}$ (see Sect. 6.1).
In contrast to $\sigma_{yx}^{\textrm{S}}$,
the magnitude and sign of 
$\sigma_{yx}^{\textrm{C}}$ can change with increasing $u$.
This may be because
the finite $\sigma_{yx}^{\textrm{C}}$ arises from the light-induced contributions.
In fact,
$\sigma_{yx}^{\textrm{C}}=0$ if only the $n_{\textrm{max}}=0$ terms are considered
[Fig. \ref{fig13}(a)].

Figures \ref{fig12}(b){--}\ref{fig12}(f)
compare $\sigma_{yx}^{\textrm{C}}$, $\bar{\sigma}_{xy}^{\textrm{C}}$,
and $\sigma_{xy}^{\textrm{C}}$ as functions of $u$
for Sr$_{2}$RuO$_{4}$ driven by BCPL at $\beta=2$
with $\theta=0$, $\frac{\pi}{4}$, $\frac{\pi}{2}$, $\frac{3\pi}{4}$, and $\pi$.
First,
$\sigma_{yx}^{\textrm{C}}=\bar{\sigma}_{xy}^{\textrm{C}}$ holds at these $\theta$'s.
Namely, the Onsager reciprocal relation 
Eq. (\ref{eq:Onsager_sigC-BCPL}) is numerically validated. 
Then, 
at $\theta=0$, $\frac{\pi}{2}$, and $\pi$,
$\sigma_{yx}^{\textrm{C}}$ for $u\leq 0.4$ is dominated by the antisymmetric part,
whereas $\sigma_{yx}^{\textrm{C}}$ for moderately larger $u$ is almost vanishing.
Meanwhile,
at $\theta=\frac{\pi}{4}$ and $\frac{3\pi}{4}$,
the antisymmetric part is the main term of $\sigma_{yx}^{\textrm{C}}$ for $u\leq 0.4$,
but the symmetric part becomes the main term for larger $u$. 
This suggests that
for Sr$_{2}$RuO$_{4}$ driven by BCPL at $\beta=2$ with some $\theta$'s,
the main term of $\sigma_{yx}^{\textrm{C}}$ can be changed
from the antisymmetric part to the symmetric part or vice versa
by tuning the magnitude of the pump field of BCPL.
This unusual crossover may be surprising, but
it does not contradict the Onsager reciprocal relation
because this relation with BCPL cannot be reduced to
the antisymmetric or symmetric part in general (see Sect. 2.1.2).

The above crossover might be due to 
the mixture of the antisymmetric and symmetric parts
and the vanishingly small antisymmetric part for large $u$.
As we can see from Figs. \ref{fig12}(a) and \ref{fig12}(b),
the $\theta$-independent terms of $\sigma_{yx}^{\textrm{C}}$ are
finite and dominated by the antisymmetric part in the range of $0\leq u\leq 0.4$,
whereas they become almost vanishing in the range of $0.5\leq u\leq 1$.
In addition, as we can see from Figs. \ref{fig12}(a) and \ref{fig12}(c){--}\ref{fig12}(f), 
the $\theta$-dependent terms become non-negligible only in the range of $0.5\leq u\leq 1$
and are dominated by the symmetric part.
These results imply that
the crossover between the antisymmetric and symmetric parts with changing $u$
might arise from
a combination of the mixture of these parts and 
the accidentally small antisymmetric part in the range of $0.5\leq u\leq 1$.
As we have explained in Sect. 2.1.2,
the Onsager reciprocal relation with BCPL does not exclude
a possibility of such a mixture.
In fact, such a mixture can be realized
in another periodically driven system with BCPL~\cite{NA-BCPL},
although the antisymmetric part is dominant
for all the $u$'s in the range of $0\leq u\leq 1$.
We should note that
the accidentally small antisymmetric part may be characteristic of this model,
but its origin is difficult to be clarified.
  
The symmetric relation between $\sigma_{yx}^{\textrm{C}}$'s 
at a couple of $\theta$'s for moderately large $u$
could be understood
in terms of the symmetry of $\mathbf{A}_{\textrm{BCPL}}(t)$. 
Figure \ref{fig12}(a) or \ref{fig14}(a) shows that 
$\sigma_{yx}^{\textrm{C}}$'s at $\theta=\frac{\pi}{4}$ and $\frac{3\pi}{4}$
at a certain $u$ in the range of $0.5\leq u\leq 1$
are of almost the same magnitude and of opposite sign.
This may be characteristic of the symmetric part of $\sigma_{yx}^{\textrm{C}}$
because $\sigma_{yx}^{\textrm{C}}$ in the range of $0.5\leq u\leq 1$
is dominated by the symmetric part [Figs. \ref{fig12}(c) and \ref{fig12}(e)].
The similar property is achieved in graphene driven by LPL
if $\sigma_{yx}^{\textrm{C}}$'s with the pump fields of LPL connected by a mirror operation
are compared~\cite{NA-Onsager}.
As we can see from Fig. \ref{fig14}(d),
the trajectories of $\mathbf{A}_{\textrm{BCPL}}(t)$'s at $\theta=\frac{\pi}{4}$ and $\frac{3\pi}{4}$
are connected by a mirror operation about the $A_{x}=0$ plane.
Therefore,
the symmetric relation between
$\sigma_{yx}^{\textrm{C}}$'s at $\theta=\frac{\pi}{4}$ and $\frac{3\pi}{4}$
in the range of $0.5\leq u\leq 1$
could be linked to the mirror symmetry of the trajectories of $\mathbf{A}_{\textrm{BCPL}}(t)$'s.
This interpretation remains valid
even if we compare $\sigma_{yx}^{\textrm{C}}$'s
at $\theta=\frac{\pi}{4}$ and -$\frac{\pi}{4}$ [Fig. \ref{fig14}(b)]
or at $\theta=\frac{3\pi}{4}$ and $\frac{5\pi}{4}$ [Fig. \ref{fig14}(c)]
in the range of $0.5\leq u\leq 1$.
The trajectories at $\theta=\frac{\pi}{4}$ and -$\frac{\pi}{4}$
or at $\theta=\frac{3\pi}{4}$ and $\frac{5\pi}{4}$
are connected by another mirror operation about the $A_{y}=0$ plane
[see Figs. \ref{fig14}(e) and \ref{fig14}(f)].

The similar properties hold at $\beta=3$.
First, 
$\sigma_{yx}^{\textrm{C}}$ is almost independent of $\theta$
in the range of $0\leq u\leq 0.4$
and dependent on it in the range of $0.5\leq u\leq 1$ [see Fig. \ref{fig15}(a)].
Second,
the light-induced terms are vital for achieving the finite $\sigma_{yx}^{\textrm{C}}$
[see Fig. \ref{fig13}(b)].
Third,
the Onsager reciprocal relation 
$\sigma_{yx}^{\textrm{C}}=\bar{\sigma}_{xy}^{\textrm{C}}$ holds
at $\theta=0$, $\frac{\pi}{4}$, $\frac{\pi}{2}$, $\frac{3\pi}{4}$,
and $\pi$ [see Figs. \ref{fig15}(b){--}\ref{fig15}(f)].
Fourth,
$\sigma_{yx}^{\textrm{C}}$ in the range of $0\leq u\leq 0.4$
is dominated by the antisymmetric part at these $\theta$'s,
whereas $\sigma_{yx}^{\textrm{C}}$ in the range of $0.5\leq u\leq 1$
is almost vanishing at $\theta=0$ and $\pi$ 
and dominated by the symmetric part at $\frac{\pi}{4}$, $\frac{\pi}{2}$, and $\frac{3\pi}{4}$
[see Figs. \ref{fig15}(b){--}\ref{fig15}(f)].
Fifth,
$\sigma_{yx}^{\textrm{C}}$'s at $\theta=\frac{\pi}{2}$ and $\frac{3\pi}{2}$,
at $\theta=\frac{\pi}{4}$ and $\frac{7\pi}{4}$,
or at $\theta=\frac{3\pi}{4}$ and $\frac{5\pi}{4}$ in the range of $0.5\leq u\leq 1$
are of almost the same magnitude and of opposite sign
[see Figs. \ref{fig16}(a){--}\ref{fig16}(c)]; 
and the trajectories of $\mathbf{A}_{\textrm{BCPL}}(t)$'s at each couple
are connected by the mirror operation
about the $A_{x}=0$ or $A_{y}=0$ plane [see Figs. \ref{fig16}(d){--}\ref{fig16}(f)].

Finally, we remark on the role of the SOC.
Figures \ref{fig17}(a) and \ref{fig17}(b) show 
the $u$ dependences of $\sigma_{yx}^{\textrm{C}}$ without SOC
for Sr$_{2}$RuO$_{4}$ driven by BCPL at $\beta=2$ and $3$, respectively,
with $\theta=0$, $\frac{\pi}{4}$, $\frac{\pi}{2}$, $\frac{3\pi}{4}$, and $\pi$.
Comparing Figs. \ref{fig17}(a) and \ref{fig12}(a) or
Figs. \ref{fig17}(b) and \ref{fig15}(a),
we find that
the effect of the SOC on $\sigma_{yx}^{\textrm{C}}$ is not large.
In particular,
$\sigma_{yx}^{\textrm{C}}$ can be finite even without the SOC.
This is in contrast to the vital role of the SOC in $\sigma_{yx}^{\textrm{S}}$
and similar to the property obtained in Sr$_{2}$RuO$_{4}$ driven by CPL~\cite{NA-FloquetSHE}.
These results suggest that 
the SOC is not important in discussing $\sigma_{yx}^{\textrm{C}}$
of the periodically driven multiorbital metals.

\section{Discussion}

\subsection{Origin of the $\theta$ dependences of $\sigma_{yx}^{\textrm{S}}$ and $\sigma_{yx}^{\textrm{C}}$}

First, we discuss the origin of the $\theta$ dependences of
$\sigma_{yx}^{\textrm{S}}$ and $\sigma_{yx}^{\textrm{C}}$
for Sr$_{2}$RuO$_{4}$ driven by BCPL at $\beta=2$ and $3$.
As we have shown
in Figs. \ref{fig7}(a), \ref{fig9}(a), \ref{fig12}(a), and \ref{fig15}(a),
$\sigma_{yx}^{\textrm{S}}$ and $\sigma_{yx}^{\textrm{C}}$
are almost independent of $\theta$ for small $u$,
whereas they depend on $\theta$ for moderately large $u$.
This property could be understood
by discussing the $u$ and $\theta$ dependences of
the kinetic energy terms $[\epsilon_{ab}(\mathbf{k})]_{mn}$'s
because
the effects of $\mathbf{A}_{\textrm{BCPL}}(t)$
are taken into account via the Peierls phase factor in the kinetic energy
[see Eqs. (\ref{eq:e_ab-kt}) and (\ref{eq:epsilon_mn})].
Let us consider $[\epsilon_{d_{yz}d_{yz}}(\mathbf{k})]_{mn}$ for example.
As shown in Appendix B,
$[\epsilon_{d_{yz}d_{yz}}(\mathbf{k})]_{mn}$ is given by
\begin{align}
  %No. XXXXV 7/31 No. 01-02
  [\epsilon_{d_{yz}d_{yz}}(\mathbf{k})]_{mn}
  =-t_{2}I_{x}^{mn}(k_{x},u,\theta,\beta)
  -t_{1}I_{y}^{mn}(k_{y},u,\theta,\beta),
\end{align}
where
\begin{align}
  %No. XXXXV 7/31 No. 01-02
  I_{x}^{mn}(k_{x},u,\theta,\beta)
  &=\sum_{l=-\infty}^{\infty}i^{n-m-(\beta-1)l}e^{-il\theta}
  \mathcal{J}_{n-m-\beta l}(u)\mathcal{J}_{l}(u)\notag\\
          &\times[e^{-ik_{x}}(-1)^{n-m-(\beta-1)l}+e^{ik_{x}}],\label{eq:Ix}\\
  I_{y}^{mn}(k_{y},u,\theta,\beta)
  &=\sum_{l=-\infty}^{\infty}e^{-il\theta}(-1)^{l}
  \mathcal{J}_{n-m-\beta l}(u)\mathcal{J}_{l}(u)\notag\\
          &\times[e^{-ik_{y}}(-1)^{n-m-(\beta-1)l}+e^{ik_{y}}],\label{eq:Iy}
\end{align}
and $\mathcal{J}_{l}(u)$ is the $l$th-order Bessel function of the first kind. 
For simplicity,
we restrict the Floquet indices $m$, $n$, and $l$ to be
$-1\leq m$, $n$, $l\leq 1$, which corresponds to the case at $n_{\textrm{max}}=1$.
After some calculation,
we obtain $I_{x}^{mn}(k_{x},u,\theta,\beta)$'s at $\beta=2$ and $3$,
\begin{align}
  %No. XXXXV 7/31 No. 09-10
  I_{x}^{mn}(k_{x},u,\theta,2)
  &\sim 2\delta_{n-m,0}\cos k_{x}\mathcal{J}_{0}(u)^{2}\notag\\
  &-2(\delta_{n-m,1}+\delta_{n-m,-1})\sin k_{x}\mathcal{J}_{0}(u)\mathcal{J}_{1}(u)\notag\\
  &-2(e^{-i\theta}\delta_{n-m,1}+e^{i\theta}\delta_{n-m,-1})\cos k_{x}\mathcal{J}_{1}(u)^{2}\notag\\
  &-2(\delta_{n-m,2}+\delta_{n-m,-2})\cos k_{x}\mathcal{J}_{0}(u)\mathcal{J}_{2}(u)\notag\\
  &-2(e^{-i\theta}\delta_{n-m,2}+e^{i\theta}\delta_{n-m,-2})\sin k_{x}
  \mathcal{J}_{0}(u)\mathcal{J}_{1}(u)\notag\\
  &+O(u^{3}),\label{eq:Ix_beta-2_Bessel}
\end{align}
and
\begin{align}
  %No. XXXXV 7/31 No. 11-12
  I_{x}^{mn}(k_{x},u,\theta,3)
  &\sim 2\delta_{n-m,0}\cos k_{x}\mathcal{J}_{0}(u)^{2}\notag\\
  &-2(\delta_{n-m,1}+\delta_{n-m,-1})\sin k_{x}\mathcal{J}_{0}(u)\mathcal{J}_{1}(u)\notag\\  
  &-2(\delta_{n-m,2}+\delta_{n-m,-2})\cos k_{x}\mathcal{J}_{0}(u)\mathcal{J}_{2}(u)\notag\\ 
  &-2(e^{-i\theta}\delta_{n-m,2}+e^{i\theta}\delta_{n-m,-2})\cos k_{x}\mathcal{J}_{1}(u)^{2}\notag\\
  &+O(u^{3}).\label{eq:Ix_beta-3_Bessel}
\end{align}
These equations show that
the $\theta$ dependences arise from
the light-induced corrections due to the finite-order Bessel functions.
Note that 
$I_{y}^{mn}(k_{y},u,\theta,\beta)$'s at $\beta=2$ and $3$ have 
the same property. 
By using the series expansions of the Bessel functions,
we can express Eqs. (\ref{eq:Ix_beta-2_Bessel}) and (\ref{eq:Ix_beta-3_Bessel})
as follows:
\begin{align}
  %No. XXXXV 7/31 No. 09-10
  I_{x}^{mn}(k_{x},u,\theta,2)
  &\sim \delta_{n-m,0}(2-u^{2})\cos k_{x}\notag\\
  &-(\delta_{n-m,1}+\delta_{n-m,-1})u\sin k_{x}\notag\\
  &-(e^{-i\theta}\delta_{n-m,1}+e^{i\theta}\delta_{n-m,-1})\frac{u^{2}}{2}\cos k_{x}\notag\\
  &-(\delta_{n-m,2}+\delta_{n-m,-2})\frac{u^{2}}{4}\cos k_{x}\notag\\
  &-(e^{-i\theta}\delta_{n-m,2}+e^{i\theta}\delta_{n-m,-2})u\sin k_{x}\notag\\
  &+O(u^{3}),\label{eq:Ix_beta-2}
\end{align}
and
\begin{align}
  %No. XXXXV 7/31 No. 09-10
  I_{x}^{mn}(k_{x},u,\theta,3)
  &\sim \delta_{n-m,0}(2-u^{2})\cos k_{x}\notag\\
  &-(\delta_{n-m,1}+\delta_{n-m,-1})u\sin k_{x}\notag\\
  &-(\delta_{n-m,2}+\delta_{n-m,-2})\frac{u^{2}}{4}\cos k_{x}\notag\\
  &-(e^{-i\theta}\delta_{n-m,2}+e^{i\theta}\delta_{n-m,-2})\frac{u^{2}}{2}\cos k_{x}\notag\\
  &+O(u^{3}).
  \label{eq:Ix_beta-3}
\end{align}
These equations can explain
the $\theta$-independent $\sigma_{yx}^{\textrm{S}}$ for small $u$
because the terms for $n-m=0$ are primary (if they give the finite contribution). 
Although the terms for $n-m\neq 0$ are necessary for $\sigma_{yx}^{\textrm{C}}\neq 0$,
these equations can also explain 
the $\theta$-independent $\sigma_{yx}^{\textrm{C}}$ for small $u$,
at which the terms for $n-m=\pm 1$ are dominant.
This is because in the small-$u$ region, 
the $\theta$-independent $u^{1}$ terms are more important
than the $\theta$-dependent $u^{2}$ terms.
For moderately large $u$,
the $\theta$-dependent $u^{2}$ terms for $n-m=\pm 1$ may become non-negligible.
Therefore,
the $\theta$ dependences of
$\sigma_{yx}^{\textrm{S}}$ and $\sigma_{yx}^{\textrm{C}}$
could be interpreted in terms of
the $u$ and $\theta$ dependences of
the light-induced corrections to the kinetic energy.

\subsection{Comparisons with other relevant studies}

We now compare our results about the Onsager reciprocal relations
with other relevant studies.
As described in Sect. 1,
there are several studies~\cite{Onsager-PD1,Onsager-PD2,Onsager-PD3,NA-Onsager}
about the Onsager reciprocal relations in periodically driven systems.
In two of them~\cite{Onsager-PD1,Onsager-PD2},
the time-periodic field was treated as perturbation.
Meanwhile, in our theory,
the time-periodic pump field has been treated nonperturbatively. 
In general, the time-periodic field should be treated nonperturbatively
to describe periodically driven systems
because its nonperturbative effects often change the electronic states drastically.
Therefore, the systems studied in these papers~\cite{Onsager-PD1,Onsager-PD2}
are insufficient to discuss the standard periodically driven systems.
Then, in another~\cite{Onsager-PD3},
the Onsager reciprocal relations
about the matter or energy transport
between a periodically driven system and one of the three reservoirs were studied.
This is distinct from the transport phenomena
within periodically driven systems.
Therefore, except our previous paper~\cite{NA-Onsager},
there has been no study of the Onsager reciprocal relations
for any transport within periodically driven systems
in which the nonperturbative effects of the driving field
are taken into account.
In that paper~\cite{NA-Onsager},
we numerically studied
the time-averaged charge off-diagonal dc conductivities
$\sigma_{yx}^{\textrm{C}}$ and $\sigma_{xy}^{\textrm{C}}$
in graphene driven by CPL or LPL 
and found that $\sigma_{yx}^{\textrm{C}}=-\sigma_{xy}^{\textrm{C}}$
or $\sigma_{yx}^{\textrm{C}}=\sigma_{xy}^{\textrm{C}}$ holds with CPL or LPL, respectively.
However, 
we did not discuss the generality of these relations,
i.e., we just showed that they are satisfied in our numerical results.
Meanwhile, in this paper,
we have made general arguments
about the Onsager reciprocal relations for spin and charge transport
within the periodically driven systems.
These arguments are applicable to many periodically driven systems
and extendable to others.
Therefore,
this work is the first systematic study of the Onsager reciprocal relations
in periodically driven systems.
Most importantly, 
this paper is the first work demonstrating the Onsager reciprocal relations
for the spin transport in periodically driven systems. 

Next, we compare our results in the case of BCPL at $\beta=3$ with
our previous study~\cite{NA-BCPL} for graphene driven by BCPL at $\beta=3$.
(Although we have also studied the case at $\beta=2$ there~\cite{NA-BCPL},
its results are not directly comparable to the results obtained in this paper
due to the vital role of the valley degree of freedom
in that case studied in Ref. \cite{NA-BCPL}.)
There are three similarities between 
our results shown in Sect. 5.2.2
and those obtained in this previous study~\cite{NA-BCPL}:
$\sigma_{yx}^{\textrm{C}}$ is almost independent of
$\theta$ for small $u$
and depends on it for moderately large $u$;
$\sigma_{yx}^{\textrm{C}}$ for small $u$ is dominated by the antisymmetric part;
and the antisymmetric and symmetric parts are mixed for moderately large $u$
at $\theta=\frac{\pi}{2}$.
Therefore, these properties may be characteristic properties
of $\sigma_{yx}^{\textrm{C}}$ in systems driven by BCPL. 
Then, the difference is that
the antisymmetric part of $\sigma_{yx}^{\textrm{C}}$
becomes almost vanishing for moderately large $u$
in Sr$_{2}$RuO$_{4}$ driven by BCPL at $\theta=0$ and $\pi$,
whereas
the antisymmetric part remains dominant even for moderately large $u$
in graphene driven by BCPL at these $\theta$'s~\cite{NA-BCPL}.

We turn to the comparisons with 
some theoretical studies~\cite{BCPL-pert1,BCPL-pert2,BCPL-pert3}
about transport properties with
BCPL at $\beta=2$. 
In these studies, 
the effects of the BCPL on charge or spin transport 
have been analyzed in perturbation theories. 
First,
our results about the effects of the SOC are similar to
those on the charge and spin conductivities
of another three-orbital electron system having the $z$ component of the SOC
in the second-order perturbation theory against the BCPL field~\cite{BCPL-pert2}.
This similarity may be reasonable 
because the SOC is treated nonperturbatively in both cases.
Then, our $\theta$-dependent $\sigma_{yx}^{\textrm{C}}$
is similar to the $\theta$-dependent charge conductivity obtained
in the third-order perturbation theory against the BCPL field~\cite{BCPL-pert3}.
However,
a critical value of $u$ above which
$\sigma_{yx}^{\textrm{C}}$ depends on $\theta$
exists only in our case [see Fig. \ref{fig12}(a)].
This critical value may arise from
the nonlinear $u$ terms of the light-induced corrections,
as we have discussed in Sect. 6.1.
Therefore,
the crossover between
the $\theta$-independent and $\theta$-dependent $\sigma_{yx}^{\textrm{C}}$
may be characteristic of nonperturbative effects of BCPL.
This interpretation remains valid in graphene driven by BCPL~\cite{NA-BCPL}. 
Although the authors of Ref. \cite{BCPL-pert3}
have claimed that a nonperturbative effect is also discussed,
their discussions are insufficient 
because they have used an approximation for the Hamiltonian, 
$H_{0}[k+A(t)]\approx H_{0}(k)+A(t)\frac{\partial H_{0}(k)}{\partial k}$
[see Eq. (29) of Supplemental Material of Ref. \cite{BCPL-pert3}],
which is valid only if $A(t)$, the vector potential of the BCPL,
can be treated perturbatively. 
Namely, their theory can analyze only the perturbative effects of the BCPL.
We do not use such an approximation; 
instead, we have analyzed the nonperturbative effects of BCPL
in the standard Floquet linear-response theory~\cite{Tsuji,Mikami,Eckstein,NA-FloquetSHE}.

\subsection{Experimental realization}

Finally, we comment on experimental realization of our results. 
We have supposed that
our periodically driven open system can reach 
a nonequilibrium steady state 
due to the damping $\Gamma$.
The AHE predicted in such a periodically driven open system~\cite{Oka-PRB}
was experimentally observed~\cite{Light-AHE-exp2}.
Therefore,
our time-averaged charge off-diagonal dc conductivities
could be experimentally observed in the pump-probe measurements
for periodically driven Sr$_{2}$RuO$_{4}$.
In addition,
our SHE could be detected via the inverse SHE~\cite{InvSHE1,InvSHE2} 
because we have demonstrated that
$\sigma_{yx}^{\textrm{S}}$ satisfies the Onsager reciprocal relation,
which is similar to that in nondriven systems,
and is dominated by the antisymmetric part.
Note that
$\tilde{\sigma}_{xy}^{\textrm{S}}$ could be experimentally observed
by measuring the charge current perpendicular to
the probe spin field [see Fig. \ref{fig4}(f)].
As the probe spin field,
we can use, for example, 
the gradient of the Zeeman field,
which could be experimentally realized~\cite{MagFieldGrad}.
In the case of Sr$_{2}$RuO$_{4}$,
%No. XXXIX 7/5 No. 01-02
$u=0.1$ for $\Omega=6$ eV corresponds to $E_{0}\approx 15.4$ MV/cm,
where we have used $u=ea_{\textrm{NN}}A_{0}=ea_{\textrm{NN}}E_{0}/\Omega$
and $a_{\textrm{NN}}\approx 0.39$ nm~\cite{Sr2RuO4-aNN}.
From an experimental point of view,
the pump field of the order of $10$ MV/cm can be realized~\cite{Iwai-review}.
Therefore,
it may be possible to observe
the $\theta$-dependent $\sigma_{yx}^{\textrm{C}}$ and $\sigma_{yx}^{\textrm{S}}$
for moderately large $u$
as well as the $\theta$-independent $\sigma_{yx}^{\textrm{C}}$ and $\sigma_{yx}^{\textrm{S}}$
for small $u$. 
Although 
there is a possibility that
the similar $\theta$-dependent $\sigma_{yx}^{\textrm{C}}$ and $\sigma_{yx}^{\textrm{S}}$ will be realized
even for smaller $\Omega$ and $E_{0}$,
it is difficult to check this possibility
due to the huge cost of the numerical calculations.
Note that
in the case of CPL~\cite{NA-FloquetSHE},
the $u$ dependence of 
the $\sigma_{yx}^{\textrm{S}}$ or $\sigma_{yx}^{\textrm{C}}$ obtained for $\Omega=6$ eV
is similar to that obtained for smaller $\Omega$.

\section{Conclusions}

In summary, we have theoretically established
the Onsager reciprocal relations for the charge and spin transport
in the periodically driven systems.
We have made general arguments about these relations
for $\sigma_{yx}^{\textrm{C}}$ and $\sigma_{yx}^{\textrm{S}}$
in the periodically driven systems with CPL, LPL, and BCPL.
We have shown that
$\sigma_{yx}^{\textrm{C}}$ and $\sigma_{yx}^{\textrm{S}}$ satisfy
the Onsager reciprocal relations in all the cases considered
and that
their main terms depend on the polarization of light.
In the case with CPL or LPL,
$\sigma_{yx}^{\textrm{C}}$ is dominated by the antisymmetric or symmetric part, respectively,
whereas $\sigma_{yx}^{\textrm{S}}$ is dominated by the antisymmetric part.
Meanwhile,
in the case with BCPL,
$\sigma_{yx}^{\textrm{C}}$ and $\sigma_{yx}^{\textrm{S}}$ are not restricted
to either the antisymmetric or symmetric part generally. 
Then,
we have numerically analyzed 
$\sigma_{yx}^{\textrm{C}}$, $\sigma_{yx}^{\textrm{S}}$, and
the other time-averaged off-diagonal dc conductivities appearing in the Onsager reciprocal relations
by applying the Floquet linear-response theory to
the model of periodically driven Sr$_{2}$RuO$_{4}$.
We have demonstrated the validity of our arguments.
In addition,
we have shown in the cases with BCPL that
the main term of $\sigma_{yx}^{\textrm{S}}$ is given by the antisymmetric part,
whereas that of $\sigma_{yx}^{\textrm{C}}$ depends on the magnitude of the pump field.
More precisely, 
$\sigma_{yx}^{\textrm{C}}$ for small $u$ is dominated by the antisymmetric part,
whereas 
$\sigma_{yx}^{\textrm{C}}$ for moderately large $u$
is almost vanishing or dominated by the symmetric part. 

Our arguments and numerical calculations have demonstrated
that $\sigma_{yx}^{\textrm{S}}$ in the periodically driven systems
satisfies the Onsager reciprocal relation. 
This means that
the finite $\sigma_{yx}^{\textrm{S}}$ in the periodically driven systems
can be indirectly observed
by measuring the time-averaged dc conductivity of the inverse SHE.
Therefore, 
our results open the way for detecting the spin current in periodically driven systems
via the inverse SHE.
Since such an indirect detection method has been widely used in
many spintronics phenomena of nondriven systems~\cite{Bauer},
this achievement is useful to develop and observe many spintronics phenomena
in periodically driven systems.
Therefore, our results provide 
a vital step
towards comprehensive understanding and further development
of spintronics in periodically driven systems.

Our arguments have shown an essential difference between the Onsager reciprocal relations
for $\sigma_{yx}^{\textrm{C}}$ and $\sigma_{yx}^{\textrm{S}}$.
The difference is that
$\sigma_{yx}^{\textrm{C}}$ possesses the antisymmetric part only without time-reversal symmetry,
whereas 
$\sigma_{yx}^{\textrm{S}}$ possesses the antisymmetric part even with it.
This is consistent with the fact that
the AHE is possible with broken time-reversal symmetry,
whereas the SHE is possible even with time-reversal symmetry.
This difference is due to the difference between
the time-reversal symmetries of 
the charge and spin currents.

Moreover, our arguments have resolved
the contradictory statements about the Onsager reciprocal relation
for the spin transport
in nondriven systems~\cite{IncorrectOnsager-JS1,IncorrectOnsager-JS2,CorrectOnsager-JS1,CorrectOnsager-JS2}.
There is a previous study claiming the violation of the Onsager reciprocal relation
for the spin transport~\cite{IncorrectOnsager-JS2},
whereas there are other studies claiming
its existence~\cite{IncorrectOnsager-JS1,CorrectOnsager-JS1,CorrectOnsager-JS2}.
Furthermore, among the latter studies,
the expression of the Onsager reciprocal relation is different:
in one of them~\cite{IncorrectOnsager-JS1}
the spin off-diagonal dc conductivity is dominated by
the symmetric part,
whereas in the others~\cite{CorrectOnsager-JS1,CorrectOnsager-JS2}
it is dominated by the antisymmetric part.
As we have shown in Sect. 2.2.1,
our results are consistent with the results of
the last two studies~\cite{CorrectOnsager-JS1,CorrectOnsager-JS2},
i.e.,
the spin off-diagonal dc conductivity satisfies 
the Onsager reciprocal relation even with time-reversal symmetry
and is dominated by the antisymmetric part. 
Therefore,
our arguments support the validity of the interpretation of
the inverse SHE~\cite{InvSHE1,InvSHE2}
as the existence of the spin current in nondriven systems.

Then, our numerical calculations in the cases with BCPL
have indicated that
$\sigma_{yx}^{\textrm{C}}$ cannot necessarily be regarded as the anomalous Hall conductivity
even if time-reversal symmetry is broken.
In general,
we can regard $\sigma_{yx}^{\textrm{C}}$ as the anomalous Hall conductivity
if and only if $\sigma_{yx}^{\textrm{C}}$ is dominated by the antisymmetric part.
In addition,
time-reversal symmetry can be broken by BCPL~\cite{NA-BCPL}.
As shown in Sect. 5.2.2,
$\sigma_{yx}^{\textrm{C}}$ with BCPL for moderately large $u$
is almost vanishing or dominated by the symmetric part,
although that for small $u$ is dominated by the antisymmetric part.
This unusual property does not contradict the Onsager reciprocal relation
because this relation in the case with BCPL restricts $\sigma_{yx}^{\textrm{C}}$
to neither the antisymmetric nor symmetric part generally (see Sect. 2.1.2).
It may be due to the lack of a simple relation between
the pump field of BCPL
and its time-reversal counterpart.
In contrast,
the pump field of CPL has the simple relation 
$\mathbf{A}_{\textrm{LCPL}}(-t)=\mathbf{A}_{\textrm{RCPL}}(t)$; 
as a result, 
$\sigma_{yx}^{\textrm{C}}$ with CPL is restricted to the antisymmetric part (see Sect. 2.1.2).
These results suggest that
even if time-reversal symmetry can be broken by the pump field,
it is highly required to check whether or not 
the main term of $\sigma_{yx}^{\textrm{C}}$ is given by the antisymmetric part
in discussing the AHE.
It is also necessary to check the main term of $\sigma_{yx}^{\textrm{S}}$
to discuss the SHE
because in some cases such as the cases with BCPL 
its main term is not restricted by the Onsager reciprocal relation
to either the antisymmetric or symmetric part.
These suggestions 
are useful for future studies of the AHE and SHE.

This paper will stimulate many future studies of transport phenomena
in periodically driven systems.
First,
our results allow the experimental detection of the spin current in periodically driven systems
via the inverse SHE.
The impact of our paper is not restricted to the SHE,
but it will encourage future studies of other spintronics phenomena
such as the spin Seebeck effect~\cite{Saitoh-SSE,Saitoh-SSE2,Bauer,Adachi}
in periodically driven systems.
This is because the inverse SHE is often used to convert the spin current
into an electrical signal. 
Then,
our results will provide a useful guideline
when studying
the Onsager reciprocal relations for
the time-averaged charge and spin off-diagonal dc conductivities
in the systems driven by CPL, LPL, or BCPL.
Moreover,
our general arguments and theory can be extended to
the other transport coefficients including the ac conductivities
in periodically driven systems.
The extension to the transport coefficients in the non-linear regime
is also an important future study. 

\begin{acknowledgments}
  This work was supported by
  JST CREST Grant No. JPMJCR1901, 
  JSPS KAKENHI Grant No. JP22K03532,
  and MEXT Q-LEAP Grant No. JP-MXS0118067426.
\end{acknowledgments}

\begin{widetext}
\appendix

\section{Derivation of Eq. (\ref{eq:sigyx^Q-w})}

We derive Eq. (\ref{eq:sigyx^Q-w}).
Since we have explained this derivation in Ref. [\cite{NA-FloquetSHE}],
we here describe its main points.
Substituting Eq. (\ref{eq:sigStt'}) or (\ref{eq:sigCtt'})
into Eq. (\ref{eq:t-av-sig-w}),
we obtain
\begin{align}
  %No. XXXIV 11/8 No. 04
  \sigma_{yx}^{\textrm{Q}}(\omega)
  =\textrm{Re}[\sigma_{yx}^{\textrm{Q}(1)}(\omega)+\sigma_{yx}^{\textrm{Q}(2)}(\omega)],
  \label{eq:sig-w-start}
\end{align}
where
\begin{align}
  %No. XXXIV 11/8 No. 04
  &\sigma_{yx}^{\textrm{Q}(1)}(\omega)
  =\int_{0}^{T_{\textrm{p}}}\frac{dt_{\textrm{av}}}{T_{\textrm{p}}}
  \int_{-\infty}^{\infty}dt_{\textrm{rel}}e^{i\omega t_{\textrm{rel}}}
  \sigma_{yx}^{\textrm{Q}(1)}
  (t_{\textrm{av}}+\frac{t_{\textrm{rel}}}{2},t_{\textrm{av}}-\frac{t_{\textrm{rel}}}{2}),\label{eq:t-av-sig1-w}\\
  &\sigma_{yx}^{\textrm{Q}(2)}(\omega)
  =\int_{0}^{T_{\textrm{p}}}\frac{dt_{\textrm{av}}}{T_{\textrm{p}}}
  \int_{-\infty}^{\infty}dt_{\textrm{rel}}e^{i\omega t_{\textrm{rel}}}
  \sigma_{yx}^{\textrm{Q}(2)}
  (t_{\textrm{av}}+\frac{t_{\textrm{rel}}}{2},t_{\textrm{av}}-\frac{t_{\textrm{rel}}}{2}).\label{eq:t-av-sig2-w}
\end{align}
We calculate the right-hand sides of
Eqs. (\ref{eq:t-av-sig1-w}) and (\ref{eq:t-av-sig2-w})
using
Eqs. (\ref{eq:sigDM-start}) and (\ref{eq:sigPM-rewrite}), 
the Floquet representation of the Green's functions, 
and two relations.
One of the two relations is that
the Floquet representation of 
a function $A(t,t^{\prime})=\int dt^{\prime\prime}B(t,t^{\prime\prime})C(t^{\prime\prime},t^{\prime})$
is given by
\begin{align}
  %No. XXXIV 11/5 No. 02
  [A(\omega)]_{mn}=\sum_{l=-\infty}^{\infty}[B(\omega)]_{ml}[C(\omega)]_{ln}. 
\end{align}
The other is that 
a product $a(t)D(t,t^{\prime})$ is expressed in the Floquet representation as 
\begin{align}
  %No. XXXIV 11/10 No. 01-02
  [aD(\omega)]_{mn}=\sum_{l=-\infty}^{\infty}[a]_{ml}[D(\omega)]_{ln}.
\end{align}
By substituting Eqs. (\ref{eq:sigDM-start}) and (\ref{eq:sigPM-rewrite})
into Eqs. (\ref{eq:t-av-sig1-w}) and (\ref{eq:t-av-sig2-w}), respectively, 
and 
using the Floquet representation and these relations,
we obtain
\begin{align}
  %No. XXXV 11/10 No. 11-14
  \sigma_{yx}^{\textrm{Q}(1)}(\omega)
  &=\int_{0}^{T_{\textrm{p}}}\frac{dt_{\textrm{av}}}{T_{\textrm{p}}}
  \int_{-\infty}^{\infty}dt_{\textrm{rel}}e^{i\omega t_{\textrm{rel}}}
  \sigma_{yx}^{\textrm{Q}(1)}
  (t_{\textrm{av}}+\frac{t_{\textrm{rel}}}{2},t_{\textrm{av}}-\frac{t_{\textrm{rel}}}{2})\notag\\
  &=-\frac{1}{\omega N}\sum_{\mathbf{k}}\sum_{a,b}\sum_{\sigma}
  \int_{-\Omega/2}^{\Omega/2}\frac{d\omega^{\prime}}{2\pi}
  \sum_{m,l=-\infty}^{\infty}
  [M_{ab\sigma}^{(\textrm{Q})yx}(\mathbf{k})]_{ml}
  [G_{b\sigma a\sigma}^{<}(\mathbf{k},\omega^{\prime})]_{lm}\notag\\
  %No. XXXV 11/10 No. 11-14
  &=-\frac{1}{\omega N}\sum_{\mathbf{k}}\sum_{a,b}\sum_{\sigma}
  \int_{-\Omega/2}^{\Omega/2}\frac{d\omega^{\prime}}{2\pi}
  \textrm{tr}[M_{ab\sigma}^{(\textrm{Q})yx}(\mathbf{k})
    G_{b\sigma a\sigma}^{<}(\mathbf{k},\omega^{\prime})],\label{eq:sig^1-w}\\
  \sigma_{yx}^{\textrm{Q}(2)}(\omega)
  &=\int_{0}^{T_{\textrm{p}}}\frac{dt_{\textrm{av}}}{T_{\textrm{p}}}
  \int_{-\infty}^{\infty}dt_{\textrm{rel}}e^{i\omega t_{\textrm{rel}}}
  \sigma_{yx}^{\textrm{Q}(2)}
  (t_{\textrm{av}}+\frac{t_{\textrm{rel}}}{2},t_{\textrm{av}}-\frac{t_{\textrm{rel}}}{2})\notag\\
  &=\frac{1}{\omega N}\sum_{\mathbf{k}}\sum_{a,b,c,d}\sum_{\sigma,\sigma^{\prime}}
  \int_{-\Omega/2}^{\Omega/2}\frac{d\omega^{\prime}}{2\pi}
  \sum_{m,n,l,q=-\infty}^{\infty}\notag\\
  &\times 
  \{[v_{ab\sigma}^{(\textrm{Q})y}(\mathbf{k})]_{ml}
  [G_{b\sigma c\sigma^{\prime}}^{\textrm{R}}(\mathbf{k},\omega^{\prime}+\omega)]_{ln}    
  [v_{cd\sigma^{\prime}}^{(\textrm{C})x}(\mathbf{k})]_{nq}
  [G_{d\sigma^{\prime}a\sigma}^{<}(\mathbf{k},\omega^{\prime})]_{qm}\notag\\
  &+[v_{ab\sigma}^{(\textrm{Q})y}(\mathbf{k})]_{ml}
  [G_{b\sigma c\sigma^{\prime}}^{<}(\mathbf{k},\omega^{\prime})]_{ln}    
  [v_{cd\sigma^{\prime}}^{(\textrm{C})x}(\mathbf{k})]_{nq}
  [G_{d\sigma^{\prime}a\sigma}^{\textrm{A}}(\mathbf{k},\omega^{\prime}-\omega)]_{qm}
  \}\notag\\
  %No. XXXV 11/10 No. 11-14
  &=\frac{1}{\omega N}\sum_{\mathbf{k}}\sum_{a,b,c,d}\sum_{\sigma,\sigma^{\prime}}
  \int_{-\Omega/2}^{\Omega/2}\frac{d\omega^{\prime}}{2\pi}
  \{\textrm{tr}[v_{ab\sigma}^{(\textrm{Q})y}(\mathbf{k})
    G_{b\sigma c\sigma^{\prime}}^{\textrm{R}}(\mathbf{k},\omega^{\prime}+\omega)
    v_{cd\sigma^{\prime}}^{(\textrm{C})x}(\mathbf{k})
    G_{d\sigma^{\prime}a\sigma}^{<}(\mathbf{k},\omega^{\prime})]\notag\\
  &\ \ \ \ \ \ \ \ \ \ \ \ \ \ \ \ \ \ \ \ \ \ \ \ \ \ \ \ \ \ \ \ \ \ \ \ \ \ \ \
  +\textrm{tr}[v_{ab\sigma}^{(\textrm{Q})y}(\mathbf{k})
    G_{b\sigma c\sigma^{\prime}}^{<}(\mathbf{k},\omega^{\prime})   
    v_{cd\sigma^{\prime}}^{(\textrm{C})x}(\mathbf{k})
    G_{d\sigma^{\prime}a\sigma}^{\textrm{A}}(\mathbf{k},\omega^{\prime}-\omega)]
  \},\label{eq:sig^2-w}
\end{align}
where 
\begin{align}
  %No. XXXV 11/10 No. 11-12
  [M_{ab\sigma}^{(\textrm{Q})yx}(\mathbf{k})]_{ml}
  &=\int_{0}^{T_{\textrm{p}}}\frac{dt}{T_{\textrm{p}}}e^{i(m-l)\Omega t}
  \frac{\delta v_{ab\sigma}^{(\textrm{Q})y}(\mathbf{k},t)}{\delta A_{\textrm{prob}}^{x}(t)}.
\end{align}
Since $\sigma_{yx}^{\textrm{Q}(1)}(\omega)$ has only a pure imaginary part~\cite{Eckstein},
we have
\begin{align}
  %No. XXXV 11/10 No. 13-14; No. XXXIV 11/12 No. 05-06
  \sigma_{yx}^{\textrm{Q}}(\omega)
  &=\frac{1}{\omega N}\sum_{\mathbf{k}}\sum_{a,b,c,d}\sum_{\sigma,\sigma^{\prime}}
  \int_{-\Omega/2}^{\Omega/2}\frac{d\omega^{\prime}}{2\pi}
  \textrm{Re}
  \{\textrm{tr}[v_{ab\sigma}^{(\textrm{Q})y}(\mathbf{k})
    G_{b\sigma c\sigma^{\prime}}^{\textrm{R}}(\mathbf{k},\omega^{\prime}+\omega)
    v_{cd\sigma^{\prime}}^{(\textrm{C})x}(\mathbf{k})
    G_{d\sigma^{\prime}a\sigma}^{<}(\mathbf{k},\omega^{\prime})]\notag\\
  &\ \ \ \ \ \ \ \ \ \ \ \ \ \ \ \ \ \ \ \ \ \ \ \ \ \ \ \ \ \ \ \ \ \ \ \ \ \ \ \ \ \ \ \
  +\textrm{tr}[v_{ab\sigma}^{(\textrm{Q})y}(\mathbf{k})
    G_{b\sigma c\sigma^{\prime}}^{<}(\mathbf{k},\omega^{\prime})   
    v_{cd\sigma^{\prime}}^{(\textrm{C})x}(\mathbf{k})
    G_{d\sigma^{\prime}a\sigma}^{\textrm{A}}(\mathbf{k},\omega^{\prime}-\omega)]
  \}.\label{eq:sig-w}
\end{align}
Moreover, we can rewrite this expression
by using the identity,
\begin{align}
  %XXXVI 11/30 No. 01-02
  \textrm{Re}[\textrm{tr}(AB)]=\frac{1}{2}\{\textrm{tr}(AB)+\textrm{tr}[(AB)^{\dagger}]\},
  \label{eq:AB}
\end{align}
and the symmetry relations of quantities in the Floquet representation,
\begin{align}
%No. XXXVI 11/30 No. 01-02
  &[G_{d\sigma^{\prime}a\sigma}^{<}(\mathbf{k},\omega^{\prime})^{\dagger}]_{ml}
  =-[G_{a\sigma d\sigma^{\prime}}^{<}(\mathbf{k},\omega^{\prime})]_{ml},\label{eq:sym-G^l}\\
  &[v_{cd\sigma^{\prime}}^{(\textrm{C})x}(\mathbf{k})^{\dagger}]_{ln}
  =[v_{dc\sigma^{\prime}}^{(\textrm{C})x}(\mathbf{k})]_{ln},\label{eq:sym-v^Cx}\\
  &[G_{b\sigma c\sigma^{\prime}}^{\textrm{R}}(\mathbf{k},\omega^{\prime}+\omega)^{\dagger}]_{nq}
  =[G_{c\sigma^{\prime}b\sigma}^{\textrm{A}}(\mathbf{k},\omega^{\prime}+\omega)]_{nq},\label{eq:sym-G^R}\\
  &[v_{ab\sigma}^{(\textrm{Q})y}(\mathbf{k})^{\dagger}]_{qm}
  =[v_{ba\sigma}^{(\textrm{Q})y}(\mathbf{k})]_{qm}.\label{eq:sym-v^Qy}
\end{align}
Note that Eqs. (\ref{eq:sym-G^l}) and (\ref{eq:sym-G^R}) are obtained by using
\begin{align}
  %No. XXXVI 11/30 No. 01-02
  G_{d\sigma^{\prime}a\sigma}^{<}
  (\mathbf{k};t_{\textrm{av}}+\frac{t_{\textrm{rel}}}{2},t_{\textrm{av}}-\frac{t_{\textrm{rel}}}{2})^{\dagger}
  =-G_{a\sigma d\sigma^{\prime}}^{<}
  (\mathbf{k};t_{\textrm{av}}-\frac{t_{\textrm{rel}}}{2},t_{\textrm{av}}+\frac{t_{\textrm{rel}}}{2}),
\end{align}
and
\begin{align}
  %No. XXXVI 11/30 No. 01-02
  G_{b\sigma c\sigma^{\prime}}^{\textrm{R}}
  (\mathbf{k};t_{\textrm{av}}+\frac{t_{\textrm{rel}}}{2},t_{\textrm{av}}-\frac{t_{\textrm{rel}}}{2})^{\dagger}
  =G_{c\sigma^{\prime}b\sigma}^{\textrm{A}}
  (\mathbf{k};t_{\textrm{av}}-\frac{t_{\textrm{rel}}}{2},t_{\textrm{av}}+\frac{t_{\textrm{rel}}}{2}),
\end{align}
respectively.
Using Eqs. (\ref{eq:AB}){--}(\ref{eq:sym-v^Qy}),
we obtain
\begin{align}
  %XXXVI 11/30 No. 01-02; No. XXXVII 11/30 No/ 01-02
  &\frac{1}{\omega}\textrm{Re}\{
  \textrm{tr}[v_{ab\sigma}^{(\textrm{Q})y}(\mathbf{k})
    G_{b\sigma c\sigma^{\prime}}^{\textrm{R}}(\mathbf{k},\omega^{\prime}+\omega)
    v_{cd\sigma^{\prime}}^{(\textrm{C})x}(\mathbf{k})
    G_{d\sigma^{\prime}a\sigma}^{<}(\mathbf{k},\omega^{\prime})]\}\notag\\
  =&\frac{1}{2\omega}
  \Bigl\{
  \textrm{tr}[v_{ab\sigma}^{(\textrm{Q})y}(\mathbf{k})
    G_{b\sigma c\sigma^{\prime}}^{\textrm{R}}(\mathbf{k},\omega^{\prime}+\omega)
    v_{cd\sigma^{\prime}}^{(\textrm{C})x}(\mathbf{k})
    G_{d\sigma^{\prime}a\sigma}^{<}(\mathbf{k},\omega^{\prime})]\notag\\
  &\ \ \ \ \ 
  -\textrm{tr}[G_{a\sigma d\sigma^{\prime}}^{<}(\mathbf{k},\omega^{\prime})
    v_{dc\sigma^{\prime}}^{(\textrm{C})x}(\mathbf{k})
    G_{ c\sigma^{\prime} b\sigma}^{\textrm{A}}(\mathbf{k},\omega^{\prime}+\omega)
    v_{ba\sigma}^{(\textrm{Q})y}(\mathbf{k})]
  \Bigr\},\label{eq:rewrite1}
\end{align}
and
\begin{align}
  %No. XXXVII 11/30 No. 01-02
  &\frac{1}{\omega}
  \textrm{Re}\{\textrm{tr}[v_{ab\sigma}^{(\textrm{Q})y}(\mathbf{k})
    G_{b\sigma c\sigma^{\prime}}^{<}(\mathbf{k},\omega^{\prime})   
    v_{cd\sigma^{\prime}}^{(\textrm{C})x}(\mathbf{k})
    G_{d\sigma^{\prime}a\sigma}^{\textrm{A}}(\mathbf{k},\omega^{\prime}-\omega)]\}\notag\\
  =&\frac{1}{2\omega}
  \Bigl\{
  \textrm{tr}[v_{ab\sigma}^{(\textrm{Q})y}(\mathbf{k})
    G_{b\sigma c\sigma^{\prime}}^{<}(\mathbf{k},\omega^{\prime})   
    v_{cd\sigma^{\prime}}^{(\textrm{C})x}(\mathbf{k})
    G_{d\sigma^{\prime}a\sigma}^{\textrm{A}}(\mathbf{k},\omega^{\prime}-\omega)]\notag\\
  &\ \ \ \ \ 
  -\textrm{tr}[G_{a\sigma d\sigma^{\prime}}^{\textrm{R}}(\mathbf{k},\omega^{\prime}-\omega)
    v_{dc\sigma^{\prime}}^{(\textrm{C})x}(\mathbf{k})
    G_{c\sigma^{\prime}b\sigma}^{<}(\mathbf{k},\omega^{\prime})   
    v_{ba\sigma}^{(\textrm{Q})y}(\mathbf{k})]
  \Bigr\}.\label{eq:rewrite2}
\end{align}
Substituting Eqs. (\ref{eq:rewrite1}) and (\ref{eq:rewrite2}) into Eq. (\ref{eq:sig-w}),
we obtain Eq. (\ref{eq:sigyx^Q-w}).

\section{$[\epsilon_{ab}(\mathbf{k})]_{mn}$ of Sr$_{2}$RuO$_{4}$ driven by BCPL}

We can calculate $[\epsilon_{ab}(\mathbf{k})]_{mn}$ of Sr$_{2}$RuO$_{4}$ driven by BCPL
by using Eqs. (\ref{eq:Ax-BCPL}) and (\ref{eq:Ay-BCPL}) and Eq. (\ref{eq:epsilon_mn}).
To do this,
we use identities,
\begin{align}
  %No. XXXVIII 8/30 No. 01
  &e^{iu\sin(\Omega t+\theta)}
  =\sum_{l=-\infty}^{\infty}\mathcal{J}_{l}(u)e^{il(\Omega t+\theta)},\\
  &e^{iu\cos(\Omega t+\theta)}
  =\sum_{l=-\infty}^{\infty}i^{l}\mathcal{J}_{l}(u)e^{il(\Omega t+\theta)},\\
  %No. XXXVI 12/30 No. 02
  &\mathcal{J}_{l}(u)=(-1)^{l}\mathcal{J}_{l}(-u)
  =(-1)^{l}\mathcal{J}_{-l}(u),
\end{align}
where $\mathcal{J}_{l}(u)$ is the Bessel function of the first kind,
and $l$ is its order. 
After some calculation,
we obtain
\begin{align}
  %No. XXXVIII 8/30 No. 02-05, 8/31 No. 01-04
  [\epsilon_{d_{yz}d_{yz}}(\mathbf{k})]_{mn}
  &=(-t_{2})\sum_{l=-\infty}^{\infty}i^{n-m-(\beta-1)l}e^{-il\theta}
  \mathcal{J}_{n-m-\beta l}(u)\mathcal{J}_{l}(u)
          [e^{-ik_{x}}(-1)^{n-m-(\beta-1)l}+e^{ik_{x}}]\notag\\
  &+(-t_{1})\sum_{l=-\infty}^{\infty}e^{-il\theta}(-1)^{l}
  \mathcal{J}_{n-m-\beta l}(u)\mathcal{J}_{l}(u)
          [e^{-ik_{y}}(-1)^{n-m-(\beta-1)l}+e^{ik_{y}}],\label{eq:e_yzyz_mn}\\
  [\epsilon_{d_{zx}d_{zx}}(\mathbf{k})]_{mn}
  &=(-t_{1})\sum_{l=-\infty}^{\infty}i^{n-m-(\beta-1)l}e^{-il\theta}
  \mathcal{J}_{n-m-\beta l}(u)\mathcal{J}_{l}(u)
          [e^{-ik_{x}}(-1)^{n-m-(\beta-1)l}+e^{ik_{x}}]\notag\\
  &+(-t_{2})\sum_{l=-\infty}^{\infty}e^{-il\theta}(-1)^{l}
  \mathcal{J}_{n-m-\beta l}(u)\mathcal{J}_{l}(u)
          [e^{-ik_{y}}(-1)^{n-m-(\beta-1)l}+e^{ik_{y}}],\label{eq:e_zxzx_mn}\\
  [\epsilon_{d_{yz}d_{zx}}(\mathbf{k})]_{mn}
  &=[\epsilon_{d_{zx}d_{yz}}(\mathbf{k})]_{mn}
  =t_{5}\sum_{l,l^{\prime},l^{\prime\prime}=-\infty}^{\infty}i^{l+l^{\prime}}e^{-i(l^{\prime}+l^{\prime\prime})\theta}
  \mathcal{J}_{l}(u)\mathcal{J}_{l^{\prime}}(u)
  \mathcal{J}_{n-m-l-(l^{\prime}+l^{\prime\prime})\beta}(u)\mathcal{J}_{l^{\prime\prime}}(u)\notag\\
  \times &
  [e^{-ik_{x}}e^{-ik_{y}}(-1)^{n-m-(l^{\prime}+l^{\prime\prime})\beta+l^{\prime}}
    +e^{ik_{x}}e^{ik_{y}}(-1)^{l^{\prime\prime}}\notag\\
    &-e^{-ik_{x}}e^{ik_{y}}(-1)^{l+l^{\prime}+l^{\prime\prime}}
  -e^{ik_{x}}e^{-ik_{y}}(-1)^{n-m-l-(l^{\prime}+l^{\prime\prime})\beta}],\label{eq:e_yzzx_mn}\\
  [\epsilon_{d_{xy}d_{xy}}(\mathbf{k})]_{mn}
  &=(-t_{3})\sum_{l=-\infty}^{\infty}e^{-il\theta}
  \mathcal{J}_{n-m-\beta l}(u)\mathcal{J}_{l}(u)
          [i^{n-m-(\beta-1)l}e^{-ik_{x}}(-1)^{n-m-(\beta-1)l}
          +i^{n-m-(\beta-1)l}e^{ik_{x}}\notag\\
          &\ \ \ \ \ \ \ \ \ \ \ \ \ \ \ \ \ \ \ \ \ \ \ \ \ \ \ \ \ \ \ \ \ \ \ \ \ \ \ 
          \ \ \ \ \ \ \ \
          +e^{-ik_{y}}(-1)^{n-m-\beta l}
          +e^{ik_{y}}(-1)^{l}]\notag\\
  &+(-t_{4})\sum_{l,l^{\prime},l^{\prime\prime}=-\infty}^{\infty}
   i^{l+l^{\prime}}e^{-i(l^{\prime}+l^{\prime\prime})\theta}
   \mathcal{J}_{l}(u)\mathcal{J}_{l^{\prime}}(u)
   \mathcal{J}_{n-m-l-(l^{\prime}+l^{\prime\prime})\beta}(u)\mathcal{J}_{l^{\prime\prime}}(u)\notag\\
  \times &
  [e^{-ik_{x}}e^{-ik_{y}}(-1)^{n-m-(l^{\prime}+l^{\prime\prime})\beta+l^{\prime}}
  +e^{ik_{x}}e^{ik_{y}}(-1)^{l^{\prime\prime}}
  +e^{-ik_{x}}e^{ik_{y}}(-1)^{l+l^{\prime}+l^{\prime\prime}}\notag\\
  &
  +e^{ik_{x}}e^{-ik_{y}}(-1)^{n-m-l-(l^{\prime}+l^{\prime\prime})\beta}],\label{eq:e_xyxy_mn}
\end{align}
where
\begin{align}
  %No. XXXIX 8/30 No. 01-02
  u=ea_{\textrm{NN}}A_{0}=eA_{0}.\label{eq:Floquet-para}
\end{align}
As described in Sect. 3,
$t_{1}$, $t_{2}$, and $t_{3}$ are the nearest-neighbor hopping integrals
on the square lattice,
and $t_{4}$ and $t_{5}$ are the next-nearest-neighbor ones~\cite{NA-FloquetSHE} [Fig. \ref{fig5}(a)]. 

We make three comments about Eqs. (\ref{eq:e_yzyz_mn}){--}(\ref{eq:e_xyxy_mn}).
First,
the relative phase difference in BCPL, $\theta$, causes
the phase factors, such as $e^{-il\theta}$ and $e^{-i(l^{\prime}+l^{\prime\prime})\theta}$,
only for the terms of the finite-order Bessel functions. 
Second,
the number of the Bessel functions
is twice that in the case of CPL~\cite{NA-FloquetSHE}.
This is because the $x$ or $y$ component of the pump field
contains two trigonometric functions in the case of BCPL,
whereas it contains one in the case of CPL~\cite{NA-FloquetSHE}.
Third,
even if the Floquet indices $m$, $n$, $l$, $l^{\prime}$, and $l^{\prime\prime}$ are restricted to
the range of $-n_{\textrm{max}}\leq m, n, l, l^{\prime}, l^{\prime\prime} \leq n_{\textrm{max}}$,
$[\epsilon_{ab}(\mathbf{k})]_{mn}$'s include 
not only the Bessel functions the order of which is within this range,
but also the Bessel functions the order of which is outside of it.
For example,
if $\beta=2$ and $-1\leq m, n, l \leq 1$ (i.e., $n_{\textrm{max}}=1$),
the order of the first Bessel function in Eq. (\ref{eq:e_yzyz_mn}), $\mathcal{J}_{n-m-\beta l}(u)$,
takes the value outside of that range.
Since the third property is also related to
the number of the trigonometric functions appearing in the Peierls phase factors,
this suggests that
the high-order Bessel functions play a more important role in the case of BCPL
than in that of CPL. 
In fact, this is consistent with the difference between 
the $n_{\textrm{max}}$ dependences of $\sigma_{yx}^{\textrm{S}}$
obtained for BCPL and CPL.

\section{$[\bar{\epsilon}_{ab}(\mathbf{k})]_{mn}$ of Sr$_{2}$RuO$_{4}$ driven by BCPL}

We can calculate $[\bar{\epsilon}_{ab}(\mathbf{k})]_{mn}$ of Sr$_{2}$RuO$_{4}$ driven by BCPL
in a similar way to the calculation explained in Appendix B.
After such calculation
using Eqs. (\ref{eq:Ax-BCPL_-t}), (\ref{eq:Ay-BCPL_-t}), and (\ref{eq:bar-epsilon_mn}),
we obtain
\begin{align}
  %No. XXXXIII 9/13 No. 01-06
  [\bar{\epsilon}_{d_{yz}d_{yz}}(\mathbf{k})]_{mn}
  &=(-t_{2})\sum_{l=-\infty}^{\infty}i^{n-m-(\beta-1)l}e^{il\theta}
  \mathcal{J}_{n-m-\beta l}(u)\mathcal{J}_{l}(u)
          [e^{-ik_{x}}(-1)^{n-m-(\beta-1)l}+e^{ik_{x}}]\notag\\
  &+(-t_{1})\sum_{l=-\infty}^{\infty}e^{il\theta}(-1)^{n-m-\beta l}
  \mathcal{J}_{n-m-\beta l}(u)\mathcal{J}_{l}(u)
          [e^{-ik_{y}}(-1)^{n-m-(\beta-1)l}+e^{ik_{y}}],\label{eq:bar-e_yzyz_mn}\\
  [\bar{\epsilon}_{d_{zx}d_{zx}}(\mathbf{k})]_{mn}
  &=(-t_{1})\sum_{l=-\infty}^{\infty}i^{n-m-(\beta-1)l}e^{il\theta}
  \mathcal{J}_{n-m-\beta l}(u)\mathcal{J}_{l}(u)
          [e^{-ik_{x}}(-1)^{n-m-(\beta-1)l}+e^{ik_{x}}]\notag\\
  &+(-t_{2})\sum_{l=-\infty}^{\infty}e^{il\theta}(-1)^{n-m-\beta l}
  \mathcal{J}_{n-m-\beta l}(u)\mathcal{J}_{l}(u)
          [e^{-ik_{y}}(-1)^{n-m-(\beta-1)l}+e^{ik_{y}}],\label{eq:bar-e_zxzx_mn}\\
  [\bar{\epsilon}_{d_{yz}d_{zx}}(\mathbf{k})]_{mn}
  &=[\bar{\epsilon}_{d_{zx}d_{yz}}(\mathbf{k})]_{mn}
  =t_{5}\sum_{l,l^{\prime},l^{\prime\prime}=-\infty}^{\infty}i^{l+l^{\prime}}e^{i(l^{\prime}+l^{\prime\prime})\theta}
  \mathcal{J}_{l}(u)\mathcal{J}_{l^{\prime}}(u)
  \mathcal{J}_{n-m-l-(l^{\prime}+l^{\prime\prime})\beta}(u)\mathcal{J}_{l^{\prime\prime}}(u)\notag\\
  \times &
  [e^{-ik_{x}}e^{-ik_{y}}(-1)^{l+l^{\prime}+l^{\prime\prime}}
  +e^{ik_{x}}e^{ik_{y}}(-1)^{n-m-l-(l^{\prime}+l^{\prime\prime})\beta}
  -e^{-ik_{x}}e^{ik_{y}}(-1)^{n-m-(l^{\prime}+l^{\prime\prime})\beta+l^{\prime}}\notag\\
  &-e^{ik_{x}}e^{-ik_{y}}(-1)^{l^{\prime\prime}}],\label{eq:bar-e_yzzx_mn}\\
  [\bar{\epsilon}_{d_{xy}d_{xy}}(\mathbf{k})]_{mn}
  &=(-t_{3})\sum_{l=-\infty}^{\infty}e^{il\theta}
  \mathcal{J}_{n-m-\beta l}(u)\mathcal{J}_{l}(u)
          [i^{n-m-(\beta-1)l}e^{-ik_{x}}(-1)^{n-m-(\beta-1)l}
          +i^{n-m-(\beta-1)l}e^{ik_{x}}\notag\\
          &\ \ \ \ \ \ \ \ \ \ \ \ \ \ \ \ \ \ \ \ \ \ \ \ \ \ \ \ \ \ \ \ \ \ \ \ \ \ \ 
          \ \ \ \ \ \ \ \
          +e^{-ik_{y}}(-1)^{l}
          +e^{ik_{y}}(-1)^{n-m-\beta l}]\notag\\
  &+(-t_{4})\sum_{l,l^{\prime},l^{\prime\prime}=-\infty}^{\infty}
   i^{l+l^{\prime}}e^{i(l^{\prime}+l^{\prime\prime})\theta}
   \mathcal{J}_{l}(u)\mathcal{J}_{l^{\prime}}(u)
   \mathcal{J}_{n-m-l-(l^{\prime}+l^{\prime\prime})\beta}(u)\mathcal{J}_{l^{\prime\prime}}(u)\notag\\
  \times &
  [e^{-ik_{x}}e^{-ik_{y}}(-1)^{l+l^{\prime}+l^{\prime\prime}}
  +e^{ik_{x}}e^{ik_{y}}(-1)^{n-m-l-(l^{\prime}+l^{\prime\prime})\beta}
  +e^{-ik_{x}}e^{ik_{y}}(-1)^{n-m-(l^{\prime}+l^{\prime\prime})\beta+l^{\prime}}\notag\\
  &+e^{ik_{x}}e^{-ik_{y}}(-1)^{l^{\prime\prime}}].\label{eq:bar-e_xyxy_mn}
\end{align}
These equations and Eqs. (\ref{eq:e_yzyz_mn}){--}(\ref{eq:e_xyxy_mn}) show that
the differences between $[\bar{\epsilon}_{ab}(\mathbf{k})]_{mn}$'s
and $[\epsilon_{ab}(\mathbf{k})]_{mn}$'s
are the differences in the phase factors due to $\theta$
and the sign factors such as $(-1)^{l}$ and $(-1)^{n-m-\beta l}$.
These differences are reasonable because
$\mathbf{A}_{\textrm{BCPL}}(-t)$ and $\mathbf{A}_{\textrm{BCPL}}(t)$
are given by Eqs. (\ref{eq:Ax-BCPL_-t}) and (\ref{eq:Ay-BCPL_-t})
and by Eqs. (\ref{eq:Ax-BCPL}) and (\ref{eq:Ay-BCPL}), respectively.

\section{Additional numerical results}

We show additional numerical results
to discuss the validity of our choice of the value of $n_{\textrm{max}}$.
Figures \ref{fig18}(a) and \ref{fig18}(b)
show the $n_{\textrm{max}}$ dependences of $\sigma_{yx}^{\textrm{S}}$ and $\sigma_{yx}^{\textrm{C}}$
for Sr$_{2}$RuO$_{4}$ driven by BCPL at $\beta=2$, $\theta=\frac{\pi}{4}$, and $\Omega=6$ eV.
In obtaining these results,
we chose some parameters to be different from the values
used in the numerical results shown in Sect. 5:
we set $N_{x}=N_{y}=64$ and $\Delta\omega^{\prime}=0.01$ eV
to reduce the cost of the numerical calculations.
These figures show that
the results obtained for $n_{\textrm{max}}=1$ and $2$ are qualitatively the same.
Therefore, 
$n_{\textrm{max}}=1$ may be reasonable to study qualitative properties of
$\sigma_{yx}^{\textrm{S}}$ and $\sigma_{yx}^{\textrm{C}}$
for Sr$_{2}$RuO$_{4}$ driven by BCPL at $\Omega=6$ eV.

\begin{figure}
  \begin{center}
    \includegraphics[width=120mm]{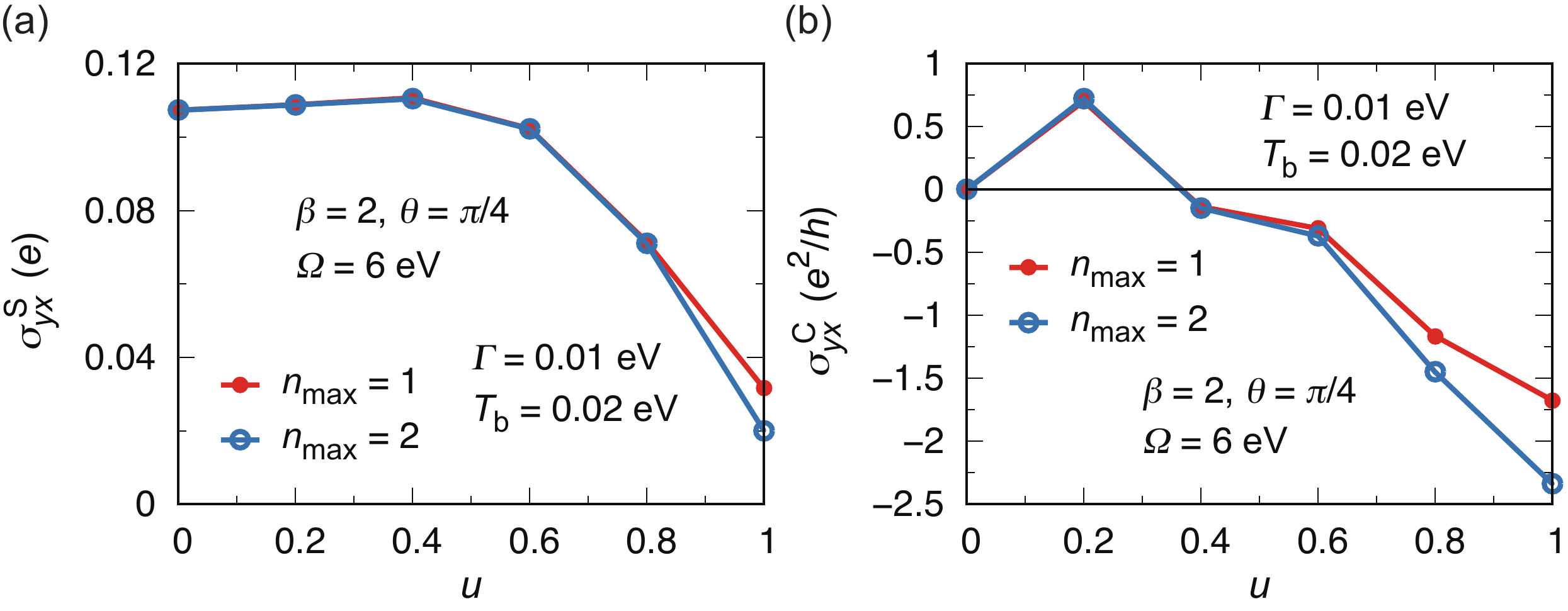}
    \end{center}
  \caption{\label{fig18}
    (Color online) The $n_{\textrm{max}}$ dependences of
    (a) $\sigma_{yx}^{\textrm{S}}$ and (b) $\sigma_{yx}^{\textrm{C}}$
    as functions of $u(=eA_{0})$
    for Sr$_{2}$RuO$_{4}$ driven by BCPL at $\beta=2$, $\theta=\frac{\pi}{4}$, and $\Omega=6$ eV.
    Here $n_{\textrm{max}}$ is the upper limit of the summation over the Floquet indices.
    The values of $N_{x}=N_{y}$ and $\Delta\omega^{\prime}$
    used in their numerical calculations 
    are different from
    those used in the numerical calculations for the results shown in Sect. 5:
    $N_{x}=N_{y}=64$ and $\Delta\omega^{\prime}=0.01$ eV were used
    in the former calculations,
    whereas $N_{x}=N_{y}=100$ and $\Delta\omega^{\prime}=0.005$ eV
    were used in the latter calculations.
    Meanwhile, the values of $\Gamma$ and $T_{\textrm{b}}$ are the same.
  }
\end{figure}
\end{widetext}

\end{document}